\documentclass[3p]{elsarticle}

\usepackage[utf8]{inputenc}
\usepackage{a4wide}
\usepackage{upgreek}
\usepackage{graphicx}
\usepackage{amssymb}
\usepackage{amsmath}
\usepackage{stmaryrd}
\usepackage{mathtools}
\usepackage{ifthen}
\usepackage{mathrsfs}
\usepackage{bm}
\usepackage{bbm}
\usepackage{floatrow}
\usepackage{subfig}
\usepackage{caption}
\usepackage{epsfig}
\usepackage{amsfonts}
\usepackage{framed}
\usepackage{color}
\usepackage{xcolor}
\usepackage{enumitem}
\usepackage{mathalfa}
\usepackage{algorithm}
\usepackage{algpseudocode}
\usepackage{soul}

\newcommand{\bs}[1]{\boldsymbol{#1}}
\usepackage[toc,page]{appendix}

\usepackage{tabularx}

\newcommand{\realset}{\mathbb{R}}

\newcommand{\vsp}{\bs{\mathcal{V}}}

\newcommand{\dsp}{\bs{\mathcal{D}}}
\newcommand{\psp}{\mathcal{W}}

\newcommand{\elemset}{\mathcal{T}}

\newcommand{\fS}{\text{s}}
\newcommand{\fSr}{\tilde{\text{s}}}
\newcommand{\fF}{\text{f}}

\newcommand{\Om}{\mathit{\Omega}}
\newcommand{\Gm}{\mathit{\Gamma}}

\newcommand{\zerom}{\textcolor{lightgray}{\bs{\mathsf{0}}}}

\newcommand{\ROP}{\boldsymbol{\mathsf{r}}}
\newcommand{\bsf}[1]{\boldsymbol{\mathsf{#1}}}

\newcommand{\uf}{\boldsymbol{u}_{\fF}} 
\newcommand{\lmz}{\mathit{\Lambda}} 
\newcommand{\zd}{\boldsymbol{\mathsf{y}}} 
\newcommand{\lmzt}{\delta\lmz} 
\newcommand{\zdt}{\delta\zd} 

\newcommand{\LMZ}{\boldsymbol{\mathsf{\Lambda}}}

\newcommand{\VFr}{\tilde{\boldsymbol{\mathsf{v}}}}

\newcommand{\lmzi}{\mathit{\Lambda}} 

\mathchardef\mhyphen="2D

\begin{document}

\begin{frontmatter}

\title{Effective Block Preconditioners for Fluid Dynamics Coupled to Reduced Models of a Non-Local Nature}

\author[mox,kcl]{Marc Hirschvogel\corref{CA}}
\ead{marc.hirschvogel@polimi.it}

\author[umichbme]{Mia Bonini}
\ead{mbonini@umich.edu}

\author[imperial,kcl]{Maximilian Balmus}
\ead{mbalmus@turing.ac.uk}

\author[umichbme,umichcard]{David Nordsletten} 
\cortext[CA]{Corresponding author} 
\ead{nordslet@umich.edu}

\address[mox]{MOX, Dipartimento di Matematica, Politecnico di Milano, Milan, Italy}
\address[kcl]{Department of Biomedical Engineering, School of Biomedical Engineering \& Imaging Sciences, King's College London, London, United Kingdom}
\address[umichbme]{Department of Biomedical Engineering, University of Michigan, MI, USA}
\address[umichcard]{Department of Cardiac Surgery, University of Michigan, MI, USA}
\address[imperial]{TRIC-DT, The Alan Turing Institute, National Heart and Lung Institute, Imperial College London, London, UK}

\begin{abstract}
Modeling cardiovascular blood flow is central to many applications in biomedical engineering. To accommodate the complexity of the cardiovascular system, in terms of boundary conditions and surrounding vascular tissue, computational fluid dynamics (CFD) often are coupled to reduced circuit and/or solid mechanics models. These allow for realistic simulations of hemodynamics in the heart or the aorta, but come at additional computational cost and complexity. In this contribution, we design a novel block preconditioner for the solution of the stabilized Navier-Stokes equations coupled to reduced-order models of a non-local nature. These models encompass lumped-parameter systems that impose flux-dependent boundary tractions, and Galerkin reduced-order models that can be used to account for outlying mechanical structures. Here we propose a 3\texttimes 3 preconditioner derived from the block factorization and approximation to the Schur complement(s). The solver performance is demonstrated for a series of examples with increasing complexity, culminating in a reduced FSI simulation in a patient-specific contracting left heart model. For all test cases, we show that our proposed approach is superior to other frequently presented 2\texttimes 2 schemes that merge stiffness contributions from reduced models into the fluid Jacobian or consolidate some variables for the purpose of efficiency---with an up to six times shorter overall computing time and/or only half as many linear iterations.
\end{abstract}


\end{frontmatter}

\clearpage
\section{Introduction}\label{sec:introduction}

\begin{figure}[!htp]
\centering
\includegraphics[width=0.9\textwidth]{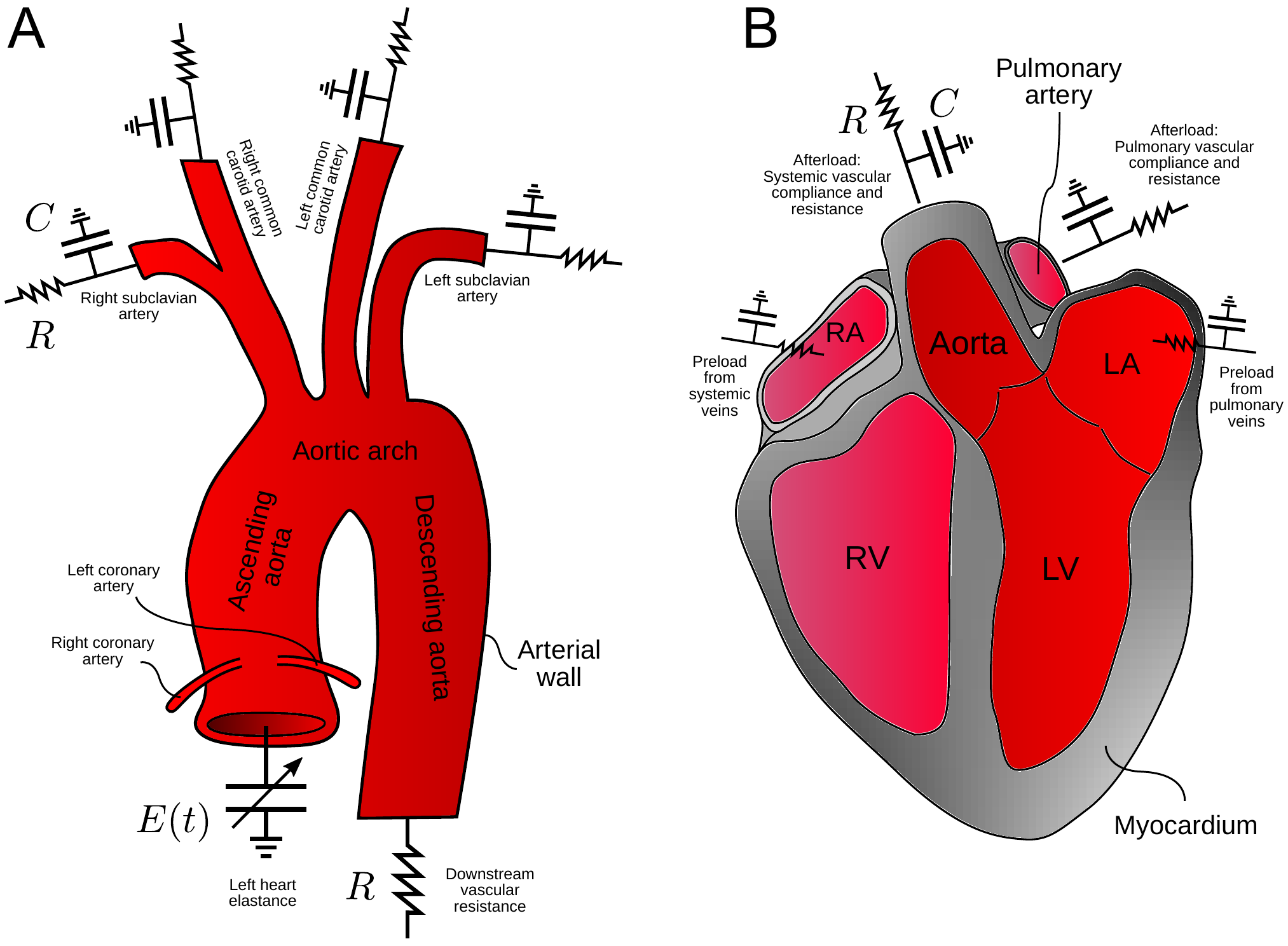}
\caption{Modeling blood flow in the aorta (\textbf{A}) and the heart (\textbf{B}): Applications of 3-dimensional (3D) and reduced-dimensional (RD/0D) model coupling. 3D models describe the fluid dynamics in the aorta or heart cavities, and 0D models account for the up- and downstream vascular architecture represented by elastances ($E$), compliances ($C$), and/or resistances ($R$), imposing boundary conditions of a non-local nature (flux-dependent tractions). The arterial wall may be represented by a RD model of co-dimension 2 \cite{colciago2014}, possibly of non-local nature (Galerkin ROM) when considering the myocardium \cite{hirschvogel2024-frsi}, or by means of 3D FSI approaches \cite{nordsletten2011-fsi,bucelli2023}.}\label{fig:aorta_heart_red_models}
\end{figure}

Predicting hemodynamics in the cardiovascular system by means of advanced computational fluid dynamics (CFD) models represents a central and important task in many biomedical engineering applications \cite{schwarz2023}. 
Blood flow simulations most often involve solving the stabilized Navier-Stokes equations \cite{taylor1998,franca1992}, either in (fixed) Eulerian or (moving) Arbitrary Lagrangian-Eulerian (ALE) reference frames \cite{duarte2004,donea1982}. 
The latter case becomes particularly relevant for modeling hemodynamics in the heart cavities or big elastic vessels like the aorta, where either known motion states are imposed \cite{bonini2022-suppl,zingaro2023} or the elasticity is considered by means of fluid-solid interaction (FSI) methods \cite{nordsletten2011-fsi,mayr2015,bucelli2023}.
Physiologically meaningful hemodynamic modeling additionally requires techniques that account for adjacent portions of the vascular system. 
These components are often represented by 0-dimensional (0D) lumped-parameter models that mimic the down- and upstream vascular architecture \cite{quarteroni2016,quarteroni2001}, or reduced-dimensional (RD) wall/FSI models that describe the elastic and viscous properties of the vessel \cite{figueroa2006,colciago2014,lan2022,hirschvogel2024-frsi}. 
Figure~\ref{fig:aorta_heart_red_models} illustrates an exemplary model of 3D blood flow in the aorta and heart (3D fluid dynamics), where multiple lumped models (0D systems of resistances, compliances, and time-varying elastances) mimic the up- and downstream vascular pressures. Reduced (RD) or fully resolved 3D wall models account for the flexibility of the vessel/myocardium.\\

The coupling between 3D CFD and RD/0D models, however, requires stable nonlinear solution schemes that ideally update all fields simultaneously in order to prevent coupling-related instabilities \cite{kuettler2006}. 
This necessitates efficient iterative solvers for the resulting linearized system of equations, with adequate block preconditioners that accelerate the convergence of stationary algorithms like GMRES \cite{saad1986}. 
Block preconditioners for the (stabilized) Navier-Stokes equations, where a 2\texttimes 2 system of velocity and pressure variables arises, are numerous, amongst them variants based on the Schur complement of the velocity block \cite{elman2008,zanetti2020,nordsletten2010-prec,silvester2001}, with a broader overview given in \cite{cyr2012}. 
Similar approaches also have shown to be effective for (single-field compressible) solid mechanics coupled to 0D models \cite{augustin2021,hirschvogel2017}, where the Schur complement reduces to a small-sized block which can be solved using direct methods. 
However, introducing a 0D/RD model to Navier-Stokes comes with the addition of a new set of equations, expanding the 2\texttimes 2 to a 3\texttimes 3 coupled matrix system to solve. 
Often, 0D systems in fluid dynamics are introduced as decoupled in-/outflow boundary conditions, whose stiffness contribution is condensed into the main fluid matrix \cite{liu2020prec,esmailymoghadam2013-3d0d,esmailymoghadam2013-prec,seo2019}. 
However, these approaches introduce dense fill-ins into the Jacobian, which require an extra level of consideration to mitigate potential ill-conditioning of the momentum stiffness matrix \cite{esmailymoghadam2013-prec}. 
Other solution approaches for hemodynamics with so-called flow-rate ``defective'' boundary conditions, which are non-local constraints to impose a prescribed flux value at one or more fluid boundaries, similarly result in 3\texttimes 3 systems, but may be adequately reduced to facilitate usage of 2\texttimes 2 preconditioners \cite{veneziani2005,formaggia2002}. 
In a comparable way, reduced solid mechanics wall models, such as the recently introduced fluid-reduced-solid interaction (FrSI) technique~\cite{hirschvogel2024-frsi}, may have a decisive effect on the performance of preconditioners. 
While it may be feasible to treat these models as if they were part of the fluid, as it is done for the (non-reduced or reduced-dimensional) solid component in cardiovascular FSI \cite{seo2019}, the projection-based FrSI models will require an extra level of consideration due to the non-locality of their degrees of freedom as a result of the Galerkin projection of the boundary space.\\

In this contribution, we present a novel Schur complement-based block preconditioner for efficiently solving fluid mechanics coupled to reduced systems of a non-local nature. 
This is achieved by separately considering the sub-systems that govern fluid velocity, pressure, and reduced variables. We derive a preconditioner based on the exact block factorization of the resulting 3\texttimes 3 matrix system, where subsequent approximations to Schur complements are aligned with common approaches \cite{elman2008} but equivalently can be designed in a flexible way. 
It therefore allows to use field-specific solution schemes and preconditioners respecting the individual physics, without compromising the sparsity patterns of the momentum or continuity/Schur complement-related Jacobians. The solver is exemplified on two types of models: fluid-0D systems of blood flow in an Eulerian reference frame, and ALE fluid-0D systems using FrSI to account for solid mechanical contributions. 
For the latter type, we propose a monolithic nonlinear solution scheme for the coupled fluid-RD/0D and ALE domain motion problem, where the 3\texttimes 3 Schur preconditioner is embedded into a (forward) Block Gauss-Seidel method \cite{gee2011} that effectively preconditions the 4\texttimes 4 block matrix system of fluid velocities, pressures, reduced variables, and domain displacements. 
For fluid-0D models, we show that the proposed solver is superior to 2\texttimes 2 block preconditioners that operate on either a condensed system (eliminating 0D variables from the main system) or a merged system (consolidating 0D and fluid pressure variables), both in terms of run time and memory.
For ALE fluid-0D FrSI models, where the wall model lives in a POD space linking all boundary degrees of freedom to a few modes, we show that classical 2\texttimes 2 block approaches fail in generating a feasible Schur complement operator in terms of memory needs due to the inherent link of reduced velocity variables to the fluid pressure space. 
Our results indicate that the proposed preconditioners are effective, displaying the least amount of linear iterations and reduced computational demand compared to the alternative approaches. 
Results are generated using two different solver frameworks: the multi-physics solver $\mathcal{C}\text{Heart}$ \cite{lee2016} as well as the FEniCSx-based \cite{baratta2023-dolfinx,logg2012-fenics} open-source multi-physics cardiovascular solver Ambit \cite{hirschvogel2024-ambit}.\\

The remainder of this article is organized as follows. 
In Sec.~\ref{sec:methods}, we present the nonlinear solution approaches of the finite element-discretized problems, briefly recapture established 2\texttimes 2 preconditioning approaches, and subsequently introduce the new block preconditioners for our specific 3\texttimes 3 and 4\texttimes 4 systems. In Sec.~\ref{sec:examples_fluidpipe}---\ref{sec:examples_frsiheart}, we demonstrate our novel approaches for four different examples, with detailed results and benchmarks against other preconditioning methods. 
In Sec.~\ref{sec:discussion}, we present a discussion of our results, followed by a conclusion in Sec.~\ref{sec:conclusion}.

\clearpage 
\section{Methods}\label{sec:methods}
In this section, we focus on preconditioners for computational fluid dynamics (CFD) problems in Eulerian and Arbitrary Lagrangian-Eulerian (ALE) \cite{donea1982,duarte2004} description that are coupled to reduced models of a \textit{non-local} nature. 
Here, these encompass lumped-parameter models (referred to as ``0D'' models) and Galerkin reduced-order models (referred to as ``RD'' models), which both introduce a new (set of) variable(s) that disrupt the typical local support common to finite element discretizations of the fluid problem.
We present the discretized systems of equations that arise upon spatio-temporal discretization (\emph{e.g.}, using finite elements in space and finite differences in time), cf. Sec.~\ref{subsec:eulerian} and Sec.~\ref{subsec:frsi} for Eulerian and ALE CFD, respectively. In Sec.~\ref{subsec:prec}, we present the linear solution along with different block preconditioning techniques that exploit the specific nature of the underlying matrix systems. Subsequently, we introduce the different test example problems along with a motivation for their specific design in Sec.~\ref{subsec:testexamples}, and conclude this section with details on software and hardware architectures, Sec.~\ref{subsec:software}.

\subsection{Discretized Form for Eulerian Fluids Coupled to 0D Models}\label{subsec:eulerian}

Here, we consider the finite element solution to the stabilized Navier-Stokes equations coupled to reduced order 0D models. 
Details of the governing equations and discrete weak form systems are provided in the \emph{Supplementary Material}, Sec.~\ref{app:eulerian_strong} and Sec.~\ref{app:eulerian_weak}.
In general, these problems can be written as root-finding problem at the current time step indexed by $n+1$, where we seek to find the discrete velocity $\bsf{v}_{n+1}$, pressure $\bsf{p}_{n+1}$, and 0D model variables $\LMZ_{n+1}$ satisfying
\begin{equation}
\begin{aligned}
    \ROP_{n+1} = \begin{bmatrix} \ROP_{v} (\bsf{v},\bs{\mathsf{p}},\LMZ) \\ 
                               \ROP_{p} (\bs{\mathsf{p}},\bs{\mathsf{v}}) \\ 
                               \ROP_{\mathit{\Lambda}} (\LMZ,\bs{\mathsf{v}}) 
               \end{bmatrix}_{n+1}
               = 
               \bs{\mathsf{0}},
               \label{eq:nonlin_sys_fluid}
\end{aligned}
\end{equation}
where $\ROP_{v}$, $\ROP_{p}$, and $\ROP_{\mathit{\Lambda}}$ denote the discrete momentum, continuity, and 0D constraint equations, respectively.
The solution to Eq.~(\ref{eq:nonlin_sys_fluid}) is commonly solved using a Newton-Raphson scheme, where iterative guesses at the current iteration indexed by $k+1$ are updated until convergence is met \cite{dedieu2015-newton}.
Typically, the update to the current guess of the state variables $\bsf{x}_{n+1}^{k+1}$ follows the 
\begin{equation}
\begin{aligned}
    \textbf{solve: }
    \bsf{K}_{n+1}^{k} \Delta\bsf{x}_{n+1}^{k+1} = - \bsf{r}_{n+1}^{k},
    \quad
    \textbf{update: }
    \bsf{x}_{n+1}^{k+1} = \bsf{x}_{n+1}^{k} + \Delta\bsf{x}_{n+1}^{k+1}
               \label{eq:nriteration}
\end{aligned}
\end{equation}
procedure, where $ \bsf{K}_{n+1}^{k} $ denotes the Jacobian (\emph{e.g.}, the directional derivative of $ \bsf{r}_{n+1}^{k}$ with respect to each state variable in the direction of each trial function). In the case of fluid systems coupled to 0D models, the system for the Newton scheme Eq.~(\ref{eq:nriteration})$_{\text{1}}$ here takes the form 
\begin{align}
    \begin{bmatrix} \bsf{K}_{vv} & \bsf{K}_{vp} & \bsf{K}_{v\lmzi} \\ \\ \bsf{K}_{pv} & \bsf{K}_{pp} & \zerom \\ \\  \bsf{K}_{\lmzi v} & \zerom & \bsf{K}_{\lmzi\lmzi} \end{bmatrix}_{n+1}^{k}\begin{bmatrix} \Delta\bsf{v} \\ \\ \Delta\bsf{p} \\ \\ \Delta\LMZ \end{bmatrix}_{n+1}^{k+1}=-\begin{bmatrix} \ROP_{v} \\ \\ \ROP_{p} \\ \\ \ROP_{\lmzi} \end{bmatrix}_{n+1}^{k} \label{eq:lin_sys_fluid}.
\end{align}

Sub-block matrices of $\bsf{K}$ denote the linearization of Eq.~(\ref{eq:nonlin_sys_fluid}), emerging from the directional derivatives of the discrete residuals in the direction of each trial function. 

\paragraph{Two-field System of Velocity and Pressure}
In some applications, the three-field approach is simplified into a two-variable system by static condensation of the 0D variables \cite{brown2024}, resulting in a two-variable system where state variables are found by solving the linearized system
\begin{align}
    \begin{bmatrix} \bsf{K}_{vv} - \bsf{K}_{v \lmzi} \bsf{K}_{\lmzi\lmzi}^{-1} \bsf{K}_{\lmzi v} & \bsf{K}_{vp} \\ \\ \bsf{K}_{pv} & \bsf{K}_{pp} \end{bmatrix}_{n+1}^{k}
    \begin{bmatrix} \Delta\bsf{v} \\ \\ \Delta\bsf{p}  \end{bmatrix}_{n+1}^{k+1} 
    =
    -\begin{bmatrix} \ROP_{v} - \bsf{K}_{v\lmzi}\bsf{K}_{\lmzi\lmzi}^{-1}\ROP_{\lmzi} \\ \\ \ROP_{p} \end{bmatrix}_{n+1}^{k} \label{eq:lin_sys_fluid_cnd}.
\end{align}
This linear system may be similarly obtained when integrating any flux-dependent boundary conditions directly into the fluid's momentum equation in the first place, cf. \cite{esmailymoghadam2013-3d0d,liu2020prec,seo2019}. Furthermore, we investigate the condensed linear system where the stiffness contribution resulting from the coupling of two boundaries via the 0D model is neglected, leading to an improved sparsity pattern and negligible effects on nonlinear solver convergence \cite{esmailymoghadam2013-3d0d,brown2024}. This system reads
\begin{align}
    \begin{bmatrix} \bsf{K}_{vv} - \bsf{K}_{v \lmzi} \mathrm{diag}(\bsf{K}_{\lmzi\lmzi})^{-1} \bsf{K}_{\lmzi v} & \bsf{K}_{vp} \\ \\ \bsf{K}_{pv} & \bsf{K}_{pp} \end{bmatrix}_{n+1}^{k}
    \begin{bmatrix} \Delta\bsf{v} \\ \\ \Delta\bsf{p}  \end{bmatrix}_{n+1}^{k+1} 
    =
    -\begin{bmatrix} \ROP_{v} - \bsf{K}_{v\lmzi}\bsf{K}_{\lmzi\lmzi}^{-1}\ROP_{\lmzi} \\ \\ \ROP_{p} \end{bmatrix}_{n+1}^{k} \label{eq:lin_sys_fluid_cndII}.
\end{align}

Advantages and possible constraints of this approach will be discussed further in later sections. Preconditioning either the three-field or two-field systems requires an appropriate approximation to the action of the Jacobian's inverse as will be discussed in Sec.~\ref{subsec:prec}.

\subsection{Discretized Form for ALE Fluid Coupled to 0D and RD FrSI Models}\label{subsec:frsi}

Considering the problem of a discrete ALE fluid coupled to both 0D and RD FrSI models adds two key conceptual differences compared to the Eulerian fluid-0D system introduced in the previous section.
First, introduction of the Arbitrary Lagrangian-Eulerian (ALE) reference frame requires the solution of a domain motion problem, typically found using a harmonic extension or pseudo-mechanical formulation \cite{shamanskiy2021}.
This can be integrated by introducing the discrete domain motion problem $\ROP_{d}$ to the root-finding problem.
Secondly, the reduced-dimensional wall model representing the elasticity of the boundary is confined to a subspace spanned by POD modes that either have been precomputed from an external data source (another high-dimensional model, observations from real world) or are synthetically generated, depending on the application. 
We define the operator that projects the full-dimensional model to a boundary-reduced model by
\begin{align}
    \bsf{V}_{v}^{\Gm} \in \mathbb{R}^{n_v \times (r_v + n_{v}^{\Om})}, \label{eq:Vgamma}
\end{align}
where $n_v$ is the size of the full velocity space, $n_{v}^{\Om}$ the size of the space of bulk (non-boundary) velocities, and $r_v$ the number of reduced boundary modes (degrees of freedom). This Galerkin projector, thereby, confines all velocity degrees of freedom of $\Gm_{0}^{\mathrm{\fF\mhyphen\fSr}}$ to the lower-dimensional subspace but provides an identity mapping for the velocity degrees of freedom associated to the bulk domain.
Details of the governing equations and discrete weak form systems are provided in the \emph{Supplementary Material}, Sec.~\ref{app:frsi_strong} and Sec.~\ref{app:frsi_weak}.\\

Similar to the Eulerian fluid-0D system, these problems can be written as root-finding problem at the current time step indexed by $n+1$, where we seek to find the discrete velocity $\bsf{v}_{n+1}$, pressure $\bsf{p}_{n+1}$, 0D model variables $\LMZ_{n+1}$, and domain displacements $\bsf{d}_{n+1}$ satisfying
\begin{align}
    \ROP_{n+1} = \begin{bmatrix} 
                  \bs{\mathsf{V}}_{v}^{\Gm^\mathrm{T}}\ROP_{v}(\bs{\mathsf{V}}_{v}^{\Gm}\VFr,\bsf{p},\LMZ,\bsf{d}) \\
                  \ROP_{p}(\bsf{p},\bs{\mathsf{V}}_{v}^{\Gm}\VFr,\bsf{d}) \\ 
                  \ROP_{\lmzi} (\LMZ,\bs{\mathsf{V}}_{v}^{\Gm}\VFr,\bsf{d}) \\ 
                  \ROP_{d} (\bsf{d},\bs{\mathsf{V}}_{v}^{\Gm}\VFr)
               \end{bmatrix}_{n+1}
               = 
               \bs{\mathsf{0}},
               \label{eq:res_nonlin_frsi}
\end{align}
with the trial space projection
\begin{align}
    \bsf{v} = \bs{\mathsf{V}}_{v}^{\Gm} \VFr, \label{eq:trial_proj_rom_surf}
\end{align}
where $\VFr \in \mathbb{R}^{r_v + n_{v}^{\Om}}$ is the (partly) reduced-dimensional velocity vector. 
Further details on the FrSI technique can be found in \cite{hirschvogel2024-frsi}.\\

\paragraph{Monolithic FrSI}
As with the previous Eulerian fluid-0D system, approximations to the discrete system can be found by the Newton-Raphson approach in Eq.~(\ref{eq:nriteration}). In this case, the discrete state vector $ \bsf{x}_{n+1}^{k+1}$ includes fluid degrees of freedom that may be, partially, reduced by POD priors as well as the domain motion.
The system for the Newton scheme Eq.~(\ref{eq:nriteration})$_{\text{1}}$ here takes the form 
\begin{align}
    \begin{bmatrix} \bsf{V}_{v}^{\Gm^\mathrm{T}}\bsf{K}_{vv}\bsf{V}_{v}^{\Gm} & \bsf{V}_{v}^{\Gm^\mathrm{T}}\bsf{K}_{vp} & \bsf{V}_{v}^{\Gm^\mathrm{T}}\bsf{K}_{v\lmzi} & \bsf{V}_{v}^{\Gm^\mathrm{T}}\bsf{K}_{vd} \\ \\ \bsf{K}_{pv}\bsf{V}_{v}^{\Gm} & \bsf{K}_{pp} & \zerom & \bsf{K}_{pd} \\ \\  \bsf{K}_{\lmzi v}\bsf{V}_{v}^{\Gm} & \zerom & \bsf{K}_{\lmzi\lmzi} & \bsf{K}_{\lmzi d} \\ \\ \bsf{K}_{dv}\bsf{V}_{v}^{\Gm} & \zerom & \zerom & \bsf{K}_{dd} \end{bmatrix}_{n+1}^{k}\begin{bmatrix} \Delta\tilde{\bsf{v}} \\ \\ \Delta\bsf{p} \\ \\ \Delta\LMZ \\ \\ \Delta\bsf{d} \end{bmatrix}_{n+1}^{k+1}=-\begin{bmatrix} \bsf{V}_{v}^{\Gm^\mathrm{T}}\ROP_{v} \\ \\ \ROP_{p} \\ \\ \ROP_{\lmzi} \\ \\  \ROP_{d} \end{bmatrix}_{n+1}^{k} \label{eq:lin_sys_rom_frsi_mono},
\end{align}
where sub-block matrices of $\bsf{K}$ are obtained from the linearization of Eq.~(\ref{eq:res_nonlin_frsi}), emerging from the directional derivatives of the discrete residuals with respect to the trial functions.

\paragraph{Partitioned FrSI}
As an alternative to the monolithic approach, the linearized fluid-0D system and ALE domain motion can be solved in a staggered way. 
This approach allows to operate on reduced overall system sizes and re-use tailored solvers and preconditioners for fluid-0D systems. 
In this case, the fluid-0D system simplifies to a matrix problem similar to the three-field approach (though incorporating potentially dense rows/columns due to POD reduction), \emph{e.g.},
\begin{align}
    \begin{bmatrix} 
       \bsf{V}_{v}^{\Gm^\mathrm{T}}\bsf{K}_{vv}\bsf{V}_{v}^{\Gm} &  \bsf{V}_{v}^{\Gm^\mathrm{T}}\bsf{K}_{vp} & \bsf{V}_{v}^{\Gm^\mathrm{T}}\bsf{K}_{v\lmzi} \\ \\ 
       \bsf{K}_{pv}\bsf{V}_{v}^{\Gm} & \bsf{K}_{pp} & \zerom \\ \\  
       \bsf{K}_{\lmzi v}\bsf{V}_{v}^{\Gm} & \zerom & \bsf{K}_{\lmzi\lmzi} 
    \end{bmatrix}_{n+1}^{k}
    \begin{bmatrix} 
       \Delta\tilde{\bsf{v}} \\ \\ 
       \Delta\bsf{p} \\ \\ 
       \Delta\LMZ 
    \end{bmatrix}_{n+1}^{k+1}
    =
    -\begin{bmatrix} 
         \bsf{V}_{v}^{\Gm^\mathrm{T}}\ROP_{v} \\ \\ 
         \ROP_{p} \\ \\ 
         \ROP_{\lmzi} 
     \end{bmatrix}_{n+1}^{k}, \label{eq:lin_sys_rom_frsi_part}
\end{align}
which is solved prior to the ALE domain motion,
\begin{align}
    [\bsf{K}_{dd} ]_{n+1}^{k}\,[\Delta\bsf{d}]_{n+1}^{k+1} = - [\ROP_{d}]_{n+1}^{k},
    \label{eq:lin_sys_ale}
\end{align}
before advancing to the next nonlinear iteration step.

\subsection{Preconditioning and Linear Solution}\label{subsec:prec}
In the following, we present suitable preconditioning techniques to efficiently solve the linearized system of equations for fluid-0D and ALE fluid-0D FrSI problems. 
We start by reviewing established preconditioning methods for the 2\texttimes 2 systems of equations, \emph{e.g.}, Eq.~(\ref{eq:lin_sys_fluid_cnd}). 
Next, we present a novel 3\texttimes 3 block preconditioning algorithm that can deal with different types of modeling elements that introduce a non-local structure into the matrix system.\\

The solution of the linear system of equations (\emph{e.g.}, each Newton-Raphson iteration for each time step) requires the solution to the linear system
\begin{align}
    \bsf{K} \Delta \bsf{x} = -\bsf{r}, \label{eq:lin_sys_general}
\end{align}
with $\bsf{K}$ being a (block) matrix as previously discussed, $ \Delta \bsf{x} $ the combined discrete update vector of all state variables (\emph{e.g.}, velocities, pressures, multipliers), and $\bsf{r}$ the block right-hand side residual vector. Note, for ease of notation, we have dropped iterative counters. \\

Solving Eq.~(\ref{eq:lin_sys_general}) using direct methods (such as LU factorization or multi-frontal approaches~\cite{amestoy2019-mumps,amestoy2001-mumps}) becomes increasingly expensive for large systems as they typically scale $ \sim \mathcal{O}(n_{\mathrm{dof}}^2)$.
Iterative approaches, such as the Generalized Minimal Residual method (GMRES) \cite{saad1986}, can significantly reduce computing costs; however, their efficiency depends significantly on the eigenspectrum of $ \bsf{K} $.
The application of so-called \emph{right} preconditioners seeks to alter the system in order to improve the eigenspectrum and thus the performance of many iterative solution techniques.
This is achieved by altering the solution to Eq.~(\ref{eq:lin_sys_general}) into a two-step process whereby we
\begin{align}
\textbf{solve: }
\bsf{K} \bs{\mathcal{P}}^{-1} \bsf{y} = - \bsf{r}, 
\quad 
\textbf{solve: } \bs{\mathcal{P}} \Delta \bsf{x} = \bsf{y}.
\end{align}
In this case, the effectiveness of iterative techniques depends now on the eigenspectrum of $ \bsf{K} \bs{\mathcal{P}}^{-1} $. 
Here we seek to construct an effective approximation where $ \varrho (\bsf{K} \bs{\mathcal{P}}^{-1}) \sim 1 $ while the action of $ \bs{\mathcal{P}}^{-1} $ can be applied using efficient iterative solution techniques (such as algebraic multigrid or additive Schwarz decomposition).
In this case, since $ \bs{\mathcal{P}}^{-1} $ is not analytically found, but its action approximated, we use the right preconditioned Flexible Generalized Minimal Residual method (FGMRES) \cite{saad1993}.

\subsubsection{Schur Complement Preconditioner S2\texttimes 2}\label{subsec:2x2systems}
A proposal to the solution of both systems introduced would be to utilize a Schur complement preconditioner (here denoted by S2\texttimes 2), whereby the variables are lumped to construct a 2\texttimes 2 matrix system with the general form
\begin{align}
    \begin{bmatrix} \bsf{A} & \bsf{B}^{\mathrm{T}} \\ \\ \hat{\bsf{B}} & \bsf{C} \end{bmatrix}\begin{bmatrix} \Delta \bsf{x}_{A} \\ \\ \Delta \bsf{x}_{C} \end{bmatrix}=-\begin{bmatrix} \bsf{r}_{A} \\ \\ \bsf{r}_{C}\end{bmatrix}, \label{eq:lin_sys_fluid_2x2}
\end{align}
where $ \bsf{A}, \ldots, \bsf{C} $ denote sub-blocks related to different components of the physical system, $  \bsf{r}_{A}$ and $ \bsf{r}_{C} $ the equivalent discrete residual vector functions, and $\Delta \bsf{x}_{A} $ and $\Delta \bsf{x}_{C} $ the updates sought.
Typically, this form is found when solving the stabilized Navier-Stokes problems \cite{franca1992}; however, Eq.~(\ref{eq:lin_sys_fluid_2x2}) may also be representative of the case where any contributions of reduced models are implicitly integrated (condensed) into the momentum residual, $\bsf{r}_{A}$, cf. Eq.~(\ref{eq:lin_sys_fluid_cnd}), or in case some variables (\emph{e.g.}, fluid pressures and 3D-0D multipliers) are consolidated from Eq.~(\ref{eq:lin_sys_fluid}).
A well-known class of preconditioners for the Navier-Stokes system Eq.~(\ref{eq:lin_sys_fluid_2x2}) are based on finding an approximation to the Schur complement of $\bsf{A}$,
\begin{align}
    \bsf{S} = \bsf{C} - \hat{\bsf{B}}\,\bsf{A}^{-1}\bsf{B}^{\mathrm{T}},
\end{align}
and aim at providing an approximation to the block matrix inverse based on its block factorization, which can be written as
\begin{align}
    \begin{bmatrix} \bsf{A} & \bsf{B}^{\mathrm{T}} \\ \\ \hat{\bsf{B}} & \bsf{C} \end{bmatrix} =
    \begin{bmatrix} \bsf{A} & \zerom \\ \\ \zerom & \bsf{I}\end{bmatrix}
    \begin{bmatrix} \bsf{I} & \zerom \\ \\ \hat{\bsf{B}} & \bsf{I}\end{bmatrix}
    \begin{bmatrix} \bsf{A}^{-1} & \zerom \\ \\ \zerom & \bsf{S}\end{bmatrix}
    \begin{bmatrix} \bsf{I} & \bsf{B}^{\mathrm{T}} \\ \\ \zerom & \bsf{I}\end{bmatrix}
    \begin{bmatrix} \bsf{A} & \zerom \\ \\ \zerom & \bsf{I}\end{bmatrix}.
\end{align}

The approximate block inverse 
\begin{align}
    &\begin{bmatrix} \bsf{A} & \bsf{B}^{\mathrm{T}} \\ \\ \hat{\bsf{B}} & \bsf{C} \end{bmatrix}^{-1} \approx \bs{\mathcal{P}}_{\mathrm{S}2\times 2}^{-1} \\ &=
    \begin{bmatrix} \mathrm{diag}(\bsf{A})^{-1} & \zerom \\ \\ \zerom & \bsf{I}\end{bmatrix}
    \begin{bmatrix} \bsf{I} & -\bsf{B}^{\mathrm{T}} \\ \\ \zerom & \bsf{I}\end{bmatrix}
    \begin{bmatrix} \bsf{A} & \zerom \\ \\ \zerom & \mathrm{inv}(\tilde{\bsf{S}})\end{bmatrix}
    \begin{bmatrix} \bsf{I} & \zerom \\ \\ -\hat{\bsf{B}} & \bsf{I}\end{bmatrix}
    \begin{bmatrix} \mathrm{inv}(\bsf{A}) & \zerom \\ \\ \zerom & \bsf{I}\end{bmatrix} \label{eq:2x2_inverse_block}
\end{align}
would correspond to the SIMPLE (Semi-Implicit Method for Pressure Linked Equations) preconditioner \cite{elman2008}, which we here denote by $\bs{\mathcal{P}}_{\mathrm{S}2\times 2}^{-1}$. Therein, notation $\mathrm{inv}(\bsf{M})$ implies the (approximate) solution of a linear system with operator $\bsf{M}$ and a suitable preconditioner.
The SIMPLE preconditioner essentially encapsulates two approximations: First, computational expense is reduced by approximating one of the actions of $\bsf{A}^{-1}$ on a vector by the inverse of its diagonal, $\mathrm{diag}(\bsf{A})^{-1}$. Second, it approximates the Schur complement with
\begin{align}
    \tilde{\bsf{S}} = \bsf{C} - \hat{\bsf{B}}\,\mathrm{diag}(\bsf{A})^{-1}\bsf{B}^{\mathrm{T}}, \label{eq:schur_approx}
\end{align}
again replacing $\bsf{A}^{-1}$ by the inverse of its diagonal. 
While there are many applications where Eq.~(\ref{eq:schur_approx}) provides a good approximation, strongly convection-dominated high Reynolds number flows may suffer from this choice. 
At this point, we refer to the literature for further discussions on the quality of Schur complement approximations \cite{elman2008,liu2020prec} and possible alternatives \cite{cyr2012}. 

The application of $\bs{\mathcal{P}}_{\mathrm{S}2\times 2}^{-1}$ on a vector $\bsf{x}$ with output $\bsf{y}$ is summarized in Alg.~\ref{alg:2x2}.
The application of the preconditioner involves two inner solves, namely in the first step to provide an approximation of $\bsf{A}^{-1}$ acting on $\bsf{x}_{A}$, and in the third step one for $\tilde{\bsf{S}}^{-1}$ on $\bsf{z}_{C}$. 
Both of these solves can be performed with efficient iterative solvers and preconditioners available in common packages like PETSc \cite{balay2022-petsc}, such as hypre \cite{falgout2002-hypre}.
Here, we use a GMRES algorithm that is preconditioned with an algebraic multigrid (AMG) method. 
Details of the specific settings of these solvers are presented in later sections.\\

\begin{algorithm}
\caption{Action of $\bs{\mathcal{P}}_{\mathrm{S}2\times 2}^{-1}$ on a vector $\bsf{x}=[\bsf{x}_{A}, \bsf{x}_{C}]^{\mathrm{T}}$ with output $\bsf{y}=[\bsf{y}_{A}, \bsf{y}_{C}]^{\mathrm{T}}$}\label{alg:schur2x2}
\begin{algorithmic}[1]
\State Solve for an intermediate velocity $\check{\bsf{y}}_{A}$:
\Statex \qquad $\bsf{A}\check{\bsf{y}}_{A} = \bsf{x}_{A}$
\Statex using a preconditioned iterative method
\Statex
\State Compute: $\bsf{z}_{C} = \bsf{x}_{C} - \hat{\bsf{B}}\check{\bsf{y}}_{A}$
\Statex
\State Solve for the pressure $\bsf{y}_{C}$:
\Statex \qquad $\tilde{\bsf{S}}\bsf{y}_{C} = \bsf{z}_{C}$
\Statex using a preconditioned iterative method
\Statex
\State Compute: $\bsf{z}_{A} = \bsf{x}_{A} - \bsf{B}^{\mathrm{T}}\bsf{y}_{C}$
\Statex
\State Update the velocity $\bsf{y}_{A}$:
\Statex \qquad $\bsf{y}_{A} = \mathrm{diag}(\bsf{A})^{-1}\bsf{z}_{A}$
\end{algorithmic} \label{alg:2x2}
\end{algorithm}

\begin{figure}[!htp]
\centering
\includegraphics[width=1\textwidth]{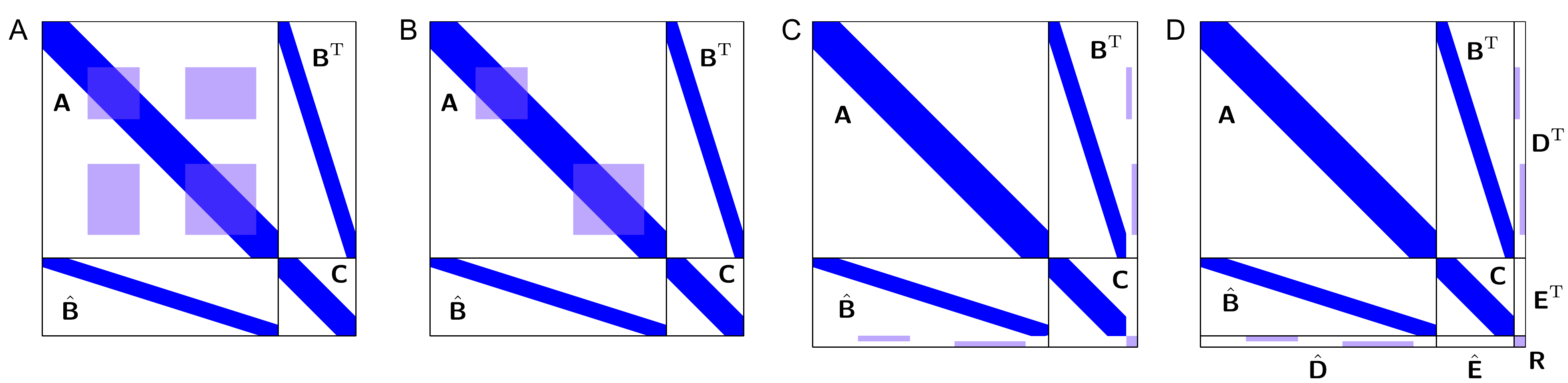}
\caption{Sketch of a sparsity pattern obtained from a fluid problem coupled to a 0D system that connects two boundaries, cf. model shown in Fig.~\ref{fig:models_all}A for instance. 
\textbf{A.} Condensed system Eq.~(\ref{eq:lin_sys_fluid_cnd}), showing two fill-ins from 0D contributions at the main diagonal, and two further fill-ins at offdiagonal positions resulting from the link between the boundaries in the 0D world. \textbf{B.} Condensed system Eq.~(\ref{eq:lin_sys_fluid_cndII}), neglecting the two offdiagonal fill-ins \cite{esmailymoghadam2013-3d0d}.
\textbf{C.} System from Eq.~(\ref{eq:lin_sys_fluid}), but with consolidating the coupling and constraint blocks in order to facilitate usage of a 2\texttimes 2 preconditioner.
\textbf{D.} System from Eq.~(\ref{eq:lin_sys_fluid}), but with separating 0D contributions to new blocks (cf. Sec.~\ref{subsec:3x3systems}).}\label{fig:s2x2c_s2x2p_sparsity}
\end{figure}

For applications of S2\texttimes 2 to fluid systems coupled to non-local reduced models, we consider three types of matrix systems based on the approach taken (see Fig.~\ref{fig:s2x2c_s2x2p_sparsity}A---C). 
First, we consider the matrix system Eq.~(\ref{eq:lin_sys_fluid_cnd}) where, for example, 0D contributions are statically condensed into the fluid system matrix. 
Figure~\ref{fig:s2x2c_s2x2p_sparsity}A shows a sketch of the sparsity pattern, where the two dense fill-ins onto the diagonal of $\bsf{A}$ result from two boundaries coupled to a 0D in- or outlet. Furthermore, the two offdiagonal fill-ins result from the boundaries being linked via the 0D model, which is a common case in hemodynamics closed-circuit models.
In the following, we denote the application of the preconditioner to this system by ``S2\texttimes 2(c)''. Second, we consider matrix system Eq.~(\ref{eq:lin_sys_fluid_cndII}), where offdiagonal fill-in blocks are neglected, cf. Fig.~\ref{fig:s2x2c_s2x2p_sparsity}B. The application of the preconditioner to this system is denoted by ``S2\texttimes 2(c$^{\star}$)''.
Third, we consider the 3\texttimes 3 matrix system Eq.~(\ref{eq:lin_sys_fluid}), but consolidate fluid pressures and constraint multipliers to one set of variables in order to apply the S2\texttimes 2 preconditioner to the system. The respective sparsity pattern is shown in Fig.~\ref{fig:s2x2c_s2x2p_sparsity}C. Due to the introduction of multiplier variables that link the 0D to the 3D model, the fill-ins are removed from the $\bsf{A}$ block and the coupling expresses itself in dense rows and columns that are grouped in $\hat{\bsf{B}}$ and $\bsf{B}^{\mathrm{T}}$, respectively. In the following, we denote the application of the preconditioner to this system by ``S2\texttimes 2(p)''.

\subsubsection{Schur Complement Preconditioner S3\texttimes 3}\label{subsec:3x3systems}
In the following, we introduce a novel type of preconditioner that directly exploits the underlying 3\texttimes 3 block structure of the system matrix and hence preserves all sparsity patterns of the fluid momentum, continuity, and coupling blocks that would arise if no reduced models were involved. 
For this, we start directly with the coupled block matrix Eq.~(\ref{eq:lin_sys_fluid}) and re-write it as follows:
\begin{align}
    \begin{bmatrix}\bsf{A} & \bsf{B}^{\mathrm{T}} & \bsf{D}^{\mathrm{T}} \\ \\ \hat{\bsf{B}} & \bsf{C} & \textcolor{lightgray}{\bsf{E}^{\mathrm{T}}} \\ \\ \hat{\bsf{D}} & \textcolor{lightgray}{\hat{\bsf{E}}} & \bsf{R} \end{bmatrix}\begin{bmatrix} \Delta \bsf{x}_{A} \\ \\ \Delta \bsf{x}_{C} \\ \\ \Delta \bsf{x}_{R} \end{bmatrix}=-\begin{bmatrix} \bsf{r}_{A} \\ \\ \bsf{r}_{C} \\ \\ \bsf{r}_{R}\end{bmatrix}. \label{eq:lin_sys_fluid_3x3}
\end{align}

The exemplary sparsity pattern is sketched in Fig.~\ref{fig:s2x2c_s2x2p_sparsity}D. We note that this system has the same structure as shown in Fig.~\ref{fig:s2x2c_s2x2p_sparsity}C but we separate the contributions of the non-local constraints. 

In order to derive the preconditioner ``S3\texttimes 3'', we start with an exact block block factorization of the matrix Eq.~(\ref{eq:lin_sys_fluid_3x3}) as follows:
\begin{equation}
\begin{aligned}
    \begin{bmatrix} \bsf{A} & \bsf{B}^{\mathrm{T}} & \bsf{D}^{\mathrm{T}} \\ \hat{\bsf{B}} & \bsf{C} & \bsf{E}^{\mathrm{T}} \\ \hat{\bsf{D}} & \hat{\bsf{E}} & \bsf{R} \end{bmatrix}  
    = &
    \begin{bmatrix} \bsf{A} & \zerom & \zerom \\ \zerom & \bsf{I} & \zerom \\ \zerom & \zerom & \bsf{I} \end{bmatrix}
    \begin{bmatrix} \bsf{I} & \zerom & \zerom \\ \hat{\bsf{B}} & \bsf{I} & \zerom \\ \zerom & \zerom & \bsf{I} \end{bmatrix}
    \begin{bmatrix} \bsf{I} & \zerom & \zerom \\ \zerom & \bsf{S} & \zerom \\ \zerom & \zerom & \bsf{I} \end{bmatrix} \\
    &
    \begin{bmatrix} \bsf{I} & \zerom & \zerom \\ \zerom & \bsf{I} & \zerom \\ \hat{\bsf{D}} & \bsf{U} & \bsf{I} \end{bmatrix}
    \begin{bmatrix} \bsf{I} & \zerom & \zerom \\ \zerom & \bsf{S}^{-1} & \zerom \\ \zerom & \zerom & \bsf{W} \end{bmatrix}
    \begin{bmatrix} \bsf{I} & \zerom & \zerom \\ \zerom & \bsf{I} & \bsf{T} \\ \zerom & \zerom & \bsf{I} \end{bmatrix} \\
    &
    \begin{bmatrix} \bsf{A}^{-1} & \zerom & \zerom \\ \zerom & \bsf{S} & \zerom \\ \zerom & \zerom & \bsf{I} \end{bmatrix}
    \begin{bmatrix} \bsf{I} & \bsf{B}^{\mathrm{T}} & \bsf{D}^{\mathrm{T}} \\ \zerom & \bsf{I} & \zerom \\ \zerom & \zerom & \bsf{I} \end{bmatrix}
    \begin{bmatrix} \bsf{A} & \zerom & \zerom \\ \zerom & \bsf{I} & \zerom \\ \zerom & \zerom & \bsf{I} \end{bmatrix},
\end{aligned}
\end{equation}
where 
\begin{align}
\bsf{S} &= \bsf{C} - \hat{\bsf{B}}\,\bsf{A}^{-1}\bsf{B}^{\mathrm{T}}, \\
\bsf{T} &= \bsf{E}^{\mathrm{T}} - \hat{\bsf{B}}\,\bsf{A}^{-1}\bsf{D}^{\mathrm{T}},\\
\bsf{U} &= \hat{\bsf{E}} - \hat{\bsf{D}}\,\bsf{A}^{-1}\bsf{B}^{\mathrm{T}}, \quad\text{and} \\
\bsf{W} &= \bsf{R} - \hat{\bsf{D}}\,\bsf{A}^{-1}\bsf{D}^{\mathrm{T}} - \bsf{U}\,\bsf{S}^{-1}\bsf{T}.
\end{align}
We note that, in this form, there are two analogous Schur complement-like matrices, $ \bsf{S} $ and $ \bsf{W} $, which must be invertible to ensure solvability. The inverse of the block factorization can be written down in a straightforward manner, readily making use of similar approximations that have been used to derive Eq.~(\ref{eq:2x2_inverse_block}). Hence, our preconditioner $\bs{\mathcal{P}}_{\mathrm{S}3\times 3}^{-1}$ reads
\begin{equation}
\begin{aligned}
    \begin{bmatrix} \bsf{A} & \bsf{B}^{\mathrm{T}} & \bsf{D}^{\mathrm{T}} \\ \hat{\bsf{B}} & \bsf{C} & \bsf{E}^{\mathrm{T}} \\ \hat{\bsf{D}} & \hat{\bsf{E}} & \bsf{R} \end{bmatrix}^{-1} \approx \bs{\mathcal{P}}_{\mathrm{S}3\times 3}^{-1} 
    = & 
    \begin{bmatrix} \mathrm{diag}(\bsf{A})^{-1} & \zerom & \zerom \\ \zerom & \bsf{I} & \zerom \\ \zerom & \zerom & \bsf{I} \end{bmatrix}
    \begin{bmatrix} \bsf{I} & -\bsf{B}^{\mathrm{T}} & -\bsf{D}^{\mathrm{T}} \\ \zerom & \bsf{I} & \zerom \\ \zerom & \zerom & \bsf{I} \end{bmatrix}
    \begin{bmatrix} \bsf{A} & \zerom & \zerom \\ \zerom & \mathrm{inv}(\tilde{\bsf{S}}) & \zerom \\ \zerom & \zerom & \bsf{I} \end{bmatrix} \\
    &
    \begin{bmatrix} \bsf{I} & \zerom & \zerom \\ \zerom & \bsf{I} & -\tilde{\bsf{T}} \\ \zerom & \zerom & \bsf{I} \end{bmatrix}
    \begin{bmatrix} \bsf{I} & \zerom & \zerom \\ \zerom & \tilde{\bsf{S}} & \zerom \\ \zerom & \zerom & \mathrm{inv}(\tilde{\bsf{W}}) \end{bmatrix}
    \begin{bmatrix} \bsf{I} & \zerom & \zerom \\ \zerom & \bsf{I} & \zerom \\ -\hat{\bsf{D}} & -\tilde{\bsf{U}} & \bsf{I} \end{bmatrix} \\
    &
    \begin{bmatrix} \bsf{I} & \zerom & \zerom \\ \zerom & \mathrm{inv}(\tilde{\bsf{S}}) & \zerom \\ \zerom & \zerom & \bsf{I} \end{bmatrix}
    \begin{bmatrix} \bsf{I} & \zerom & \zerom \\ -\hat{\bsf{B}} & \bsf{I} & \zerom \\ \zerom & \zerom & \bsf{I} \end{bmatrix}
    \begin{bmatrix} \mathrm{inv}(\bsf{A}) & \zerom & \zerom \\ \zerom & \bsf{I} & \zerom \\ \zerom & \zerom & \bsf{I} \end{bmatrix}, \label{eq:3x3_inverse_block}
\end{aligned}
\end{equation}
introducing the approximations to the Schur complement-like matrices:
\begin{align}
    \tilde{\bsf{S}} &= \bsf{C} - \hat{\bsf{B}}\,\mathrm{diag}(\bsf{A})^{-1}\bsf{B}^{\mathrm{T}}, \label{eq:schur_approx_3x3}\\
    \tilde{\bsf{T}} &= \bsf{E}^{\mathrm{T}} - \hat{\bsf{B}}\,\mathrm{diag}(\bsf{A})^{-1}\bsf{D}^{\mathrm{T}}, \\
    \tilde{\bsf{U}} &= \hat{\bsf{E}} - \hat{\bsf{D}}\,\mathrm{diag}(\bsf{A})^{-1}\bsf{B}^{\mathrm{T}}, \quad\text{and} \\
    \tilde{\bsf{W}} &= \bsf{R} - \hat{\bsf{D}}\,\mathrm{diag}(\bsf{A})^{-1}\bsf{D}^{\mathrm{T}} - \tilde{\bsf{U}}\,\mathrm{diag}(\tilde{\bsf{S}})^{-1}\tilde{\bsf{T}}.
\end{align}
Again, notation $\mathrm{inv}(\bsf{M})$ implies the (approximate) solution of a linear system with operator $\bsf{M}$ and a suitable preconditioner.\\

The application of $\bs{\mathcal{P}}_{\mathrm{S}3\times 3}^{-1}$ on a vector $\bsf{x}$ with output $\bsf{y}$ is summarized in Alg.~\ref{alg:3x3}.
Compared to Alg.~\ref{alg:2x2}, we move from a 5- to a 9-step procedure and increase the number of (approximate) solves from 2 to 4. 
However, we note that the solve in step 5, which requires the inverse of the matrix $\tilde{\bsf{W}}$ hardly will contribute to the overall computational cost, since it only updates the reduced-dimensional variables, whose dimension is typically small. 
Practically, only one additional solve of some computational expense is needed compared to S2\texttimes 2, namely providing the action of $\tilde{\bsf{S}}^{-1}$ (which has the dimension of the fluid pressure space) on $\bsf{z}_{C}$ in step 7.

\begin{algorithm}
\caption{Action of $\bs{\mathcal{P}}_{\mathrm{S}3\times 3}^{-1}$ on a vector $\bsf{x}=[\bsf{x}_{A}, \bsf{x}_{C}, \bsf{x}_{R}]^{\mathrm{T}}$ with output $\bsf{y}=[\bsf{y}_{A}, \bsf{y}_{C}, \bsf{y}_{R}]^{\mathrm{T}}$}
\begin{algorithmic}[1]
\State Solve for an intermediate velocity $\check{\bsf{y}}_{A}$:
\Statex \qquad $\bsf{A}\check{\bsf{y}}_{A} = \bsf{x}_{A}$
\Statex using a preconditioned iterative method
\Statex
\State Compute: $\bsf{z}_{C} = \bsf{x}_{C} - \hat{\bsf{B}}\check{\bsf{y}}_{A}$
\Statex
\State Solve for an intermediate pressure $\check{\bsf{y}}_{C}$:
\Statex \qquad $\tilde{\bsf{S}}\check{\bsf{y}}_{C} = \bsf{z}_{C}$
\Statex using a preconditioned iterative method
\Statex
\State Compute: $\bsf{z}_{R} = \bsf{x}_{R} - \hat{\bsf{D}}\check{\bsf{y}}_{A} - \tilde{\bsf{U}}\check{\bsf{y}}_{C}$
\Statex
\State Solve for the reduced variables $\bsf{y}_{R}$:
\Statex \qquad $\tilde{\bsf{W}}\bsf{y}_{R} = \bsf{z}_{R}$
\Statex using a direct solver
\Statex
\State Compute: $\bsf{z}_{C} = \bsf{x}_{C} - \hat{\bsf{B}}\check{\bsf{y}}_{A} - \tilde{\bsf{T}}\bsf{y}_{R}$
\Statex
\State Solve for the pressure $\bsf{y}_{C}$:
\Statex \qquad $\tilde{\bsf{S}}\bsf{y}_{C} = \bsf{z}_{C}$
\Statex using a preconditioned iterative method
\Statex
\State Compute: $\bsf{z}_{A} = \bsf{x}_{A} - \bsf{B}^{\mathrm{T}}\bsf{y}_{C} - \bsf{D}^{\mathrm{T}}\bsf{y}_{R}$
\Statex
\State Update the velocity $\bsf{y}_{A}$:
\Statex \qquad $\bsf{y}_{A} = \mathrm{diag}(\bsf{A})^{-1}\bsf{z}_{A}$
\end{algorithmic} \label{alg:3x3}
\end{algorithm}

\subsubsection{Block Preconditioners for FrSI}\label{subsec:3x3frsi}
\begin{figure}[!htp]
\centering
\includegraphics[width=0.9\textwidth]{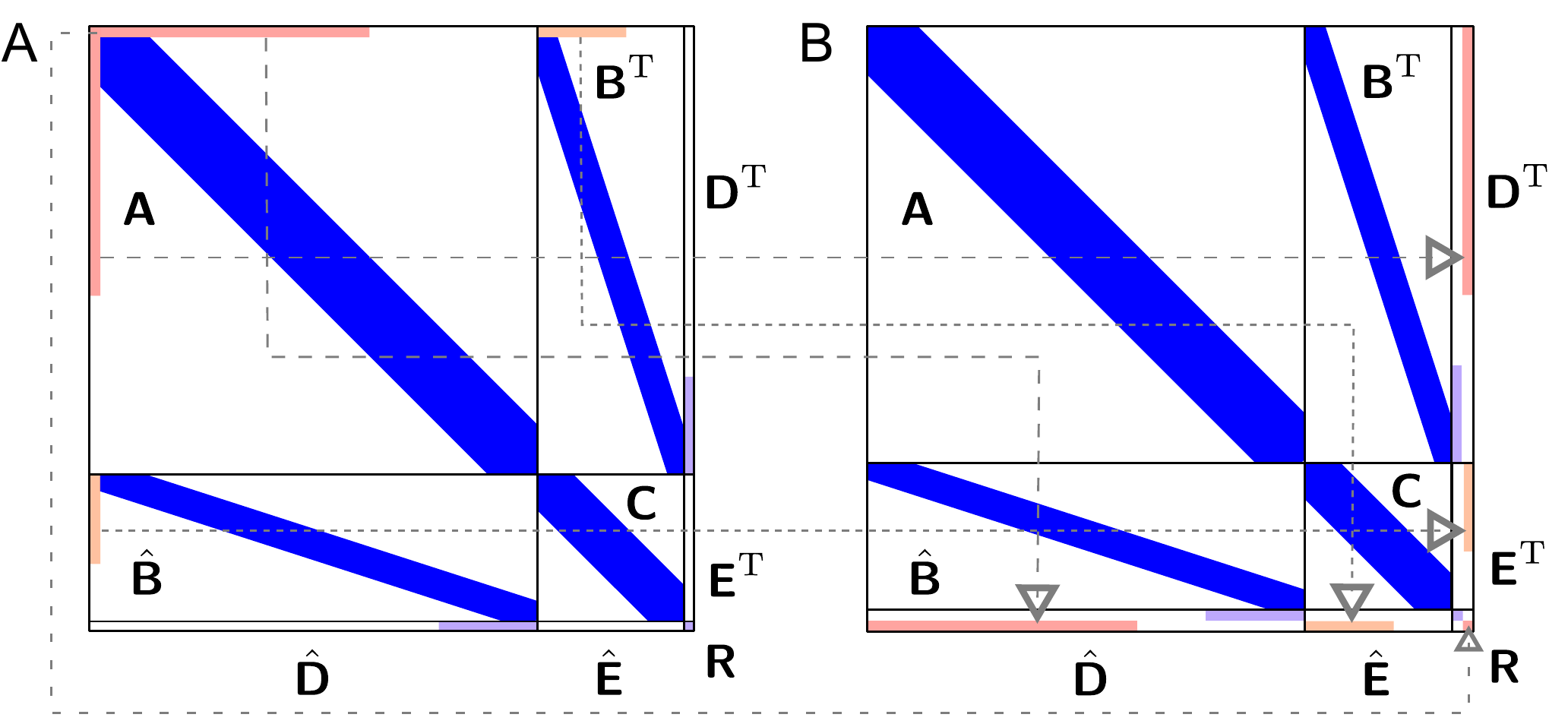}
\caption{Exemplary sketch of the sparsity pattern of a FrSI problem, i.e. of fluid flow in a deformable pipe with a resistive (0D) boundary condition, cf. the model shown in Fig.~\ref{fig:models_all}B for instance. \textbf{A.} Native sparsity pattern of fluid velocities, pressures, and multiplier variables. \textbf{B}. System after regrouping of ROM variables (projected fluid velocities) of the projected boundary solid model. These boundary velocity variables are moved out of the $\bsf{A}$ block into the $\bsf{R}$ block. Any coupling between bulk fluid and reduced solid variables hence move to the $\bsf{D}^{\mathrm{T}}$ and $\hat{\bsf{D}}$ blocks; likewise, the link between the pressure space and the reduced boundary variables is expressed in $\bsf{E}^{\mathrm{T}}$ and $\hat{\bsf{E}}$.}\label{fig:s3x3_frsi_sparsity}
\end{figure}

Fluid-reduced-solid interaction (FrSI) involves the Galerkin projection of prior data to a lower-dimensional subspace spanned by Eq.~(\ref{eq:Vgamma}), introducing a narrow-banded, dense sub-block into the fluid momentum and pressure coupling blocks. 
Similar to non-local boundary conditions, this can amplify the number of non-zeros and limit capacity in the use of efficient solvers. Figure~\ref{fig:s3x3_frsi_sparsity}A shows the native sparsity pattern of the 3\texttimes 3 fluid sub-system with reduced solid, either Eq.~(\ref{eq:lin_sys_rom_frsi_part}) or Eq.~(\ref{eq:lin_sys_rom_frsi_mono}) with omitted rows and columns from the ALE problem (for demonstrative purposes). Two main aspects that may deteriorate solver performance arise:
\begin{itemize}
    \item Narrow bands in the $\bsf{A}$ block reach out to far off-diagonal locations due to the dependence of a few projected boundary residuals on many non-projected velocity degrees of freedom belonging to the bulk, and vice versa. This may cause suboptimal performance of preconditioned iterative methods that operate on this block.
    \item Further dense narrow bands appear in the $\hat{\bsf{B}}$ and $\bsf{B}^{\mathrm{T}}$ blocks due to the dependence of many continuity residuals on a few projected boundary velocities, and vice versa. These render dense fill-ins of the size of the whole boundary pressure space when computing the product $\hat{\bsf{B}}\bsf{B}^{\mathrm{T}}$ which is involved in the Schur complement, cf. Eq.~(\ref{eq:schur_approx_3x3}). These may either cause depletion of CPUs' memory resources during matrix allocation, or may significantly slow down typical preconditioned iterative methods.
\end{itemize}
By restructuring into a 3\texttimes 3 (see Fig.~\ref{fig:s3x3_frsi_sparsity}B), these dense non-local sub-blocks can be efficiently clustered. 
This enables blocks with large rank (with degrees of freedom in the velocity and pressure spaces) maintain efficient sparsity. Further, Schur complement matrices which must be formed and their inverses approximated either retain local sparsity inherent to the finite element scheme (in the case of $\tilde{\bsf{S}}$) or are dense but low rank (in the case of $\tilde{\bsf{W}}$) due to the reduced dimension of the Galerkin projection.

\paragraph{Monolithic FrSI: Block Gauss-Seidel Schur Complement Preconditioner BGS-S3\texttimes 3}
When dealing with a fully monolithic approach to FrSI, the coupling of the grid motion described by the ALE domain problem to the fluid, and vice versa, is considered in a consistently linearized approach, yielding the system matrix Eq.~(\ref{eq:lin_sys_rom_frsi_mono}). We re-write this system in a more compact manner as follows:
\begin{align}
    \begin{bmatrix} \bsf{G} & \bsf{H}^{\mathrm{T}} \\ \\ \hat{\bsf{H}} & \bsf{F}  \end{bmatrix}\begin{bmatrix} \Delta\bsf{x}_{G} \\ \\ \Delta\bsf{x}_{F} \end{bmatrix}=-\begin{bmatrix} \bsf{r}_{G} \\ \\ \bsf{r}_{F} \end{bmatrix}, \label{eq:4x4_compact}
\end{align}
where the matrix $\bsf{F}$ represents the entire fluid-0D/RD system, i.e. Eq.~(\ref{eq:lin_sys_fluid_3x3}), and $\bsf{G}$ represents the Jacobian of the ALE problem. The coupling between the two fields is expressed in $\bsf{H}^{\mathrm{T}}$ and $\hat{\bsf{H}}$. We now propose the following (forward) Block Gauss-Seidel (BGS) preconditioner:
\begin{align}
    \bs{\mathcal{P}}_{\mathrm{BGS}(\mathrm{S}3\times 3)}^{-1} = \begin{bmatrix} \bsf{I} & \zerom \\ \\ \zerom & \bs{\mathcal{P}}_{\mathrm{S}3\times 3}^{-1}\end{bmatrix}
    \begin{bmatrix} \bsf{I} & \zerom \\ \\ -\hat{\bsf{H}} & \bsf{I}\end{bmatrix}
    \begin{bmatrix} \mathrm{inv}(\bsf{G}) & \zerom \\ \\ \zerom & \bsf{I}\end{bmatrix}.\label{eq:bgs_s3x3}
\end{align}

Likewise, the action of Eq.~(\ref{eq:bgs_s3x3}) on a vector $\bsf{x}$ with output $\bsf{y}$ can be expressed by Alg.~\ref{alg:4x4}:
\begin{algorithm}
\caption{Action of $\bs{\mathcal{P}}_{\mathrm{BGS}(\mathrm{S}3\times 3)}^{-1}$ on a vector $\bsf{x}=[\bsf{x}_{G}, \bsf{x}_{F}]^{\mathrm{T}}$ with output $\bsf{y}=[\bsf{y}_{G}, \bsf{y}_{F}]^{\mathrm{T}}$}\label{alg:bgs_schur3x3}
\begin{algorithmic}[1]
\State Solve for a grid displacement $\bsf{y}_{G}$:
\Statex \qquad $\bsf{G} \bsf{y}_{G} = \bsf{x}_{G}$
\Statex using a preconditioned iterative method
\Statex
\State Compute: $\bsf{z}_{F} = \bsf{x}_{F} - \hat{\bsf{H}} \bsf{y}_{G}$
\Statex
\State Apply $\bs{\mathcal{P}}_{\mathrm{S}3\times 3}^{-1}$ for the fluid+reduced variables $\bsf{y}_{F}$:
\Statex \qquad $\bsf{F} \bsf{y}_{F} = \bsf{z}_{F}$
\Statex using Alg.~\ref{alg:3x3}
\end{algorithmic}\label{alg:4x4}
\end{algorithm}

We note that the order of solving first the grid displacement and subsequently the fluid-0D/RD system is motivated by the intensity of coupling between the fields and the degree of nonlinearity involved. While the grid deformation impacts fluid momentum, continuity, as well as reduced models, the fluid displacement only impacts the ALE grid momentum via Dirichlet conditions at the boundary. This is in accordance with other BGS schemes that have been used for FSI \cite{gee2011}.\\

\textit{\textbf{Remark:} To reflect the bi-directional coupling between ALE and fluid inside the preconditioner, a symmetric BGS method could be employed instead, necessitating one additional solve of the grid motion problem. However, our experiences have shown only very little improvement with respect to GMRES iterations, which do not compensate for the computational cost needed to do the extra solve.}

\paragraph{Partitioned FrSI}
In case of using a partitioned approach to FrSI, that is, solving Eq.~(\ref{eq:lin_sys_rom_frsi_part}) and Eq.~(\ref{eq:lin_sys_ale}) in a staggered manner within the Newton scheme, the block preconditioner S3\texttimes 3, Eq.~(\ref{eq:3x3_inverse_block}) or Alg. \ref{alg:3x3}, along with the regrouping of ROM variables (Fig.~\ref{fig:s3x3_frsi_sparsity}B) may be used in a straightforward manner. For the subsequent solution of the ALE problem Eq.~(\ref{eq:lin_sys_ale}), a single-field preconditioned iterative algorithm can be employed, \emph{e.g.}, a GMRES algorithm with algebraic multigrid preconditioning. 

\subsection{Test Examples}\label{subsec:testexamples}
\begin{figure}[!htp]
\centering
\includegraphics[width=1\textwidth]{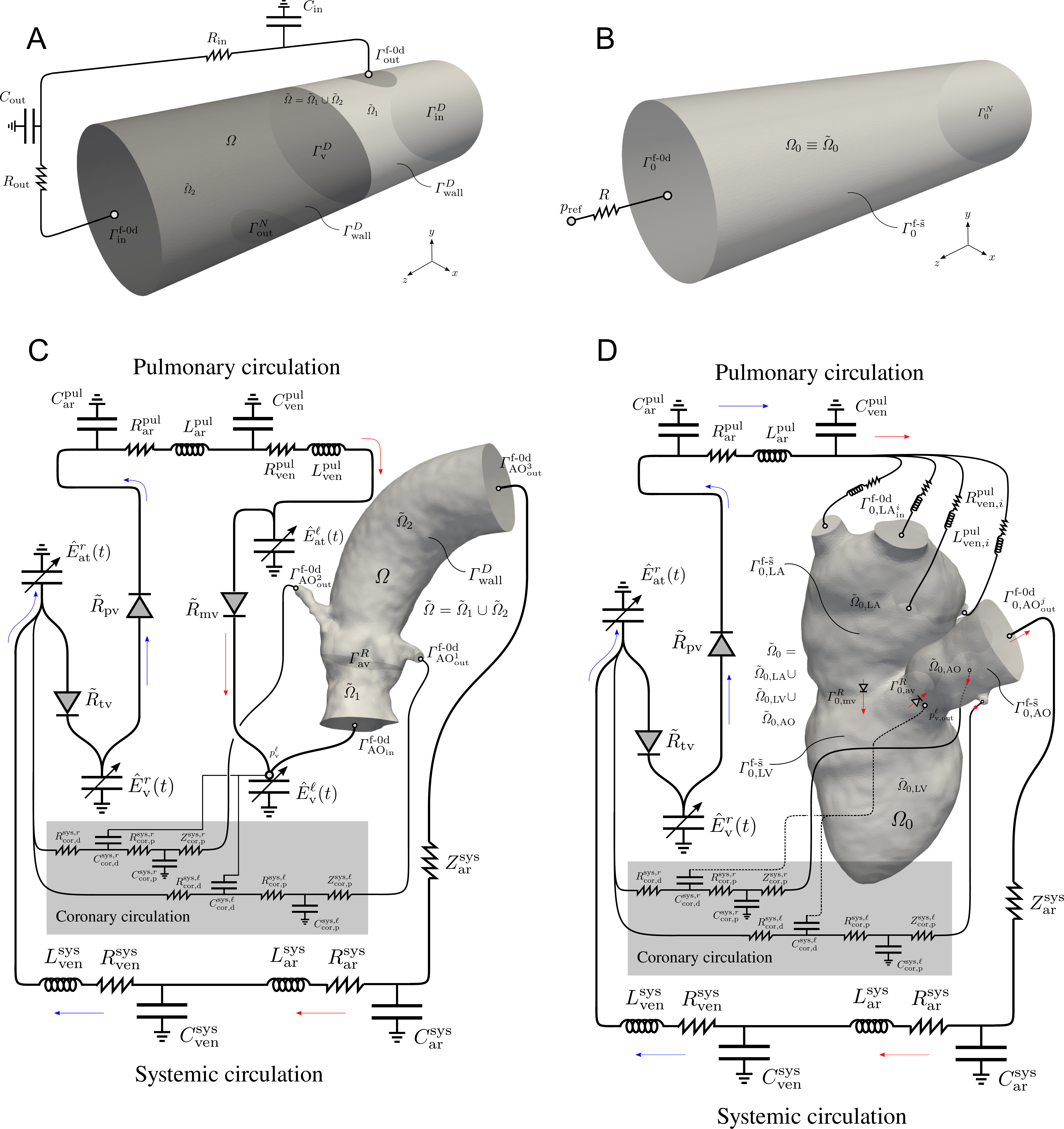}
\caption{Example problems used to evaluate different preconditioners.
\textbf{A.} Fluid flow in a blocked pipe with bypass via a 0D model. The flow entering the domain is inhibited to pass through the pipe but directed to the other end via a lumped-parameter model.
\textbf{B.} FrSI model of deformable pipe, connected to a resistive boundary condition.
\textbf{C.} Blood flow through patient-specific rigid aortic arch model, connected to a closed-loop circulatory system model.
\textbf{D.} FrSI model of the left heart (atrium, ventricle) connected to aortic outflow tract, coupled to a closed-loop circulatory system model.
For more details on system setup for all examples, see the \emph{Supplementary Material}.
}\label{fig:models_all}
\end{figure}

To test the impact of the preconditioners discussed in this section, we consider four examples of increasing complexity (see Fig.~\ref{fig:models_all}). 
More details on the specifics of each model problem formulation can be found in the \emph{Supplementary Material}, Sec.~\ref{app:examples}.
\paragraph{Blocked Pipe Flow with Bypass (Fluid-0D)}
First, we consider the academic example of Eulerian fluid flow through an idealized pipe (see Fig.~\ref{fig:models_all}A), with a blockage that is bypassed via a 0D model interconnecting two boundaries of the 3D fluid region (Sec.~\ref{sec:examples_fluidpipe}). 
The example is simple enough for reproducibility but demonstrates the influence of more interconnected coupling between 3D and 0D models, as is common to more complex cardiovascular simulations.
\paragraph{Deformable Pipe Flow with Resistive BC (FrSI-0D)}
The second example presented builds on the same pipe model, but with different boundary conditions, now accounting for the deformability of the wall by means of the fluid-reduced-solid interaction (FrSI) \cite{hirschvogel2024-frsi} method (fluid in ALE reference frame)  (see Fig.~\ref{fig:models_all}B). 
The setup is kept equally simple in order to facilitate reproducibility, using only one synthetically generated deformation mode for the lateral wall and a resistance-only 0D model at the outflow (Sec.~\ref{sec:examples_frsipipe}).
\paragraph{Blood Flow in Patient-Specific Aortic Arch (Fluid-0D)}
Increasing the problem complexity, we present a patient-specific example of Eulerian blood flow through a rigid aortic arch in Sec.~\ref{sec:examples_fluidaort}, coupled to a closed-loop circulation model that interlinks several in- and outflow boundaries of the 3D model (see Fig.~\ref{fig:models_all}C). 
It hence can be considered as a very general case of 3D-0D coupling using a complex geometry with boundary layer as well as delicate outflow structures like the coronary arteries.
\paragraph{Hemodynamics in the Left Heart (FrSI-0D)}
Finally, we present a FrSI model of a patient-specific left heart and aortic root model in Sec.~\ref{sec:examples_frsiheart}, linked to a closed-loop model of systemic, pulmonary, and coronary circulation with several in- and outflows to and from the 3D domain, including immersed valve models (see Fig.~\ref{fig:models_all}D). We further incorporate novel methods that deal with a POD space that is regionally local to atrium, ventricle, and aorta, hence allowing individual contraction behavior of the active chambers. This model is meant to demonstrate the performance of the solver even for computationally challenging large-deformation ALE fluid problems with sudden changes in the flow state (isovolumetric phases, ejection, filling).\\

\begin{figure}[!htp]
\centering
\includegraphics[width=1\textwidth]{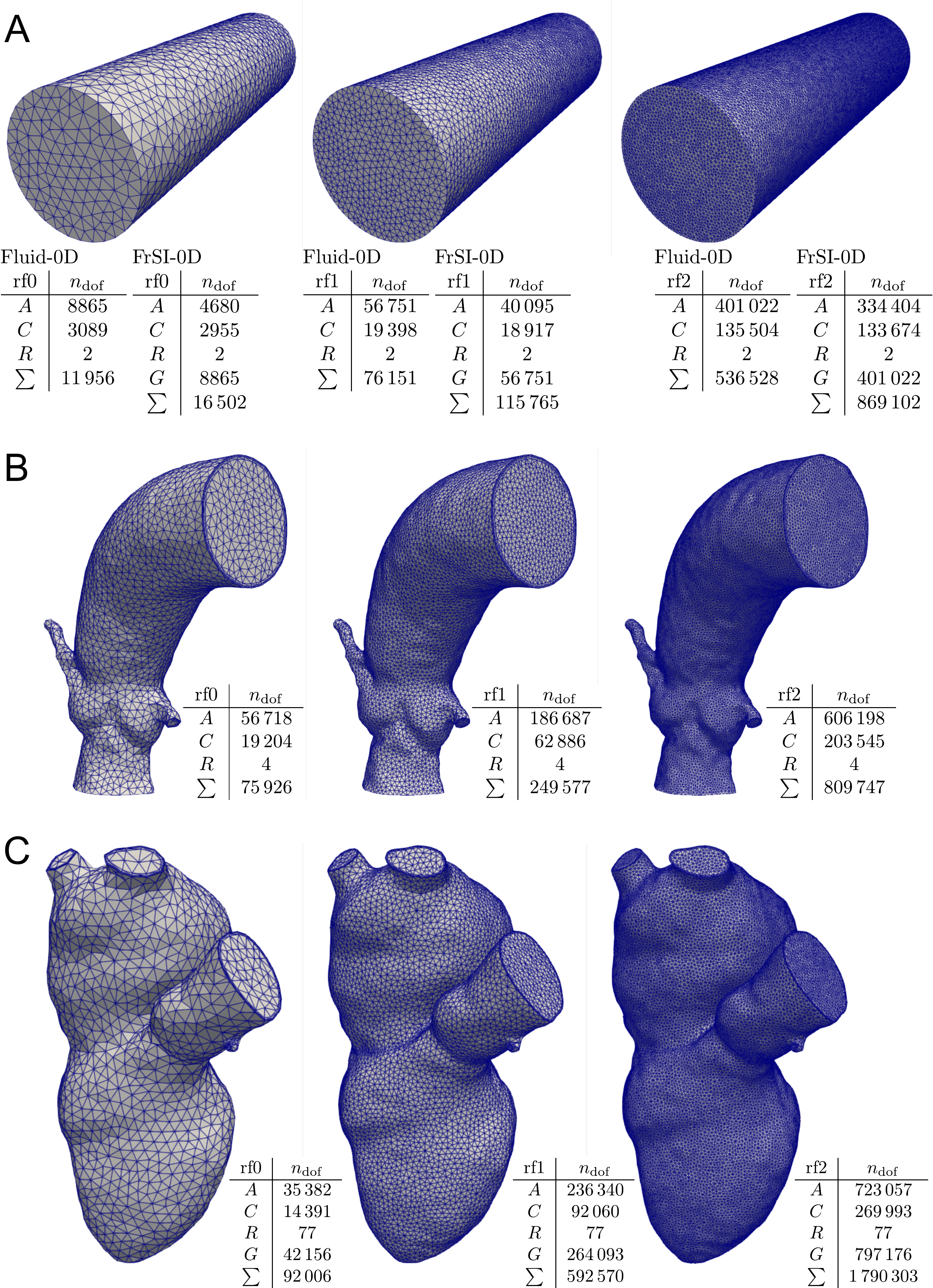}
\caption{$\mathbb{P}_1\mhyphen\mathbb{P}_1$ meshes and degree of freedom tables for the different examples presented. Number of degrees of freedom $n_{\mathrm{dof}}$, relating to $A$ (fluid momentum, velocity space), $C$ (fluid continuity, pressure space), $R$ (reduced variables), and $G$ (grid/ALE momentum, displacement space) block, as well as the sum. \textbf{A}. Pipe model, same mesh for fluid-0D as well as FrSI problem. Note the difference in the size of the pressure degrees of freedom ($C$) due to the split function space at the inner boundary for the fluid-0D problem. \textbf{B}. Aortic arch. \textbf{C}. Left heart.}\label{fig:meshes_dofs_all}
\end{figure}

For each example, the efficacy of the solvers is evaluated by considering three levels of spatial refinement (rf0, rf1, rf2), which are shown in Fig.~\ref{fig:meshes_dofs_all} for all of the examples, along with the degree of freedom tables for each of the sub-systems (velocity, pressure, multipliers/reduced variables, grid displacement).
To examine solver performance, total run time is quantified along with the number of linear solves per time step (GMRES iterations) and number of nonlinear iterations.
Critically, solvers must demonstrate parallel scalability to enable effective use for large problems.
To assess scalability, both weak and strong scaling are evaluated.
For comparability, all nonlinear and linear solver parameters remain constant for the examples and results reported in this chapter. They are summarized in Tab.~\ref{tab:params_solver}.

\subsection{Software and Tools}\label{subsec:software}
All methodologies presented are implemented in two different scientific software environments for the purpose of validation and demonstration of platform-independence. First, the Fortran-based multi-physics finite element code $\mathcal{C}\mathrm{Heart}$ \cite{lee2016} is used, which has been developed at King's College London as well as at the University of Michigan for the last several years. Second, the open-source solver Ambit \cite{hirschvogel2024-ambit} is employed, a software package written in Python building on the recent finite element backend FEniCSx \cite{baratta2023-dolfinx,logg2012-fenics} and tailored towards solving multi-physics problems in cardiac mechanics. Both solvers make use of the linear algebra backend PETSc \cite{balay2022-petsc}, either through the Fortran bindings or by means of the respective Python wrappers. All idealized models which are presented are constructed in FreeCAD \cite{freecad2012} and meshed in Gmsh \cite{gmsh2009}. The patient-specific geometries are meshed using the proprietary tool SimModeler \cite{simmodeler}. All computations are run on the Great Lakes computing cluster of the University of Michigan, whose standard partition consists of 455 computing nodes each comprising 36 cores (2 \texttimes\,3.0 GHz Intel Xeon Gold 6154 processors). To guarantee optimal performance measures, a node always is requested for exclusive usage, even if demanding less than all its 36 cores.

\begin{table}[!htp]
\begin{center}
\caption{Nonlinear and linear solver parameters that are used for all problems presented. Only parameters deviating from the default PETSc parameters are listed. PETSc release version 3.19.5 has been used.}\label{tab:params_solver}
\begin{tabular}{rl|c|l}\hline
\multicolumn{4}{l}{Nonlinear solver} \\\hline
\multicolumn{2}{l|}{Algorithm} & \multicolumn{2}{l}{Newton-Raphson \cite{dedieu2015-newton}} \\
\multicolumn{2}{l|}{Fluid momentum residual tolerance} & $10^{-7}$ & $[\frac{\mathrm{mN}}{\mathrm{s}}]$ \\
\multicolumn{2}{l|}{Fluid continuity residual tolerance} & $10^{-7}$ & $[\frac{\mathrm{mN}\cdot\mathrm{mm}}{\mathrm{s}}]$ \\
\multicolumn{2}{l|}{3D-0D constraint residual tolerance} & $10^{-7}$ & $[\frac{\mathrm{mN}\cdot\mathrm{mm}}{\mathrm{s}}]$ \\
\multicolumn{2}{l|}{ALE domain motion residual tolerance} & $10^{-5}$ & $[\mathrm{mN}]$ \\\\\hline
\multicolumn{4}{l}{Linear solver} \\\hline
\multicolumn{2}{l|}{Algorithm} & \multicolumn{2}{l}{FGMRES \cite{saad1993} with mod. Gram-Schmidt orthogonalization} \\
\multicolumn{2}{l|}{GMRES restart} & \multicolumn{2}{l}{$100$} \\
\multicolumn{2}{l|}{Relative linear tolerance} & \multicolumn{2}{l}{$10^{-5}$} \\
\multicolumn{2}{l|}{Absolute linear tolerance} & \multicolumn{2}{l}{$10^{-8}$} \\
\multicolumn{2}{l|}{Maximum iterations} & \multicolumn{2}{l}{$500$} \\
\multicolumn{2}{l|}{Preconditioner} & \multicolumn{2}{l}{BGS-S3\texttimes 3, \; S3\texttimes 3, \; S2\texttimes 2(p), \; or \; S2\texttimes 2(c/c$^{\star}$)} \\\\
\multicolumn{4}{l}{\textit{A-solve (fluid momentum)}} \\\hline
\multicolumn{2}{l|}{Algorithm} & \multicolumn{2}{l}{FGMRES \cite{saad1993}} \\
\multicolumn{2}{l|}{Relative linear tolerance} & \multicolumn{2}{l}{$10^{-3}$} \\
\multicolumn{2}{l|}{Maximum iterations} & \multicolumn{2}{l}{$100$} \\
\multicolumn{2}{l|}{Preconditioner} & \multicolumn{2}{l}{Algebraic Multigrid (hypre \cite{falgout2002-hypre} BoomerAMG \cite{henson2002-boomeramg})} \\
 & \multicolumn{3}{l}{\textit{AMG Settings}} \\\hline
 & Strong threshold & \multicolumn{2}{l}{$0.8$} \\
 & Smoother up-cycle & \multicolumn{2}{l}{$\ell_1$ Gauss-Seidel} \\
 & Smoother down-cycle & \multicolumn{2}{l}{$\ell_1$ Gauss-Seidel} \\\\
\multicolumn{4}{l}{\textit{S-solve (Schur complement)}} \\\hline
\multicolumn{2}{l|}{Algorithm} & \multicolumn{2}{l}{FGMRES \cite{saad1993}} \\
\multicolumn{2}{l|}{Relative linear tolerance} & \multicolumn{2}{l}{$10^{-3}$} \\
\multicolumn{2}{l|}{Maximum iterations} & \multicolumn{2}{l}{$100$} \\
\multicolumn{2}{l|}{Preconditioner} & \multicolumn{2}{l}{Algebraic Multigrid (hypre \cite{falgout2002-hypre} BoomerAMG \cite{henson2002-boomeramg})} \\
 & \multicolumn{3}{l}{\textit{AMG Settings}} \\\hline
 & Strong threshold & \multicolumn{2}{l}{$0.4$} \\
 & Smoother up-cycle & \multicolumn{2}{l}{$\ell_1$-scaled Jacobi} \\
 & Smoother down-cycle & \multicolumn{2}{l}{$\ell_1$-scaled Jacobi} \\\\
\multicolumn{4}{l}{\textit{W-solve (reduced variables)} (only applies to BGS-S3\texttimes 3 or S3\texttimes 3 preconditioner)} \\\hline
\multicolumn{2}{l|}{Algorithm} & \multicolumn{2}{l}{Direct solve (MUMPS \cite{amestoy2001-mumps})} \\\\
\multicolumn{4}{l}{\textit{G-solve (ALE domain motion)} (only applies to fluid in ALE formulation)} \\\hline
\multicolumn{2}{l|}{Algorithm} & \multicolumn{2}{l}{GMRES \cite{saad1986}} \\
\multicolumn{2}{l|}{Relative linear tolerance} & \multicolumn{2}{l}{$10^{-3}$} \\
\multicolumn{2}{l|}{Maximum iterations} & \multicolumn{2}{l}{$100$} \\
\multicolumn{2}{l|}{Preconditioner} & \multicolumn{2}{l}{Algebraic Multigrid (hypre \cite{falgout2002-hypre} BoomerAMG \cite{henson2002-boomeramg})} \\
 & \multicolumn{3}{l}{\textit{AMG Settings}} \\\hline
 & Strong threshold & \multicolumn{2}{l}{$0.7$} \\
\end{tabular}
\end{center}
\end{table}

\clearpage
\section{Fluid-0D: Blocked Pipe Flow with Bypass}\label{sec:examples_fluidpipe}
In this section, we consider the flow of an Eulerian fluid through a simple pipe geometry that experiences coupling through reduced order models across different non-local surfaces (see Fig.~\ref{fig:models_all}A).
Full details on the problem setup are provided in the \emph{Supplementary Materials}, Sec.~\ref{app:examples_fluidpipe}.
Briefly, a prescribed inlet flow ($\Gm_{\mathrm{in}}^D$) is ramped up over time.
Flow through the pipe is blocked by an inclined surface (Dirichlet zero) and forced through an outflow ($\Gm_{\mathrm{out}}^{\fF\mhyphen\mathrm{0d}}$) coupled using a 0D circuit model to the other end of the pipe ($\Gm_{\mathrm{in}}^{\fF\mhyphen\mathrm{0d}}$) and allowed to exit the domain through a side outlet ($\Gm_{\mathrm{out}}^{N}$). The lumped model is described in \emph{Supplementary Materials}, Sec.~\ref{app:crlink}.

\begin{figure}[!htp]
\centering
\includegraphics[width=1.0\textwidth]{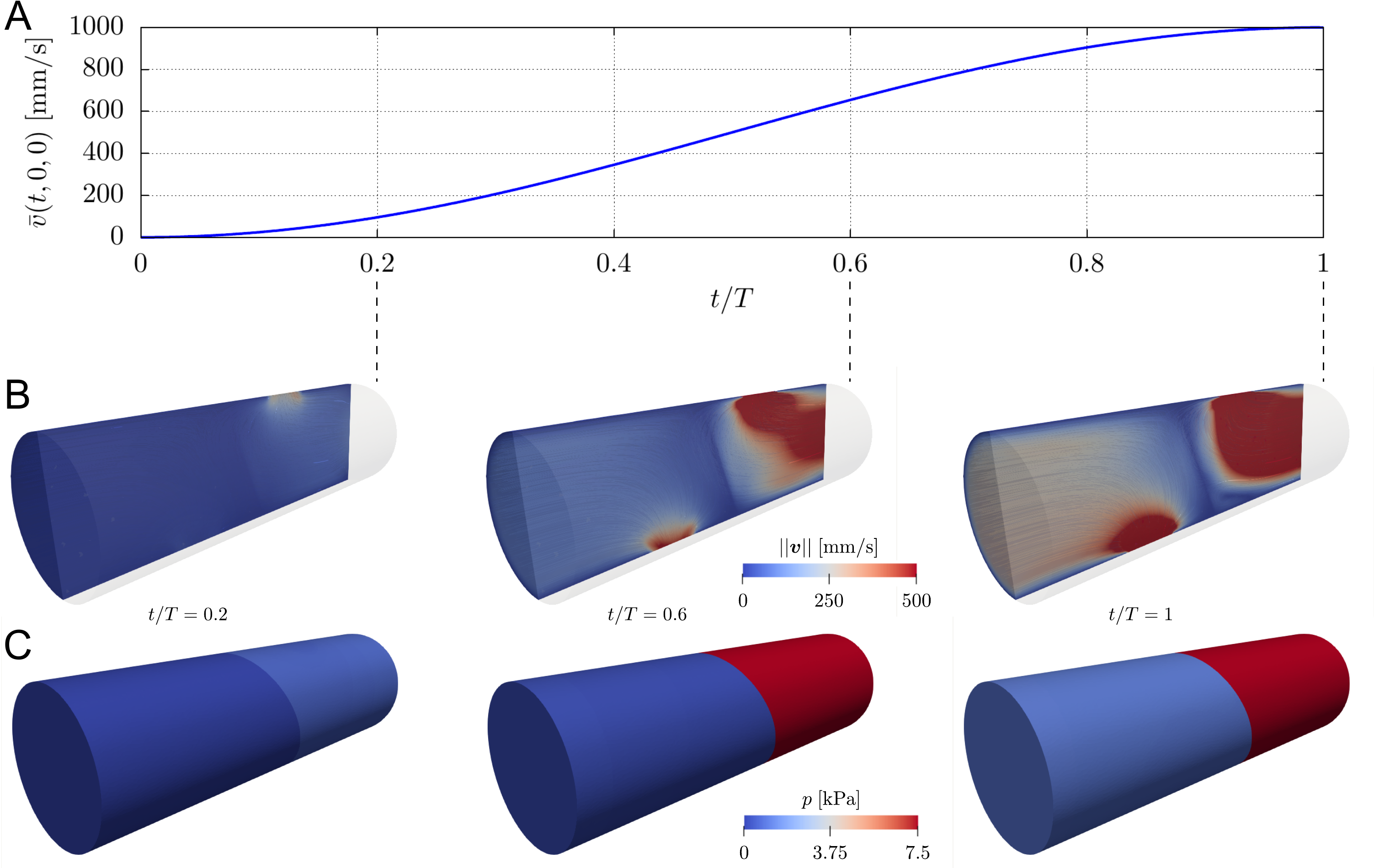}
\caption{\textit{Pipe fluid-0D} (mesh rf2): \textbf{A.} Prescribed inflow velocity over time at center of $\Gm_{\mathrm{in}}^{D}$, Eq.~(\ref{eq:fluidpipe_dbc_vbar}). \textbf{B.} Magnitude of fluid velocity $\bs{v}$ on longitudinal cut through domain, shading shows velocity streamlines. \textbf{C.} Fluid pressure $p$.}\label{fig:pipefluid_results}
\end{figure}

Figure~\ref{fig:meshes_dofs_all}A shows the three different meshes along with the degree of freedom tables for each sub-system as well as the sum. 
The simulation output for the finest mesh ($\mathrm{rf2}$) of velocity and pressure along with the prescribed inlet velocity profile are shown in Fig.~\ref{fig:pipefluid_results}.

\begin{figure}[!htp]
\centering
\includegraphics[width=1\textwidth]{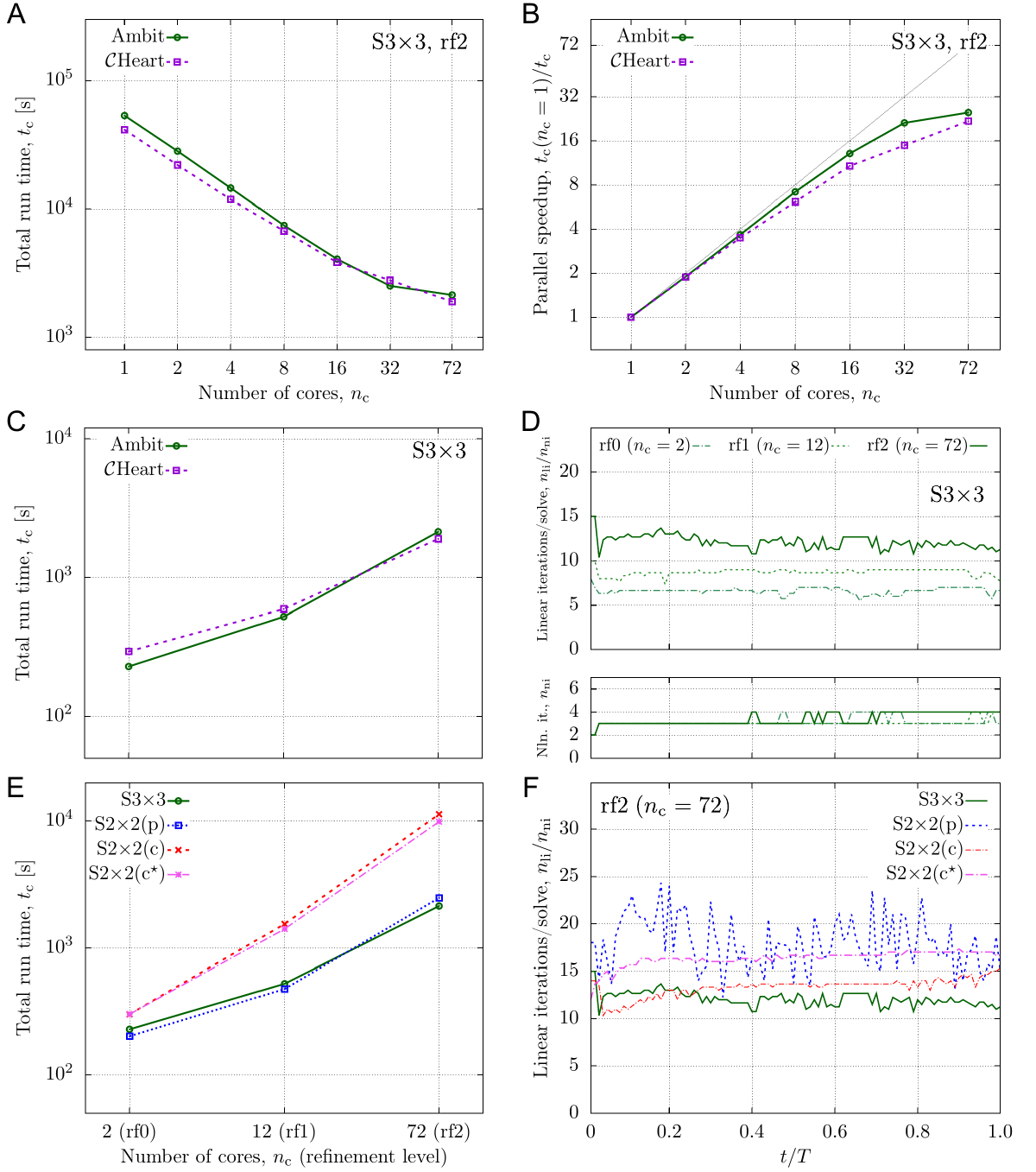}
\caption{\textit{Pipe fluid-0D:} \textit{\textbf{Top row:}} Strong scaling for S3\texttimes 3 preconditioner on mesh $\text{rf2}$. \textbf{A.} Total run time over number of cores. \textbf{B.} Parallel speedup over number of cores. Black dashed line indicates ``optimal'' speedup in absence of any inter-core communication. \textit{\textbf{Middle row:}} Weak scaling of S3\texttimes 3 for three different refinements and number of cores, with a ratio of degrees of freedom per core of $n_{\mathrm{dof}}/n_{\mathrm{c}} \approx 7000$. \textbf{C.} Total run time. \textbf{D.} Linear iterations per solve over relative physical time (top), nonlinear iterations over relative physical time (bottom) (solver Ambit). \textit{\textbf{Bottom row:}} Comparison of weak scaling for preconditioners S3\texttimes 3, S2\texttimes 2(p) (0D multipliers lumped into fluid pressure block), and S2\texttimes 2(c/c$^{\star}$) (0D multipliers condensed from linear system, without/with neglecting offdiagonal fill-in blocks) (solver Ambit). \textbf{E.} Total run time. \textbf{F.} Linear iterations per solve over relative physical time, mesh rf2.}\label{fig:pipefluid_scaling_all}
\end{figure}

\paragraph{Strong scaling}
The strong scaling properties of the S3\texttimes 3 preconditioner are investigated by comparing the run time $t_{\mathrm{c}}$ for the finest discretization $\mathrm{rf2}$ with increasing number of cores, $n_{\mathrm{c}}$. 
For optimal usage of the hardware architecture (which is organized in computing nodes of 36 cores each), $n_{\mathrm{c}}$ is incremented in multiples of 36 once more than one node is occupied. 
Figure \ref{fig:pipefluid_scaling_all}A shows the total run time $t_{\mathrm{c}}$ and Fig. \ref{fig:pipefluid_scaling_all}B the parallel speedup over the number of cores $n_{\mathrm{c}}$ for both solvers $\mathcal{C}\mathrm{Heart}$ and Ambit.
Speedups are close to linear up to 32 cores, eventually deviating from the ideal line for larger number of cores and therefore smaller per-core workloads (degrees of freedom per core ratio $n_{\mathrm{dof}}/n_{\mathrm{c}} \approx 7000$ for $n_{\mathrm{c}}=72$), which agrees with other solvers for comparable problems \cite{seo2019}.

\paragraph{Weak scaling}
The weak scaling properties of the S3\texttimes 3 preconditioner are investigated based on the three different meshes $\mathrm{rf0}$, $\mathrm{rf1}$, and $\mathrm{rf2}$, choosing the number of cores used for computing the problem to yield a ratio of degrees of freedom per core of $n_{\mathrm{dof}}/n_{\mathrm{c}} \approx 7000$. Figure \ref{fig:pipefluid_scaling_all}C shows the total run time $t_{\mathrm{c}}$ over the number of cores (the respective refinement level) for both solvers, and Fig. \ref{fig:pipefluid_scaling_all}D depicts the number of linear iterations per solve, $n_{\mathrm{li}}/n_{\mathrm{ni}}$ (top) as well as the number of nonlinear iterations $n_{\mathrm{ni}}$ (bottom) over the relative time of the simulation (note, for ease, only results for solver Ambit shown). Both solvers show comparable solution times. Linear iterations per solve increase slightly with mesh refinement, apparently due to a substantially improved resolution of the flow field. Nonlinear iterations are only mildly affected by mesh refinement, ranging within $n_{\mathrm{ni}} \in \{2,4\}$.

\paragraph{Benchmarking with Other Approaches}
The performance of S3\texttimes 3 is compared to using S2\texttimes 2 for three different types of matrix systems. The application of the preconditioner to the condensed systems Eq.~(\ref{eq:lin_sys_fluid_cnd})+(\ref{eq:lin_sys_fluid_cndII}) (Fig.~\ref{fig:s2x2c_s2x2p_sparsity}A\&B) is denoted by S2\texttimes 2(c) and S2\texttimes 2(c$^{\star}$), and its application to the matrix system Eq.~(\ref{eq:lin_sys_fluid}) with consolidated pressure and multiplier variables (Fig.~\ref{fig:s2x2c_s2x2p_sparsity}C) by S2\texttimes 2(p). 
Figure~\ref{fig:pipefluid_scaling_all}E shows the total run time $t_{\mathrm{c}}$ over different numbers of cores (different refinement levels), and Fig.~\ref{fig:pipefluid_scaling_all}F depicts the number of linear iterations per solve over the relative simulation time for mesh $\mathrm{rf2}$ ($n_{\mathrm{c}}=72$).  Iteration-wise, S2\texttimes 2(p) needs nearly twice as many per solve than S3\texttimes 3, which however can be partly compensated by reduced costs for preconditioner setup and reduced overall effort. 
Nonetheless, S3\texttimes 3 is still superior in overall run time for the finest mesh due to the comparably low iteration count. On the contrary, S2\texttimes 2(c/c$^{\star}$) exhibit linear iterations comparable to S3\texttimes 3, but with an over five-/four-fold higher total run time.

\clearpage
\section{ALE Fluid-RD/0D (FrSI): Deformable Pipe Flow with Resistive BC}\label{sec:examples_frsipipe}

A fluid-reduced-solid interaction (FrSI) model \cite{hirschvogel2024-frsi} of Arbitrary Eulerian-Lagrangian (ALE) fluid flow in a deformable pipe is considered in this section (see Fig.~\ref{fig:models_all}B).
Full details on the problem setup are provided in the \emph{Supplementary Materials}, Sec.~\ref{app:examples_frsipipe}.
Briefly, a time-controlled inlet pressure at $\Gm_{0}^{N}$ is prescribed, and the properties of lateral wall $\Gm_{0}^{\mathrm{\fF\mhyphen\fSr}}$ are governed by a reduced-dimensional solid mechanics model which exhibits hyperelastic and viscous properties. 
For ease and repeatability, only one Galerkin ROM degree of freedom, pointing in radial direction of the lateral surface, is used to define the admissible subspace of the reduced solid wall model. Hence, the global lateral surface motion is constrained in radial direction. The whole pipe domain is further fixed in axial $z$-direction. On $\Gm_{0}^{\mathrm{f\mhyphen 0d}}$, a resistive 0D model accounts for a flow-dependent pressure (lumped model description in \emph{Supplementary Materials}, Sec.~\ref{app:resist}).\\

\begin{figure}[!htp]
\centering
\includegraphics[width=1.0\textwidth]{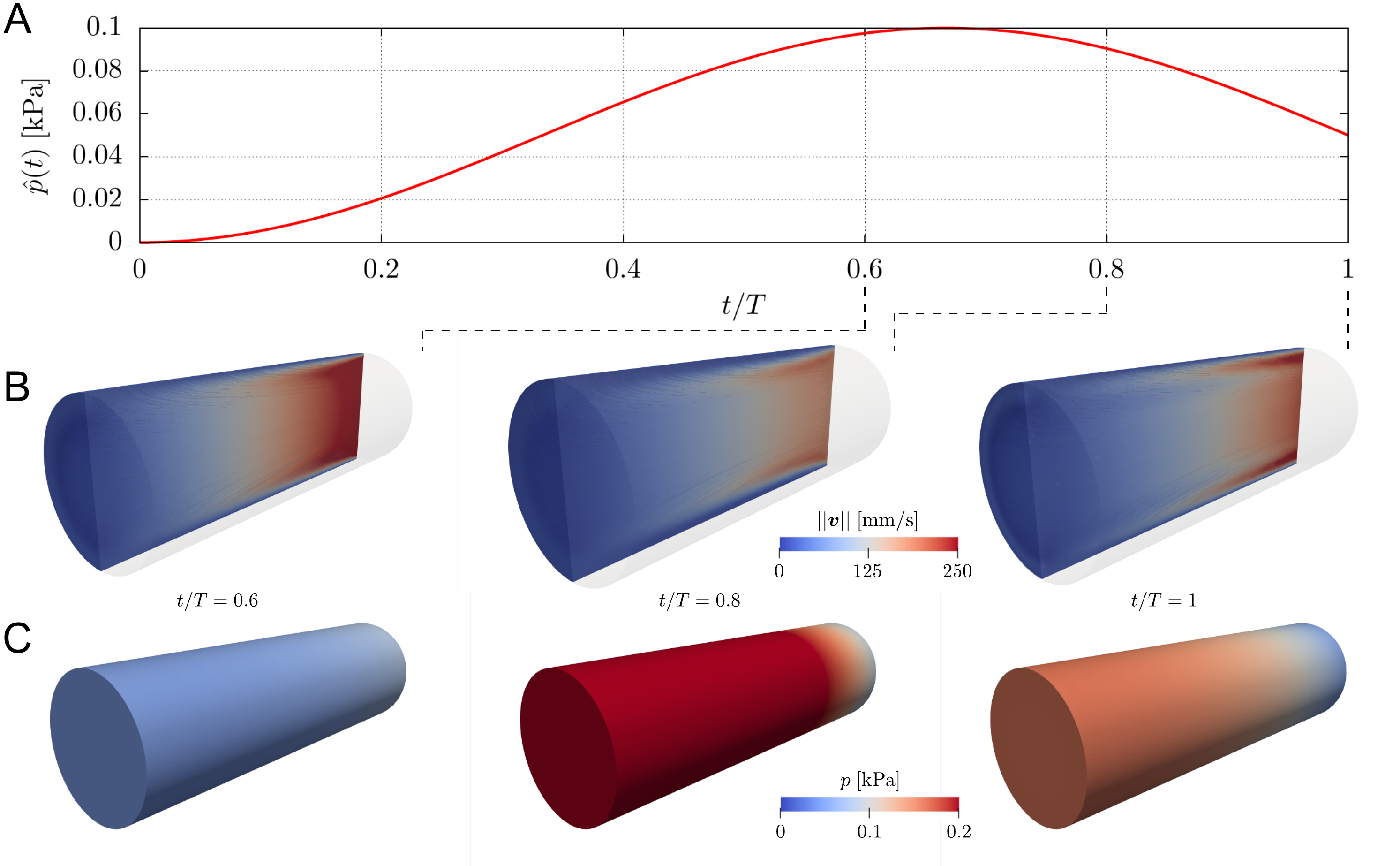}
\caption{\textit{Pipe FrSI-0D} (mesh rf2): \textbf{A.} Prescribed inflow pressure over time at $\Gm_{0}^{N}$, Eq.~(\ref{eq:frsipipe_nbc_phat}). \textbf{B.} Magnitude of fluid velocity $\bs{v}$ on longitudinal cut through deformed domain $\Om$, shading shows velocity streamlines. \textbf{C.} Fluid pressure $p$, plotted on undeformed reference domain $\tilde{\Om}_{0}$.}\label{fig:pipefrsi_results}
\end{figure}

The same three different spatial discretizations (refinement levels) as for the fluid-0D pipe model are used, cf. Fig.~\ref{fig:meshes_dofs_all}A. Despite the identity of the meshes, the velocity space and hence the dimension of the momentum ($A$) block is reduced compared to the rigid model due to the projection operator Eq.~(\ref{eq:Vgamma}). Hence, the $R$-block comprises one multiplier from 3D-0D coupling and one wall degree of freedom. The simulation output for the finest mesh ($\mathrm{rf2}$) of velocity and pressure along with the prescribed inlet pressure profile are shown in Fig.~\ref{fig:pipefrsi_results}.

\begin{figure}[!htp]
\centering
\includegraphics[width=1\textwidth]{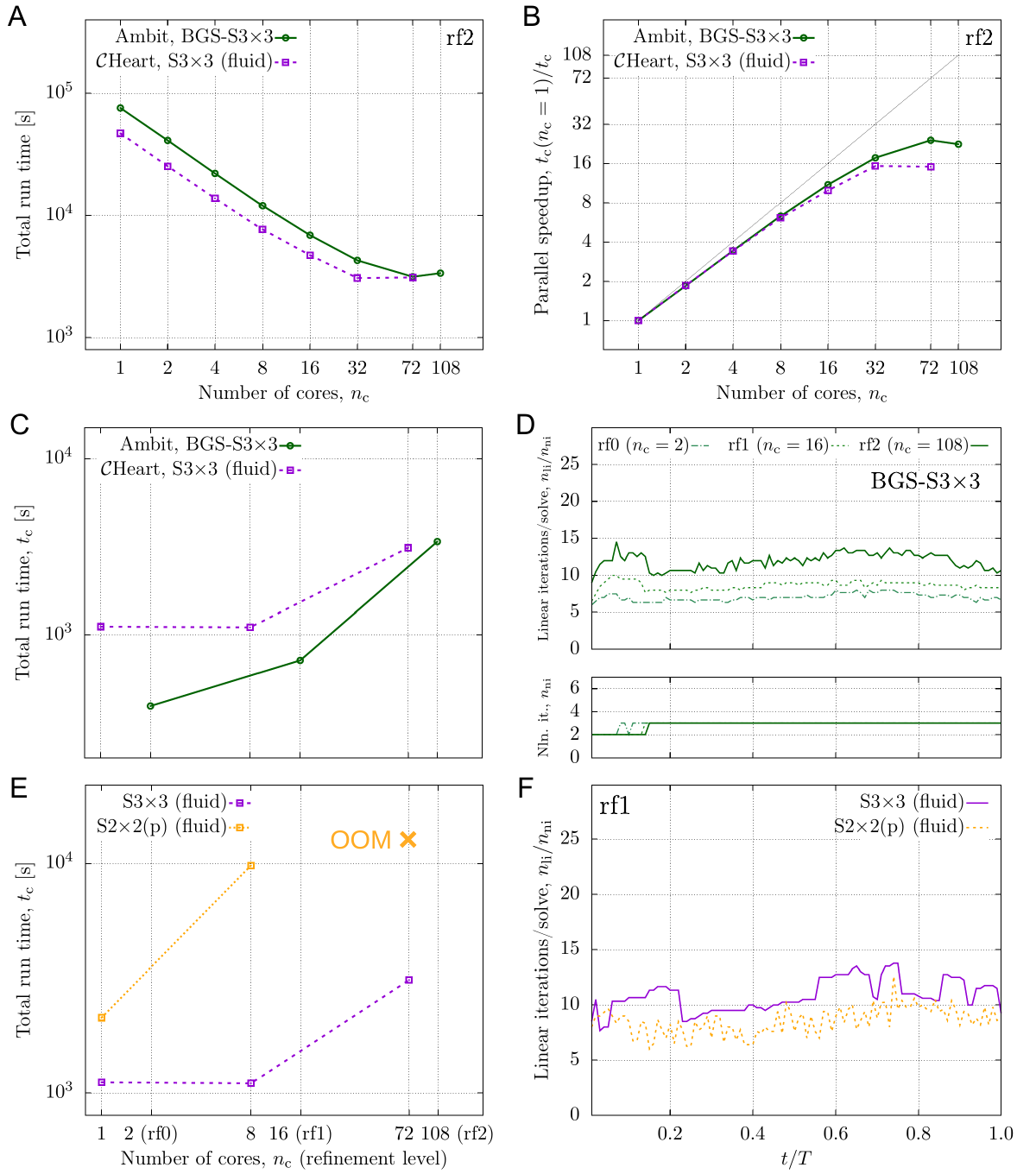}
\caption{\textit{Pipe FrSI-0D:} \textit{\textbf{Top row:}} Strong scaling for BGS-S3\texttimes 3 (monolithic FrSI) and S3\texttimes 3 (partitioned FrSI) on mesh $\text{rf2}.$ \textbf{A.} Total run time over number of cores. \textbf{B.} Parallel speedup over number of cores. Black dashed line indicates ``optimal'' speedup in absence of any inter-core communication. \textit{\textbf{Middle row:}} Weak scaling for BGS-S3\texttimes 3 (monolithic FrSI) and S3\texttimes 3 (partitioned FrSI) preconditioner for three different refinements and number of cores, with a ratio of degrees of freedom per core of $n_{\mathrm{dof}}/n_{\mathrm{c}} \approx 7000$. \textbf{C.} Total run time. \textbf{D.} Linear iterations of BGS-S3\texttimes 3 per solve over relative physical time (top), nonlinear iterations over relative physical time (bottom) (solver Ambit). \textit{\textbf{Bottom row:}} Comparison of weak scaling for preconditioners S3\texttimes 3 and S2\texttimes 2(p) (0D multipliers lumped into fluid pressure block, Galerkin ROM variables kept in velocity block) (solver $\mathcal{C}\mathrm{Heart}$). \textbf{E.} Total run time over number of cores for partitioned FrSI approach. $\mathrm{OOM}$ indicates that the problem could not be computed with S2\texttimes 2(p) due to memory limits (considering a specifiable maximum of $5\;\mathrm{GB}$ per CPU). \textbf{F.} Linear iterations per solve over relative physical time, mesh rf1.}\label{fig:pipefrsi_scaling_all}
\end{figure}

\paragraph{Strong scaling}
We investigate the strong scaling properties of both the BGS-S3\texttimes 3 preconditioner, which is suitable for a monolithic FrSI implementation (provided only by solver Ambit), as well as the S3\texttimes 3 preconditioner applied to the partitioned FrSI approach (solver $\mathcal{C}\mathrm{Heart}$). For the latter, the standalone ALE domain motion problem is solved with an AMG-preconditioned GMRES solver. The run time $t_{\mathrm{c}}$ for the finest discretization $\mathrm{rf2}$ with increasing number of cores, $n_{\mathrm{c}}$, is considered. For optimal usage of the hardware architecture (which is organized in computing nodes of 36 cores each), $n_{\mathrm{c}}$ is incremented in multiples of 36 once more than one node is occupied. Figure \ref{fig:pipefrsi_scaling_all}A shows the total run time $t_{\mathrm{c}}$ and Fig. \ref{fig:pipefrsi_scaling_all}B the parallel speedup over the number of cores $n_{\mathrm{c}}$ for both solvers $\mathcal{C}\mathrm{Heart}$ and Ambit.

\paragraph{Weak scaling}
The weak scaling properties of both the BGS-S3\texttimes 3 and the S3\texttimes 3 preconditioner are investigated based on the three different meshes $\mathrm{rf0}$, $\mathrm{rf1}$, and $\mathrm{rf2}$, choosing the number of cores used for computing the problem to yield a ratio of degrees of freedom per core of $n_{\mathrm{dof}}/n_{\mathrm{c}} \approx 7000$. Since the system size for BGS-S3\texttimes 3 is substantially bigger than for S3\texttimes 3, we compute the latter on fewer cores to maintain the approximate ratio. Figure \ref{fig:pipefrsi_scaling_all}C shows the total run time $t_{\mathrm{c}}$ over the number of cores (the respective refinement level) for both solvers, and Fig. \ref{fig:pipefrsi_scaling_all}D depicts the number of linear iterations per solve, $n_{\mathrm{li}}/n_{\mathrm{ni}}$ (top) as well as the number of nonlinear iterations $n_{\mathrm{ni}}$ (bottom) over the relative time of the simulation (note, for ease, only results for solver Ambit shown). Solution times between solvers (and hence monolithic vs. partitioned approach) differ significantly for smaller problem sizes but converge for the finest mesh $\mathrm{rf2}$. Linear iterations per solve increase slightly with mesh refinement, apparently due to a substantially improved resolution of the flow field. Nonlinear iterations are hardly affected by mesh refinement, ranging within $n_{\mathrm{ni}} \in \{2,3\}$ (monolithic FrSI).

\paragraph{Benchmarking with Other Approaches}
For the partitioned FrSI approach, we compare the preconditioner S3\texttimes 3 with the S2\texttimes 2(p) version, where fluid pressure and multiplier variables are consolidated, cf. Fig.~\ref{fig:s2x2c_s2x2p_sparsity}B for fluid-0D systems. Consequently, Jacobian entries from the boundary Galerkin ROM are kept in the $\bsf{A}$ block, cf. Fig.~\ref{fig:s3x3_frsi_sparsity}A. Figure~\ref{fig:pipefrsi_scaling_all}E shows the total run time $t_{\mathrm{c}}$ over different numbers of cores (different refinement levels), and Fig.~\ref{fig:pipefrsi_scaling_all}F the number of linear iterations per solve for mesh rf1. Run times are substantially larger for S2\texttimes 2(p) compared to S3\texttimes 3, and the problem could not be computed for the finest mesh ($\mathrm{rf2}$, 72 cores) due to depletion of memory resources (considering a specifiable maximum of $5\;\mathrm{GB}$ of memory per CPU on our hardware architecture).

\clearpage
\section{Fluid-0D: Blood Flow in Patient-Specific Aortic Arch}\label{sec:examples_fluidaort}

Blood flow in a rigid patient-specific aortic arch model is considered, cf. Fig.~\ref{fig:models_all}C, governed by Eulerian fluid dynamics.
Full details on the problem setup are provided in the \emph{Supplementary Materials}, Sec.~\ref{app:examples_fluidaort}.
Briefly, the 3D blood domain is coupled to a closed-loop lumped-parameter model of the systemic, pulmonary, and coronary circulation, detailed in \emph{Supplementary Materials} Sec. \ref{app:syspul} and Sec.~\ref{app:coronary}. The left and right heart are modeled by time-varying elastance functions. The aortic valve is modeled by an explicit Robin term depending on the pressure difference, cf. \emph{Supplementary Materials} Sec.~\ref{app:3dvalve}.

\begin{figure}[!htp]
\centering
\includegraphics[width=0.975\textwidth]{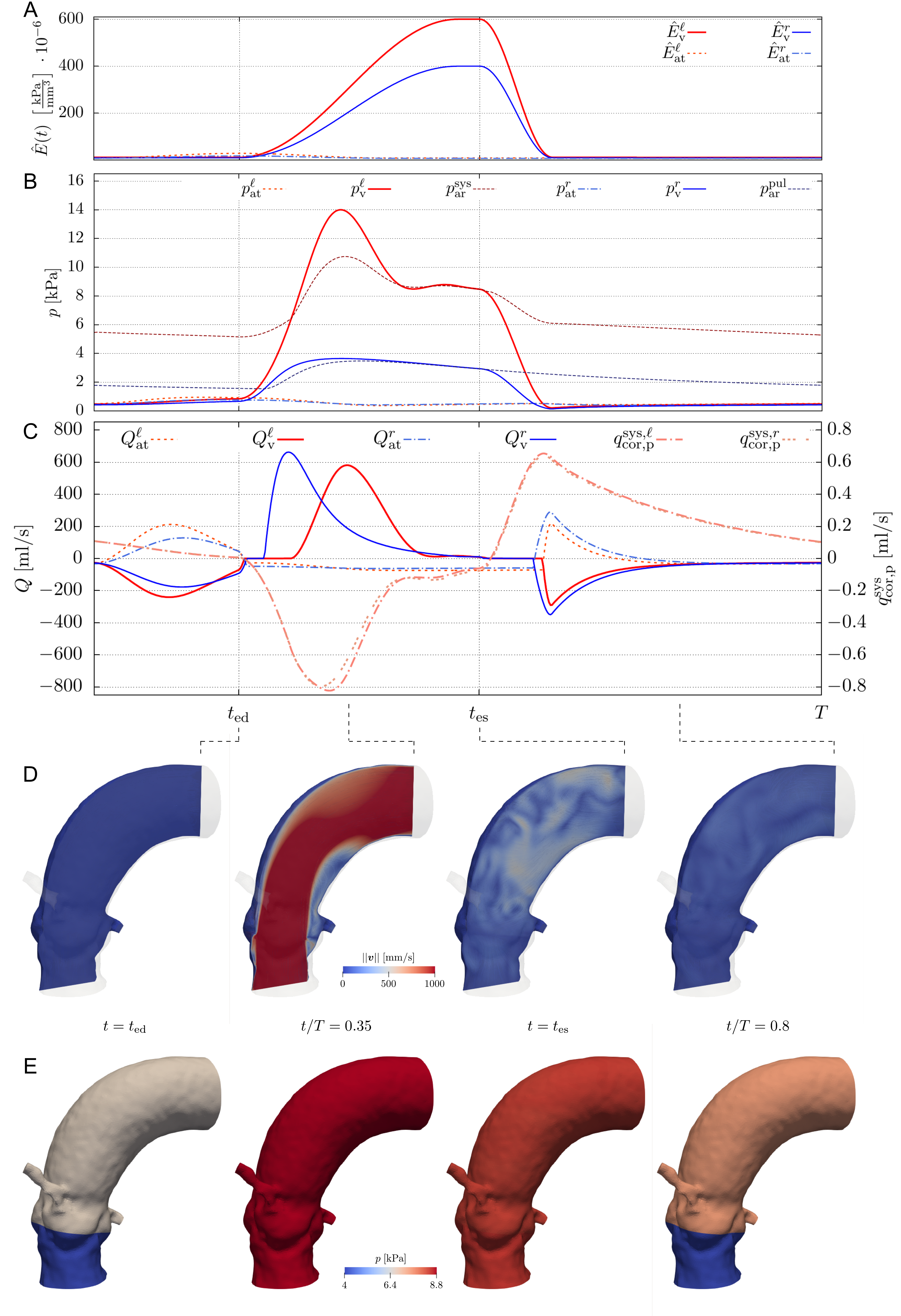}
\caption{\textit{Aorta fluid-0D} (mesh rf2): \textbf{A.} Prescribed left and right ventricular and atrial elastances. \textbf{B.} Time courses of left atrial, left ventricular, systemic arterial, right atrial, right ventricular, and pulmonary arterial pressures. \textbf{C.} Time courses of left atrial, left ventricular, right atrial, right ventricular, left and right proximal coronary fluxes. \textbf{D.} Magnitude of fluid velocity $\bs{v}$ on longitudinal cut, shading shows velocity streamlines. \textbf{E.} Fluid pressure $p$.}\label{fig:aort_results}
\end{figure}

The meshes and degree of freedom tables are shown in Fig.~\ref{fig:meshes_dofs_all}B. The simulation output for the finest mesh ($\mathrm{rf2}$) of velocity and pressure along with the integrated pressures and chamber fluxes as well as the prescribed heart chamber elastances are shown in Fig.~\ref{fig:aort_results}.

\begin{figure}[!htp]
\centering
\includegraphics[width=1\textwidth]{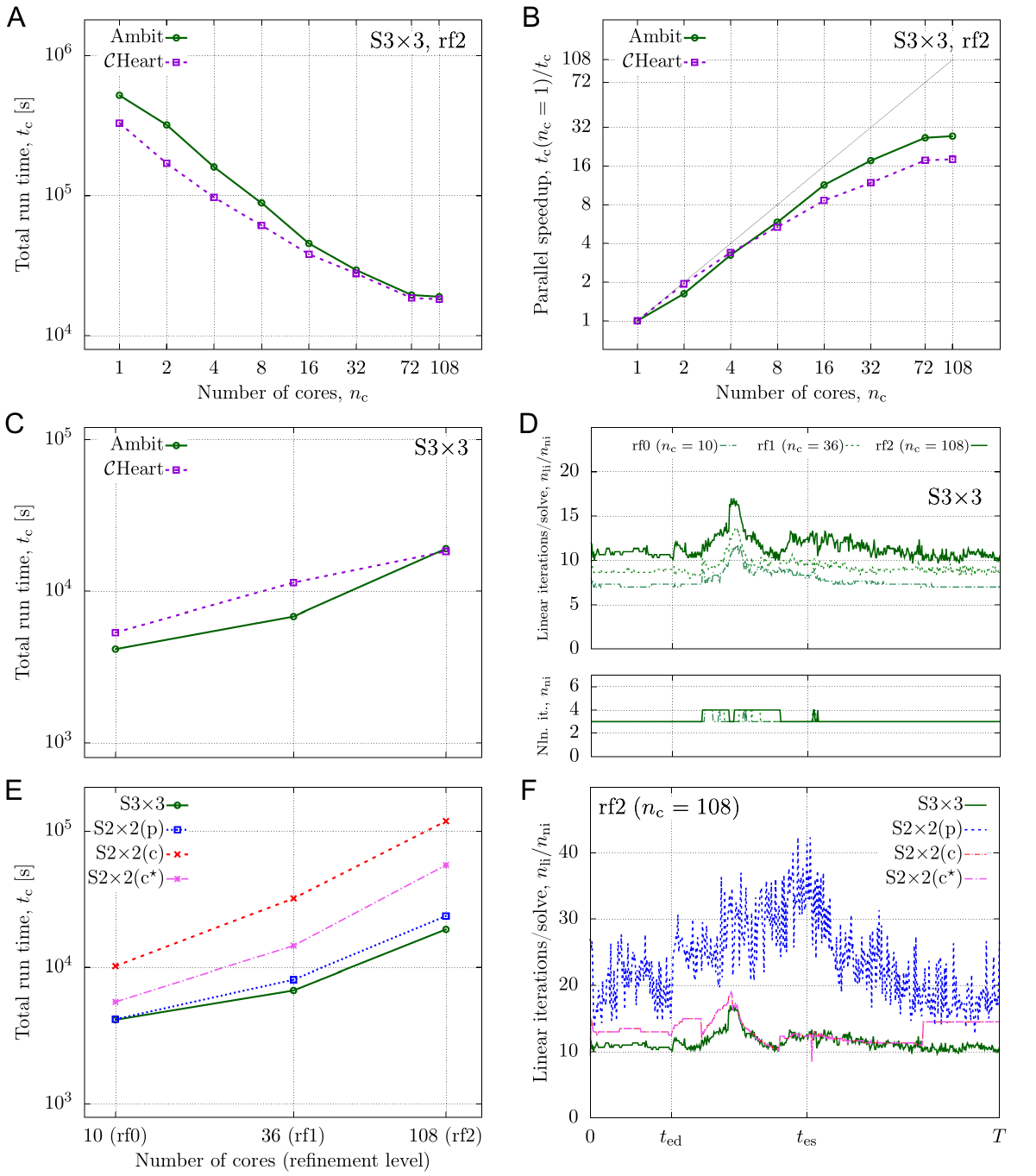}
\caption{\textit{Aorta fluid-0D:} \textit{\textbf{Top row:}} Strong scaling for S3\texttimes 3 preconditioner on mesh $\text{rf2}$. \textbf{A.} Total run time over number of cores. \textbf{B.} Parallel speedup over number of cores. Black dashed line indicates ``optimal'' speedup in absence of any inter-core communication. \textit{\textbf{Middle row:}} Weak scaling for S3\texttimes 3 preconditioner for three different refinements and number of cores, with a ratio of degrees of freedom per core of $n_{\mathrm{dof}}/n_{\mathrm{c}} \approx 7000$. \textbf{C.} Total run time. \textbf{D.} Linear iterations per solve over relative physical time (top), nonlinear iterations over relative physical time (bottom) (solver Ambit). \textit{\textbf{Bottom row:}} Comparison of weak scaling for preconditioners S3\texttimes 3, S2\texttimes 2(p) (0D multipliers lumped into fluid pressure block), and S2\texttimes 2(c/c$^{\star}$) (0D multipliers condensed from linear system, without/with neglecting offdiagonal fill-in blocks), with a ratio of degrees of freedom per core of $n_{\mathrm{dof}}/n_{\mathrm{c}} \approx 7000$ (solver Ambit). \textbf{E.} Total run time. \textbf{F.} Linear iterations per solve on mesh $\text{rf2}$ over cardiac cycle time.}\label{fig:aort_scaling_all}
\end{figure}

\paragraph{Strong scaling}
We investigate the strong scaling properties of the S3\texttimes 3 preconditioner on the finest aortic arch mesh $\mathrm{rf2}$ by comparing the run time $t_{\mathrm{c}}$ with increasing number of cores, $n_{\mathrm{c}}$. For optimal usage of the hardware architecture (which is organized in computing nodes of 36 cores each), $n_{\mathrm{c}}$ is incremented in multiples of 36 once more than one node is occupied. Figure \ref{fig:aort_scaling_all}A shows the total run time $t_{\mathrm{c}}$ and Fig. \ref{fig:aort_scaling_all}B the parallel speedup over the number of cores $n_{\mathrm{c}}$ for both solvers $\mathcal{C}\mathrm{Heart}$ and Ambit.

\paragraph{Weak scaling}
Further, we investigate the weak scaling properties of the S3\texttimes 3 preconditioner based on the three different discretizations $\mathrm{rf0}$, $\mathrm{rf1}$, and $\mathrm{rf2}$, choosing the number of cores used for computing the problem to yield a ratio of degrees of freedom per core of $n_{\mathrm{dof}}/n_{\mathrm{c}} \approx 7000$. Figure \ref{fig:aort_scaling_all}C shows the total run time $t_{\mathrm{c}}$ over the number of cores (the respective refinement level) for both solvers, and Fig. \ref{fig:aort_scaling_all}D depicts the number of linear iterations per solve, $n_{\mathrm{li}}/n_{\mathrm{ni}}$ (top) as well as the number of nonlinear iterations $n_{\mathrm{ni}}$ (bottom) over the time of the simulation, where $t_{\mathrm{ed}}$ and $t_{\mathrm{es}}$ indicate the times of end-diastole and end-systole, respectively (note, for ease, only results for solver Ambit shown). Largest number of iterations is observable during the ejection phase (ventricular systole, between $t_{\mathrm{ed}}$ and $t_{\mathrm{es}}$). Both solvers show nearly identical solution times for the finest mesh $\mathrm{rf2}$. Linear iterations per solve increase moderately with mesh refinement, apparently due to a substantially improved resolution of the flow field. Nonlinear iterations are hardly affected by mesh refinement, ranging within $n_{\mathrm{ni}} \in \{3,4\}$.

\paragraph{Benchmarking with Other Approaches}
We compare the performance of S3\texttimes 3 to using S2\texttimes 2 for three different types of matrix systems. Again, the application of the preconditioner to the condensed system Eq.~(\ref{eq:lin_sys_fluid_cnd})+(\ref{eq:lin_sys_fluid_cndII}) (Fig.~\ref{fig:s2x2c_s2x2p_sparsity}A\&B) is denoted by S2\texttimes 2(c) and S2\texttimes 2(c$^{\star}$), whereas application to the matrix system Eq.~(\ref{eq:lin_sys_fluid}) with consolidated pressure and multiplier variables (Fig.~\ref{fig:s2x2c_s2x2p_sparsity}C) by S2\texttimes 2(p). Figure~\ref{fig:aort_scaling_all}E shows the total run time $t_{\mathrm{c}}$ over different numbers of cores (different refinement levels), and Fig.~\ref{fig:aort_scaling_all}F depicts the number of linear iterations per solve over the relative simulation time for mesh $\mathrm{rf2}$ ($n_{\mathrm{c}}=108$).  Iteration-wise, S2\texttimes 2(p) needs between three and four times as many per solve than S3\texttimes 3, which however can be partly compensated by reduced costs for preconditioner setup and reduced overall effort. For all meshes, S3\texttimes 3 is equal or superior in overall run time due to the comparably low iteration count. Similar to the pipe flow example, S2\texttimes 2(c/c$^{\star}$) exhibit linear iterations in the same range compared to S3\texttimes 3, but here with an over six-/nearly three-fold higher total run time for the finest mesh.

\clearpage
\section{ALE Fluid-RD/0D (FrSI): Hemodynamics in the Left Heart}\label{sec:examples_frsiheart}

In this section, we propose a novel application to reduced modeling of left heart fluid-solid interaction. 
Full details on the problem setup are provided in the \emph{Supplementary Materials}, Sec.~\ref{app:examples_frsiheart}.
It expands the recently proposed approach of fluid-reduced-solid interaction (FrSI) for a single (idealized) ventricle \cite{hirschvogel2024-frsi} to patient-specific multi-compartment models of atrium, ventricle, and aortic outflow tract, where known motion states that are retrieved from a time series of medical imaging data are leveraged to build a subspace of POD modes constituting operator Eq.~(\ref{eq:Vgamma}). In order to model region-specific response characteristics of the wall, the global POD space further is decomposed using a partition of unity approach, along with appropriate subspaces close to the in- and outflow rims in order to facilitate the application of local boundary conditions in the ROM space. Section~\ref{app:frsi_heart_pod} of the \emph{Supplementary Materials} expands further on these concepts. Figure~\ref{fig:models_all}D depicts the whole model, where three different reduced solid models are defined on the walls of left atrium $\Gm_{0,\mathrm{LA}}^{\mathrm{\fF\mhyphen\fSr}}$, left ventricle $\Gm_{0,\mathrm{LV}}^{\mathrm{\fF\mhyphen\fSr}}$, and aortic outflow tract $\Gm_{0,\mathrm{AO}}^{\mathrm{\fF\mhyphen\fSr}}$. The space on which the fluid pressure test functions are defined is further split into three subregions, $\tilde{\Om}_{0}=\tilde{\Om}_{0,\mathrm{LA}}\cup\tilde{\Om}_{0,\mathrm{LV}}\cup\tilde{\Om}_{0,\mathrm{AO}}$ with matching but disjoint interfaces on mitral and aortic valve plane $\Gm_{0,\mathrm{mv}}^{R}$ and $\Gm_{0,\mathrm{av}}^{R}$ in order to capture discontinuities for closed valve states. The 3D regions are connected to a 0D closed-loop systemic, pulmonary, and coronary circulation system, detailed in \emph{Supplementary Materials} Sec.~\ref{app:syspul} and Sec.~\ref{app:coronary}. The right heart is modeled by time-varying elastance functions. The mitral and aortic valve are modeled by a Robin term depending on the pressure difference, cf. \emph{Supplementary Materials}, Sec.~\ref{app:3dvalve}.\\

\begin{figure}[!htp]
\centering
\includegraphics[width=0.89\textwidth]{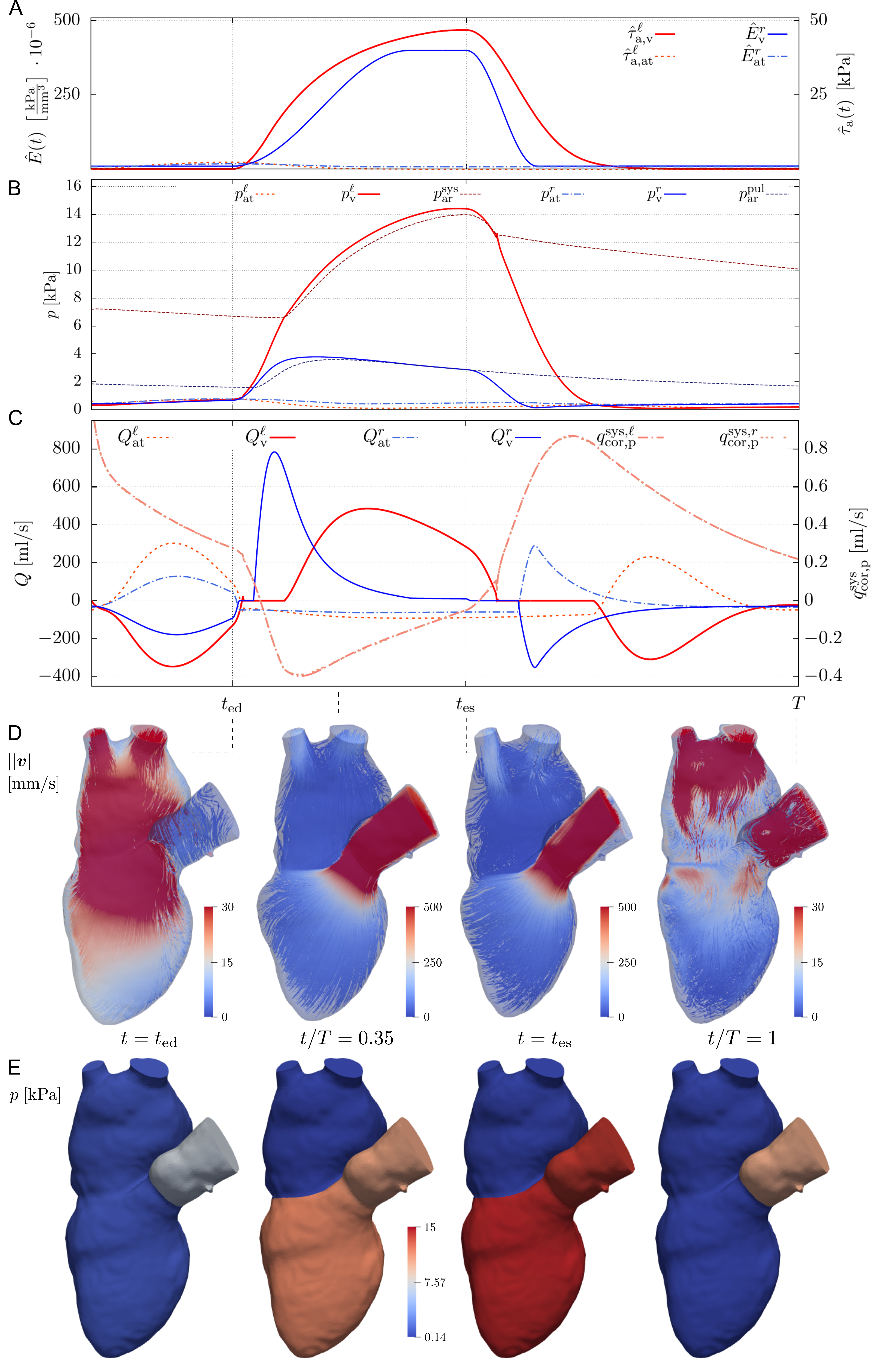}
\caption{\textit{Left heart FrSI-0D} (mesh $\mathrm{rf2}$): \textbf{A.} Forcing: Prescribed left heart active stress and right heart elastances. \textbf{B.} Time courses of left atrial, left ventricular, systemic arterial, right atrial, right ventricular, and pulmonary arterial pressures. \textbf{C.} Time courses of left atrial, left ventricular, right atrial, right ventricular, left and right proximal coronary fluxes. \textbf{D.} Magnitude of fluid velocity $\bs{v}$, streamlines on longitudinal cut through deformed domain $\Om$. Note the different scales for diastolic and systolic snapshots. \textbf{E.} Fluid pressure $p$, plotted on undeformed reference domain $\tilde{\Om}_{0}$.}\label{fig:lalvao_results}
\end{figure}

Three different spatial discretizations (refinement levels) are considered, denoted by $\mathrm{rf0}$, $\mathrm{rf1}$, and $\mathrm{rf2}$, respectively. Figure~\ref{fig:meshes_dofs_all}C illustrates the three different meshes along with the degrees of freedom per sub-system. The simulation output for the finest mesh ($\mathrm{rf2}$) of velocity and pressure along with the integrated pressures and chamber fluxes as well as the prescribed left active stresses and right heart chamber elastances are shown in Fig.~\ref{fig:lalvao_results}.

\begin{figure}[!htp]
\centering
\includegraphics[width=1\textwidth]{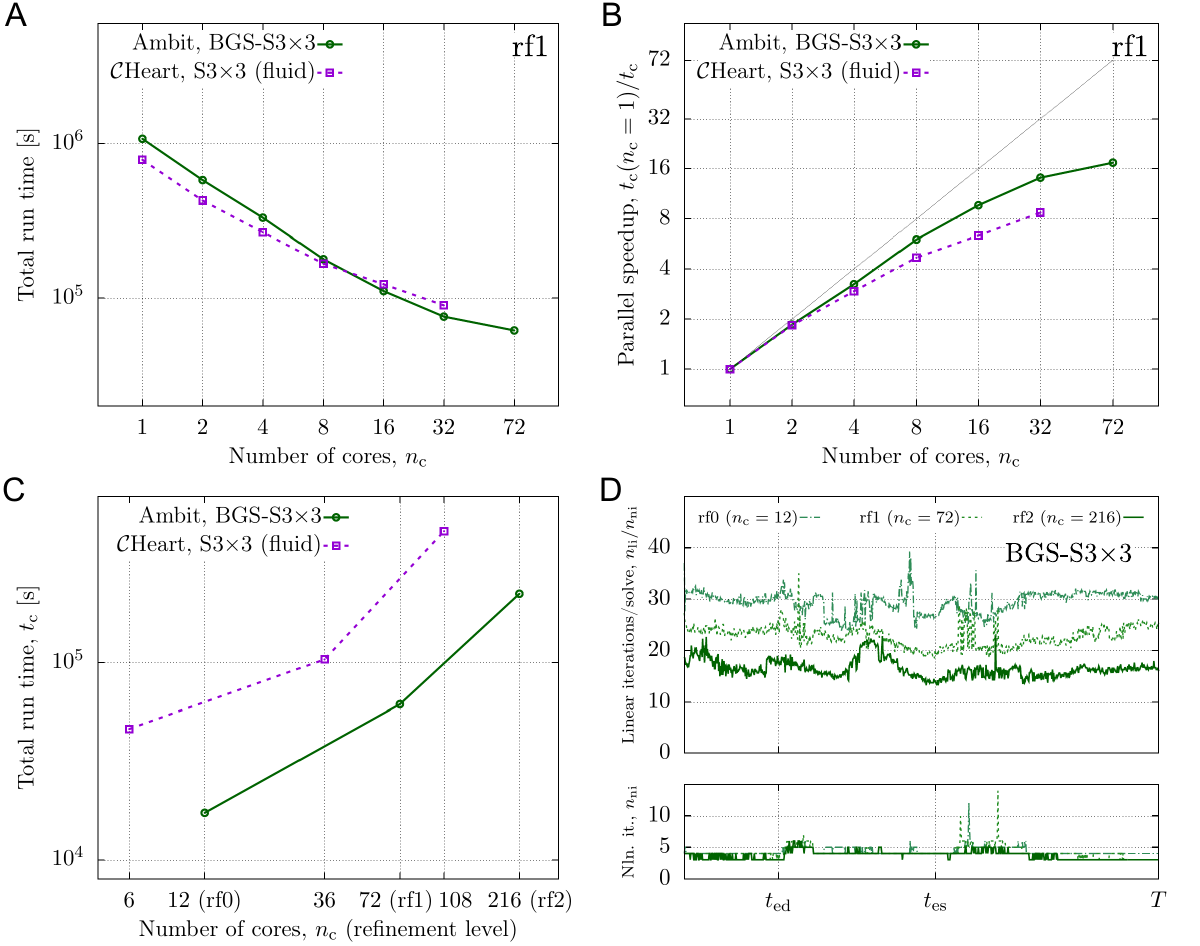}
\caption{\textit{Left heart FrSI-0D:} \textit{\textbf{Top row:}} Strong scaling for BGS-S3\texttimes 3 (monolithic FrSI) and S3\texttimes 3 (partitioned FrSI) on mesh $\text{rf1}.$ \textbf{A.} Total run time over number of cores. \textbf{B.} Parallel speedup over number of cores. Black dashed line indicates ``optimal'' speedup in absence of any inter-core communication. \textit{\textbf{Bottom row:}} Weak scaling for S3\texttimes 3 preconditioner for three different refinements and number of cores, with a ratio of degrees of freedom per core of $n_{\mathrm{dof}}/n_{\mathrm{c}} \approx 7000$. \textbf{C.} Total run time. \textbf{D.} Linear iterations per solve over physical time (top), nonlinear iterations over physical time (bottom) (solver Ambit).}\label{fig:lalvao_scaling_all}
\end{figure}

\paragraph{Strong scaling}
The strong scaling properties of both the BGS-S3\texttimes 3 preconditioner, which is suitable for a monolithic FrSI implementation (provided only by solver Ambit), as well as the S3\texttimes 3 preconditioner applied to the partitioned FrSI approach (solver $\mathcal{C}\mathrm{Heart}$) are investigated. For the latter, the standalone ALE domain motion problem is solved with an AMG-preconditioned GMRES solver. The run time $t_{\mathrm{c}}$ for the medium-sized discretization $\mathrm{rf1}$ with increasing number of cores, $n_{\mathrm{c}}$, is considered. For optimal usage of the hardware architecture (which is organized in computing nodes of 36 cores each), $n_{\mathrm{c}}$ is incremented in multiples of 36 once more than one node is occupied. Figure \ref{fig:lalvao_scaling_all}A shows the total run time $t_{\mathrm{c}}$ and Fig. \ref{fig:lalvao_scaling_all}B the parallel speedup over the number of cores $n_{\mathrm{c}}$ for both solvers $\mathcal{C}\mathrm{Heart}$ and Ambit.

\paragraph{Weak scaling}
Weak scaling for both the BGS-S3\texttimes 3 and the S3\texttimes 3 preconditioner is investigated based on the three different meshes $\mathrm{rf0}$, $\mathrm{rf1}$, and $\mathrm{rf2}$, choosing the number of cores used for computing the problem to yield a ratio of degrees of freedom per core of $n_{\mathrm{dof}}/n_{\mathrm{c}} \approx 7000$. Due to the substantially larger system size for BGS-S3\texttimes 3 compared to S3\texttimes 3, we compute the latter on fewer cores to maintain the approximate ratio. Figure \ref{fig:lalvao_scaling_all}C shows the total run time $t_{\mathrm{c}}$ over the number of cores (the respective refinement level) for both solvers, and Fig. \ref{fig:lalvao_scaling_all}D depicts the number of linear iterations per solve, $n_{\mathrm{li}}/n_{\mathrm{ni}}$ (top) as well as the number of nonlinear iterations $n_{\mathrm{ni}}$ (bottom) over the time of the simulation, where $t_{\mathrm{ed}}$ and $t_{\mathrm{es}}$ indicate the times of end-diastole and end-systole, respectively (note, for ease, only results for solver Ambit shown). As for the aortic arch flow model, the largest number of iterations is observable during the ejection phase (ventricular systole, between $t_{\mathrm{ed}}$ and $t_{\mathrm{es}}$). Solution times between solvers (and hence monolithic vs. partitioned approach) differ significantly, since the partitioned approach needs substantially more nonlinear iterations compared to the monolithic one. In contrast to the other models, linear iterations per solve here decrease with mesh refinement. For most of the time, nonlinear iterations are hardly affected by mesh refinement, while some peaks appear at the end of isovolumetric relaxation for the coarsest mesh. For a majority of the time, nonlinear iterations range within $n_{\mathrm{ni}} \in \{3,5\}$ (monolithic FrSI).

\clearpage
\section{Discussion}\label{sec:discussion}

\begin{figure}[!htp]
\centering
\includegraphics[width=1\textwidth]{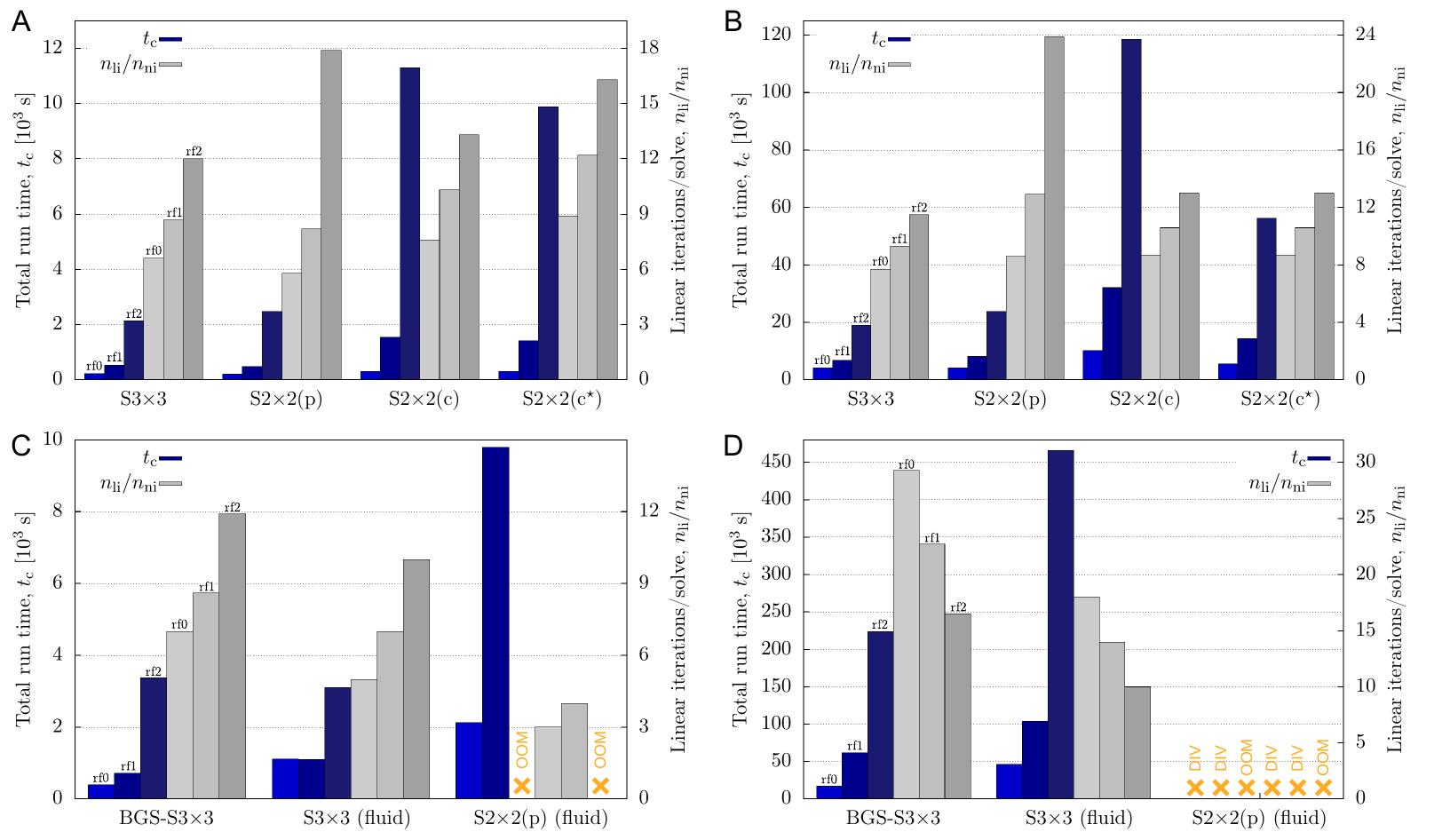}
\caption{Visualization of values in Tab.~\ref{tab:res_summary_fluid}: Total run time $t_{\mathrm{c}}$, number of linear iterations per nonlinear iteration $n_{\mathrm{li}}/ n_{\mathrm{ni}}$ for the four different models. OOM indicates that problem yielded an out-of-memory error, while DIV means that the nonlinear solver was unable to converge given the linear solver settings (Tab.~\ref{tab:params_solver}). \textbf{A.} Blocked pipe. \textbf{B.} Aortic arch. \textbf{C.} Deformable pipe. \textbf{D.} Left heart.}\label{fig:scaling_all_bars}
\end{figure}

This paper introduces a novel block preconditioner tailored towards efficiently solving stabilized Navier-Stokes equations with strong coupling to reduced models of a non-local nature. 
The preconditioner exploits the 3\texttimes 3 block structure of the monolithic problem and allows for use of efficient field-specific solvers for the fluid momentum, continuity of mass, and constraint/reduced equations. 
It may be seen as an extension to the concept of Schur complement preconditioners for 2\texttimes 2 systems \cite{elman2008}, retaining the flexibility for different types of approximations to the Schur complement \cite{cyr2012}. 
The proposed methodology preserves the underlying 3\texttimes 3 matrix structure without the need for updates to the Jacobian originating from integration of reduced models as is needed in other commonly used strategies \cite{esmailymoghadam2013-3d0d,seo2019}. We demonstrate that the proposed preconditioner is highly suitable and effective for solving computational fluid dynamics, both with strong coupling to 0D lumped-parameter models as well as projection-based reduced wall models of co-dimension 2 accommodating the elasticity of the fluid's boundary.
The preconditioner is evaluated over four examples, testing efficacy in simple, repeatable tests as well as complex patient-specific cardiovascular applications.
Our main findings for the 3\texttimes 3 block preconditioner in terms of solver performance and comparisons to alternative 2\texttimes 2 approaches are summarized in Tab.~\ref{tab:res_summary_fluid} and Fig.~\ref{fig:scaling_all_bars}, showing total run time and number of linear iterations per nonlinear iteration for the two fluid-0D as well as FrSI-0D problems, respectively. \\

\begin{table*}[!h]
\begin{center}
\caption{Total run time $t_{\mathrm{c}}$, number of linear iterations per nonlinear iteration $n_{\mathrm{li}}/ n_{\mathrm{ni}}$. Comparison of fluid-0D models (\emph{blocked pipe}, \emph{aortic arch}) for preconditioners S3\texttimes 3, S2\texttimes 2(p) and S2\texttimes 2(c/c$^{\star}$) (solved using Ambit) as well as FrSI-0D models (\emph{deformable pipe}, \emph{left heart}) for preconditioners BGS-S3\texttimes 3 (solver Ambit), S3\texttimes 3 (fluid) (solver $\mathcal{C}\mathrm{Heart}$), and S2\texttimes 2(p) (fluid) (solver $\mathcal{C}\mathrm{Heart}$). 
Bold numbers indicate lowest simulation time/least number of average linear iterations per nonlinear iteration. 
OOM indicates that problem yielded an out-of-memory error, while DIV means that the nonlinear solver was unable to converge given the linear solver settings (Tab.~\ref{tab:params_solver}). Note that lower iteration counts for S\texttimes 2(p) (fluid) compared to BGS-S3\texttimes 3 stem from the fact that we only monitor iterations on the fluid-0D system (partitioned FrSI Eq.~(\ref{eq:lin_sys_rom_frsi_part})), whereas BGS-S3\texttimes 3 operates on the full system (monolithic FrSI Eq.~(\ref{eq:lin_sys_rom_frsi_mono})) including the domain motion problem.
}\label{tab:res_summary_fluid}
\begin{tabular}{rr|cc|cc|cc|cc}
\emph{Fluid-0D} &  & \multicolumn{2}{c|}{S3\texttimes 3} & \multicolumn{2}{c|}{S2\texttimes 2(p)} & \multicolumn{2}{c|}{S2\texttimes 2(c)} & \multicolumn{2}{c}{S2\texttimes 2(c$^{\star}$)}\\
             & & $t_{\mathrm{c}}\;[\mathrm{s}]$ & $n_{\mathrm{li}}/n_{\mathrm{ni}}$ & $t_{\mathrm{c}}\;[\mathrm{s}]$ & $n_{\mathrm{li}}/n_{\mathrm{ni}}$ & $t_{\mathrm{c}}\;[\mathrm{s}]$ & $n_{\mathrm{li}}/n_{\mathrm{ni}}$ & $t_{\mathrm{c}}\;[\mathrm{s}]$ & $n_{\mathrm{li}}/n_{\mathrm{ni}}$ \\
\hline
Blocked & $\mathrm{rf0}$ & 229.2 & 6.6 & \textbf{202.3} & \textbf{5.8} & 301.1 & 7.6 & 301.8 & 8.9 \\
pipe          & $\mathrm{rf1}$ & 523.4 & 8.7 & \textbf{476.1} & \textbf{8.2} & 1.54e3 & 10.3 & 1.41e3 & 12.2\\
             & $\mathrm{rf2}$ & \textbf{2.14e3} & \textbf{12.0} & 2.48e3 & 17.9 & 11.3e3 & 13.3 & 9.88e3 & 16.3 \\
\hline
Aortic  & $\mathrm{rf0}$ & \textbf{4.15e3} & \textbf{7.7} & 4.19e3 & 8.6 & 10.2e3 & 8.7 & 5.62e3 & 8.7 \\
  arch       & $\mathrm{rf1}$ & \textbf{6.81e3} & \textbf{9.3} & 8.17e3 & 12.9 & 32.1e3 & 10.6 & 14.4e3 & 10.6 \\
             & $\mathrm{rf2}$ & \textbf{19.0e3} & \textbf{11.5} & 23.8e3 & 23.9 & 118.5e3 & 13.0 & 56.4e3 & 13.0 \\
\hline
\hline
\emph{FrSI-0D} & & \multicolumn{2}{c|}{BGS-S3\texttimes 3} & \multicolumn{2}{c|}{S3\texttimes 3 (fluid)} & \multicolumn{2}{c|}{S2\texttimes 2(p) (fluid)} & \multicolumn{2}{c}{} \\
& & $t_{\mathrm{c}}\;[\mathrm{s}]$ & $n_{\mathrm{li}}/n_{\mathrm{ni}}$ & $t_{\mathrm{c}}\;[\mathrm{s}]$ & $n_{\mathrm{li}}/n_{\mathrm{ni}}$ & $t_{\mathrm{c}}\;[\mathrm{s}]$ & $n_{\mathrm{li}}/n_{\mathrm{ni}}$ & & \\\hline
Deformable & $\mathrm{rf0}$ & \textbf{394.0} & 7.0 & 1.11e3 & 5.0 & 2.13e3 & \textbf{3.0}& & \\
    pipe         & $\mathrm{rf1}$ & \textbf{716.2} & 8.6 & 1.10e3 & 7.0 & 9.80e3 & \textbf{4.0}& & \\
                & $\mathrm{rf2}$ & 3.38e3 & 11.9 & \textbf{3.11e3} & \textbf{10.0} & \multicolumn{2}{c|}{\textcolor{red}{OOM}}& &  \\\hline
Left     & $\mathrm{rf0}$ & \textbf{1.73e4} & 29.3 & 4.62e4 & \textbf{18.0} & \multicolumn{2}{c|}{\textcolor{red}{DIV}} & & \\
      heart      & $\mathrm{rf1}$ & \textbf{6.17e4} & 22.7 & 1.04e5 & \textbf{14.0} & \multicolumn{2}{c|}{\textcolor{red}{DIV}} & & \\
                & $\mathrm{rf2}$ & \textbf{2.24e5} & 16.5 & 4.66e5 & \textbf{10.0} & \multicolumn{2}{c|}{\textcolor{red}{OOM}} & & \\
\end{tabular}
\end{center}
\end{table*}

\begin{figure}[!htp]
\centering
\includegraphics[width=1\textwidth]{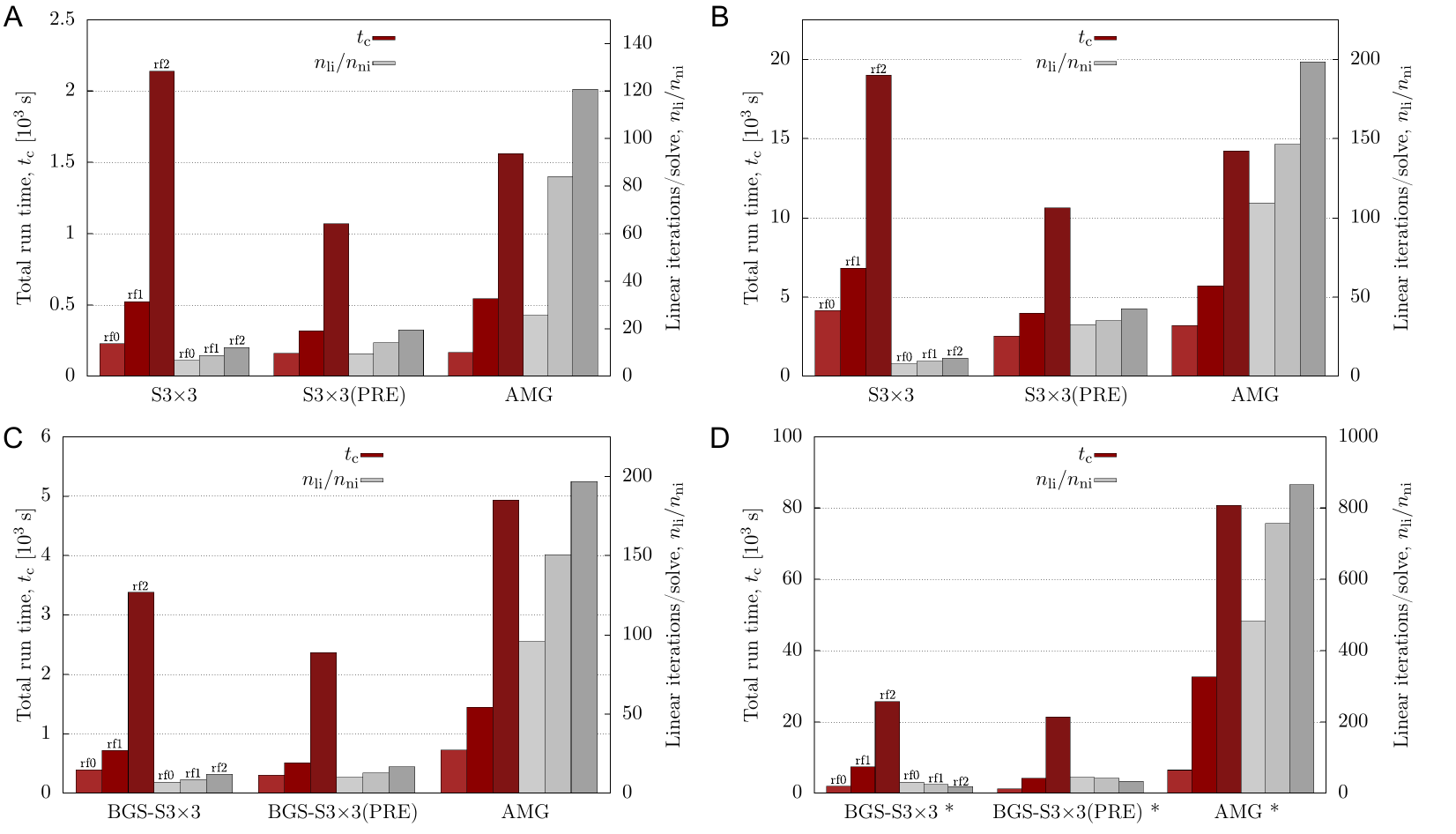}
\caption{Comparison of of 3\texttimes 3 solvers (default plus speed-optimized versions, denoted by suffix ``(PRE)'') with pure AMG on coupled problem. Total run time $t_{\mathrm{c}}$ and number of linear iterations per nonlinear iteration $n_{\mathrm{li}}/ n_{\mathrm{ni}}$ for the four different models. We changed the maximum allowed number of iterations as well as the restart parameter for the FGMRES solver (cf. Tab.~\ref{tab:params_solver}) both to $10\,000$ for this comparison due to the observed ineffectiveness of plain AMG methods with respect to linear iterations. \textbf{A.} Blocked pipe. \textbf{B.} Aortic arch. \textbf{C.} Deformable pipe. \textbf{D.} Left heart. *Only the first 100 time steps of the problem were computed due to the computational expense of AMG for this model.}\label{fig:scaling_all_bars_vs_amg}
\end{figure}

As an alternative to the block preconditioners we presented in this contribution, out-of-the-box preconditioners, such as algebraic multigrid (AMG), may be applied to the whole monolithic matrix systems Eq.~(\ref{eq:lin_sys_fluid}) or Eq.~(\ref{eq:lin_sys_rom_frsi_mono}) without special considerations on their block structure. Figure~\ref{fig:scaling_all_bars_vs_amg} shows a comparison of AMG with the proposed S3\texttimes 3/BGS-S3\texttimes 3 solver (using the AMG settings from the $A$-solve in Tab.~\ref{tab:params_solver}). 
In addition, we also present the S3\texttimes 3(PRE)/BGS-S3\texttimes 3(PRE), which prioritizes speed over iteration count by substituting iterative single field solves for $A$, $S$, $G$ to one preconditioner sweep.\\

We subsequently discuss the preconditioner in the context of alternative approaches for fluid-0D and reduced FSI systems in Sec.~\ref{subsec:disc:prec_fluid} and Sec.~\ref{subsec:disc:prec_frsi}, respectively, and conclude with a short outlook in Sec.~\ref{subsec:disc:outl}.

\subsection{Preconditioning of Fluid-0D Models}\label{subsec:disc:prec_fluid}
From Tab.~\ref{tab:res_summary_fluid} and Fig.~\ref{fig:scaling_all_bars}A\&B, we see that for the two fluid-0D test problems (\emph{e.g.}, \emph{blocked pipe}, \emph{aortic arch}), the introduced S3\texttimes3 solver compares competitively with S2\texttimes2(p) and superior to S2\texttimes2(c/c$^{\star}$).
For smaller problem sizes (rf0, rf1), S3\texttimes3 and S2\texttimes2(p) compare favorably with similar compute times and linear iterations per FGMRES solve.
However, as the system size increases (rf2), the S3\texttimes3 solver shows a 15---20\,\% reduction in compute time and a 30---50\,\% reduction in the number of linear iterations.
This separation in compute cost as the system size increases likely stems from inaccuracies of 0D model variable updates in the approximate Schur solve.
In contrast, S2\texttimes2(c) and S2\texttimes2(c$^{\star}$) perform significantly worse, increasing compute times by a factor 5---6 and 3---4 for the most refined systems, respectively.
This impact is likely caused by the introduction of dense sub-blocks into the primary fluid matrix.\\
From Fig.~\ref{fig:scaling_all_bars_vs_amg}A\&B, we see that for the two fluid-0D tests, pure AMG is superior with respect to overall computing time when comparing to S3\texttimes 3 with the original settings (using inner solves on the $A$- and $S$-block), whereas AMG linear iterations are increased (and growing significantly with problem size), limiting the effectiveness per iteration. Optimizing S3\texttimes 3 for speed---compromising on linear iteration effectiveness---by using preconditioner sweeps (``pre-only'') instead of iterative solves on the sub-problems, run times can also be decreased, eventually being superior to AMG with the given settings (with more limited growth with problem size). This is likely due to the fact that AMG may be able to effectively construct coarse grids of the 3\texttimes 3 system due to its overall sparse structure, however deteriorates in representing the strong coupling features of the system. \\

The approaches in \cite{liu2020prec,esmailymoghadam2013-3d0d,seo2019} deal with preconditioning fluid-0D systems, where the 0D model contributes by rank 1 updates to the fluid Jacobian (matrix $\bsf{A}$), retaining a 2\texttimes 2 system of equations amenable to S2\texttimes 2 preconditioners of the type Eq.~(\ref{eq:2x2_inverse_block}) with minor modifications \cite{liu2020prec}. 
While this approach has been shown efficient for a number of applications, challenges arise for applications that introduce constraints of a non-local nature (such as 0D models).
The 0D model's inherent non-locality endorses rank 1 updates to the momentum stiffness matrix that may alter its standard sparsity pattern, particularly for in-/outflow boundaries of larger size (\emph{e.g.}, increasing the rank of the non-local constraints).
This is considered even more challenging if the 0D model interlinks several boundaries of the 3D model, where a change of flow at surface $i$ due to a change in pressure at surface $j$ has to be accounted for. This coupling endorses further fill-ins at offdiagonal locations, cf. Fig.~\ref{fig:s2x2c_s2x2p_sparsity}A. However, for similar models, it was stated that these offdiagonal blocks may be neglected in the tangent matrix (resulting in a system sketched in Fig.~\ref{fig:s2x2c_s2x2p_sparsity}B), presumably improving preconditioner performance while not or only negligibly affecting nonlinear iterations \cite{esmailymoghadam2013-3d0d,brown2024}. Our results here confirm these findings, seeing no change in nonlinear solver performance and shorter run times for S2\texttimes2(c$^{\star}$) compared to S2\texttimes2(c).\\ 
The effects of preconditioning these 2\texttimes 2 systems are exemplified in Tab.~\ref{tab:res_summary_fluid} and Fig.~\ref{fig:scaling_all_bars}A\&B, where dense sub-blocks associated with the coupled boundaries are introduced into the fluid momentum block for S2\texttimes2(c/c$^{\star}$), leading to increased compute time. On the contrary, S2\texttimes 2(p) presumably suffers from the offdiagonal multiplier-velocity coupling blocks collapsing inside the Schur complement, whose approximation is limited Eq.~(\ref{eq:schur_approx}) and whose solve is only done up to a rough tolerance (cf. Tab.~\ref{tab:params_solver}), leading to increased linear iterations compared to using a direct (hence exact) solve on the reduced block like in S3\texttimes 3.\\

Nonetheless, retaining 2\texttimes 2 block systems for fluid-0D problems can be advantageous when access to the solver infrastructure is limited, providing increased modularity for coupling 0D models to fluid systems without the need for more invasive alterations of the equation/code structure. Like this, an ordinary fluid system is retained, which more easily can be interfaced to provide the needed residual and Jacobian updates in case of the statically condensed approach.

\subsection{Preconditioning of ALE Fluid-0D/RD (FrSI) Models}\label{subsec:disc:prec_frsi}
Comparing the ALE fluid-0D/RD (FrSI) models (\emph{e.g.}, \emph{deformable pipe}, \emph{left heart}),
3\texttimes 3 solvers are shown to perform superior to their 2\texttimes 2 counterparts (see Tab.~\ref{tab:res_summary_fluid} and Fig.~\ref{fig:scaling_all_bars}C\&D).
Using S2\texttimes 2(p) as a benchmark (which outperformed S2\texttimes2(c/c$^{\star}$) in the fluid-0D tests), the 3\texttimes 3 preconditioners were approximately 5 to 14 times faster in the deformable pipe example (for rf0 and rf1, respectively). 
For all other models, the size of the introduced dense sub-block into the Schur complement resulted in the solver running out of memory (``OOM'') or poor Schur complement approximations that prevented convergence of the Newton solver given the linear solver settings, cf. Tab.~\ref{tab:params_solver} (``DIV'').
In addition, we considered both S3\texttimes 3, which performs a partitioned solve of the ALE domain motion sub-problem within the nonlinear fixed point, and BGS-S3\texttimes 3, which introduces a Block Gauss-Seidel approximation to the monolithic problem including the domain motion. 
Both S3\texttimes 3 and BGS-S3\texttimes 3 perform comparably for the pipe model, likely due to the simplicity of the solid motion (which is restricted to one POD mode).
However, in the more realistic \emph{left heart} example, the impact of the grid motion yields an approximate 2-fold increase in compute time due to the increase in the number of nonlinear iterations required. Therefore, a monolithic approach preconditioned with BGS-S3\texttimes 3 is considered the most stable and effective option here.\\
From Fig.~\ref{fig:scaling_all_bars_vs_amg}C\&D, we see that for the two ALE fluid-0D/RD (FrSI) tests, pure AMG is inferior with respect to overall computing time to both BGS-S3\texttimes 3 with the original settings (using inner solves on the $A$-, $S$-, and $G$-block) as well as to the speed-optimized version, where solves on the iterative sub-problems are replaced by preconditioner sweeps (``pre-only''). As for the fluid-0D tests, AMG experiences ineffectiveness per linear iteration, in the extreme case showing on average $>\!800$ linear iterations per solve for the left heart model.\\

Preconditioners for Navier-Stokes equations that are coupled to solid mechanics models of reduced or 3D nature (fluid-solid interaction, FSI) have to deal with the added complexity of another type of field equations entering the system. 
For monolithic FSI, Block Gauss-Seidel (BGS) methods with single-field AMG-preconditioned solvers have proven to be effective \cite{deparis2016,gee2011}, similar to more intrusive approaches using AMG on the fully coupled system and BGS on each multigrid level \cite{gee2011}. Alternative methods are based on Schwarz decomposition techniques \cite{wu2014}, and some combine the former and the latter to effectively reduce FSI interface errors \cite{mayr2020}.
Other 3D FSI preconditioners, on the contrary, embed the contribution of the solid into the main fluid Jacobian in order to facilitate usage of 2\texttimes 2 Schur complement-based block preconditioners for the FSI system \cite{seo2019}. 
Therein, this avenue equally has been pursued for solid models of co-dimension 2, that is, for reduced models defined on the fluid's boundary. While the effectiveness of these approaches is shown, in our contribution, we uniquely deal with a reduced solid model of co-dimension 2 which has a \textit{non-local} structure. 
This so-called fluid-reduced-solid interaction (FrSI) methodology \cite{hirschvogel2024-frsi} performs a Galerkin projection of the fluid-tissue boundary to confine it to a subspace spanned by POD modes. 
As outlined in Sec.~\ref{subsec:3x3frsi}, the non-locality of this model introduces Jacobian contributions that deteriorate the sparsity patterns of both the $\bsf{A}$ block and the Schur complement when using standard 2\texttimes 2 preconditioning methods, i.e. S2\texttimes 2(p). 
For the flexible pipe example, Sec.~\ref{sec:examples_frsipipe}, we observe that using S2\texttimes 2(p) for discretization $\mathrm{rf2}$ fails due to depletion of the CPUs' memory resources. 
This is explained by the need to perform the matrix product $\hat{\bsf{B}}\bsf{B}^{\mathrm{T}}$ when computing the Schur complement, Eq.~(\ref{eq:schur_approx_3x3}). 
This becomes apparent in the sparsity sketch in Fig.~\ref{fig:s3x3_frsi_sparsity}A, where the reduced boundary residuals depend on the size of the (non-reduced) fluid pressure space. 
Vice versa, the non-reduced continuity residuals exhibit a dependence on the reduced set of boundary velocities. 
Hence, the respective matrix product introduces a dense fill-in on the Schur complement of the size of the entire boundary pressure space, hampering a memory-optimized matrix allocation. 
For the left heart hemodynamics model, Sec.~\ref{sec:examples_frsiheart}, the effectiveness of the provided approximate block inverses deteriorates. 
We observe that nonlinear solver convergence even for the coarser meshes ($\mathrm{rf0}$ and $\mathrm{rf1}$) cannot be achieved for the given solver settings, while the fine mesh ($\mathrm{rf2}$) similarly cannot be setup due to memory limits, cf. Tab.~\ref{tab:res_summary_fluid}. 
Therefore, for our FSI models of non-local nature, S3\texttimes 3 represents the only candidate (of considered preconditioners) that allows an effective and scalable preconditioning of the underlying coupled matrix systems.

\subsection{Outlook}\label{subsec:disc:outl}
The presented solver enables the efficient solution of complex CFD and FrSI models and will pave the way for future research on large-scale patient-specific models of blood flow in arteries or the heart cavities. It will facilitate the integration of closed-circuit 0D models as well as reduced-dimensional Galerkin ROMs that provide boundary subspaces for large-strain solid models of co-dimension 2. In order to increase the flexibility of the proposed preconditioner, more sophisticated approximations for the Schur complement readily can be integrated into S3\texttimes 3, providing robustness for strongly convection-dominated flows at high Reynolds numbers \cite{cyr2012,elman2008,liu2020prec}. Approaches presented in \cite{gee2011}, switching the order of block decomposition and (multigrid-preconditioned) single-field solves, also may be suitable candidates for improving the Schur approximation. Furthermore, the solver readily may be used and tested for inf-sup stable approximation pairs of velocity and pressure, \emph{e.g.}, Taylor-Hood finite elements, which have not been considered in this contribution.

\clearpage
\section{Conclusion}\label{sec:conclusion}
We present a novel 3\texttimes 3 block preconditioner that is designed for accelerating the convergence of solvers for the finite element-discretized 3D stabilized Navier-Stokes equations, subject to constraints of a non-local nature. 
These include coupling to flux-dependent boundary conditions at in- or outflows as well as Galerkin projection-based reduced wall/FSI models of co-dimension 2. 
The preconditioner leverages the intrinsic block matrix structure of the linearized three-equation system of fluid momentum, conservation of mass, and constraints/reduced momentum terms and extends the concept of 2\texttimes 2 Schur complement solvers. 
At the same time, it preserves the modularity of these concepts by allowing any suitable approximations for the Schur complements. 
With the presented approach, all sparsity patterns of the original (non-constrained) fluid matrices are retained, and optimal field-specific solvers (AMG-preconditioned GMRES for fluid momentum and Schur complement, direct solve for reduced variable block) can be used. 
Conclusively, superiority of this preconditioner compared to more classical 2\texttimes 2 approaches, which need some block matrix manipulations deteriorating the native sparsity patterns, is shown, with an up to 50\,\% reduction in linear iterations and up to 80\,\% reduction in overall computing time, depending on the model and the solver.

\clearpage
\section{Acknowledgments}
DN acknowledges funding from the Engineering and Physical Sciences Research Council Healthcare Technology Challenge Award (EP/R003866/1), and support from the Wellcome Trust EPSRC Centre of Excellence in Medical Engineering (WT 088641/Z/09/Z) and the NIHR Biomedical Research Centre at Guy's and St. Thomas' NHS Foundation Trust and KCL.

\clearpage
\section*{Supplementary Material}\label{sec:appendix}

\subsection{Strong Problem Statements}\label{app:general_strong}
In this section, we state the strong forms of 3D fluid dynamics coupled to reduced models of a non-local nature in a very general form. The equations governing Eulerian fluid flow with flux-dependent traction constraints are elaborated in Sec.~\ref{app:eulerian_strong}, and those for ALE fluid with a reduced-dimensional solid mechanics wall model of co-dimension 2 in Sec.~\ref{app:frsi_strong}.

\subsubsection{Eulerian Fluid Coupled to 0D Models}\label{app:eulerian_strong}

The general strong form of the governing equations for incompressible Navier-Stokes flow in an Eulerian reference frame with coupling to reduced 0D models is elaborated. The primary variables are the fluid velocity $\bs{v}$, fluid pressure $p$, and a set of multipliers that enforce flux constraints to a lumped-parameter model at $n_{\mathrm{0d}}^{\mathrm{b}}$ boundaries, $\{\lmz\}_{n_{\mathrm{0d}}^{\mathrm{b}}} = \left\{\lmz_1, \ldots, \lmz_{n_{\mathrm{0d}}^{\mathrm{b}}}\right\}$. The initial boundary value problem is stated as follows:
\begin{align}
    \rho\left(\frac{\partial \bs{v}}{\partial t} + (\nabla\bs{v})\bs{v}\right) &= \nabla\cdot\bs{\sigma}(\bs{v},p) \quad &&\text{in} \; \Om \times [0,T],\label{eq:fluid_mom_gen}\\
    \nabla\cdot\bs{v} &= 0 \quad &&\text{in} \; \tilde{\Om} \times [0,T],\label{eq:fluid_mass_gen} \\
    \bs{t}_{i}^{\mathrm{\fF\mhyphen 0d}} &= -\lmz_{i} \,\bs{n} \quad &&\text{on} \; \Gm_{i}^{\mathrm{\fF\mhyphen 0d}} \times [0,T], \quad i \in \{1,\ldots,n_{\mathrm{0d}}^{\mathrm{b}}\}, \label{eq:fluid_t0d_gen}\\
    \bs{t}^{N} &= \hat{\bs{t}}_{N}(t,\bs{v}) \quad && \text{on}\; \Gm^{N} \times [0,T], \label{eq:fluid_nbc_gen}\\
    \bs{v} &= \hat{\bs{v}} \quad && \text{on}\; \Gm^{D} \times [0,T],\label{eq:fluid_dbc_gen} \\
    \int\limits_{\Gm_{j}^{\mathrm{\fF\mhyphen 0d}}} \bs{v}\cdot\bs{n}\,\mathrm{d}A &= \alpha_{j}\,q_{j}(\{\lmz\}_{n_{\mathrm{0d}}^{\mathrm{b}}}) \quad &&\text{in} \; [0,T],  \quad j \in \{1,\ldots,n_{\mathrm{0d}}^{\mathrm{b}}\}, \label{eq:fluid_constr_gen}
\end{align}
where $\nabla$ is the gradient operator with respect to physical space, $\nabla\bs{v}:=\frac{\partial v_{i}}{\partial x_{j}}\bs{e}_{i}\otimes\bs{e}_{j}$, $\rho$ is the fluid's density, $\bs{n}$ a unit outward normal vector, and $\bs{t}^{N}(t,\bs{v})$ a generic, possibly time- or velocity-dependent Neumann or Robin-like traction term. The fluid is Newtonian (dynamic viscosity $\mu$), with the constitutive equation for the Cauchy stress
\begin{align}
    \bs{\sigma}(\bs{v},p) = -p \bs{I} + \mu \left(\nabla \bs{v} + (\nabla \bs{v})^{\mathrm{T}}\right).\label{eq:cauchy_eulerian}
\end{align}

The constraint Eq.~(\ref{eq:fluid_constr_gen}) enforces consistency between the flux over a fluid boundary and a flux $q_{j}$ emerging from the solution of the 0D model problem, possibly dependent on one or more multiplier variables. Scaling parameters $\alpha_j$ account for the directionality of flow, i.e. should take the value $-1$ if a 0D flux variable is imposed as inflow to the fluid domain, and $1$ otherwise. Specific forms of possible 0D models are given in Sec.~\ref{app:0d} and will be introduced along with the example-specific problems.

\subsubsection{ALE Fluid Coupled to 0D and RD Models (Fluid-reduced-Solid Interaction, FrSI)}\label{app:frsi_strong}

The strong form of the governing equations for incompressible Navier-Stokes flow in an ALE reference frame \cite{duarte2004,donea1982} with coupling to reduced 0D models as well as solid mechanics models of co-dimension 2 is elaborated. This constitutes the physics-related part of the FrSI method \cite{hirschvogel2024-frsi}. The primary variables here are the fluid velocity $\bs{v}$, fluid pressure $p$, a set of multipliers that enforce flux constraints to a lumped-parameter model at $n_{\mathrm{0d}}^{\mathrm{b}}$ boundaries, $\{\lmz\}_{n_{\mathrm{0d}}^{\mathrm{b}}} = \left\{\lmz_1, \ldots, \lmz_{n_{\mathrm{0d}}^{\mathrm{b}}}\right\}$, as well as the domain displacement $\bs{d}$. In order to generate this arbitrary motion field, a model of the type
\begin{align}
    \nabla_{0}\cdot \bs{\sigma}_{\mathrm{g}}(\bs{d}) &= \bs{0} \quad &&\text{in} \;\Om_{0} \times [0,T], \label{eq:ale_strong_gen}\\
    \bs{d} &= \bs{u}_{\mathrm{f}}(\bs{v}) 
    \quad &&  
    \text{on}\; \Gm_{0}^{\mathrm{\fF\mhyphen\fSr}} \times [0,T], \label{eq:ale_dbc_gen}
\end{align}
is employed. On the deformable boundary $\Gm_{0}^{\mathrm{\fF\mhyphen\fSr}}$, the model is subject to a Dirichlet condition that takes the value of the fluid displacement
\begin{align}
    \bs{u}_{\mathrm{f}}(\bs{v}) = \int\limits_{0}^{t}\bs{v}\,\mathrm{d}\bar{t}
    \label{eq:ufluid}.
\end{align}
A constitutive equation for $\bs{\sigma}_{\mathrm{g}}(\bs{d})$ may be chosen dependent on the problem and will be detailed later. We introduce the kinematics of the bijective mapping between the reference and the arbitrary domain deformation, namely the domain velocity, deformation gradient as well as its determinant, as follows:
\begin{equation}
\begin{aligned}
    \widehat{\bs{w}}=\frac{\partial\bs{d}}{\partial t}, \quad \widehat{\bs{F}}=\bs{I}+\nabla_{0}\bs{d} \quad \text{and} \quad \widehat{J}=\det\widehat{\bs{F}}.
    \label{eq:defgrad_ale}
\end{aligned}
\end{equation}

The initial boundary value problem with the tractions from the 0D constraint and reduced solid model, Neumann as well as Dirichlet conditions reads
\begin{align}
    \rho\left(\left.\frac{\partial \bs{v}}{\partial t}\right|_{\bs{x}_{0}} + (\nabla_{0}\bs{v}\,\widehat{\bs{F}}^{-1})(\bs{v}-\widehat{\bs{w}})\right) &= \nabla_{0}\bs{\sigma}(\bs{v},p,\bs{d}) : \widehat{\bs{F}}^{-\mathrm{T}}\quad &&\text{in} \; \Om_{0} \times [0,T],\label{eq:frsi_mom_gen}\\
    \nabla_{0}\bs{v} : \widehat{\bs{F}}^{-\mathrm{T}} &= 0 \quad &&\text{in} \; \tilde{\Om}_{0} \times [0,T], \label{eq:frsi_mass_gen} \\
    \bs{t}_{0,i}^{\mathrm{\fF\mhyphen 0d}} &= -\lmz_{i} \,\widehat{J}\widehat{\bs{F}}^{-\mathrm{T}}\bs{n}_{0} \quad &&\text{on} \; \Gm_{0,i}^{\mathrm{\fF\mhyphen 0d}} \times [0,T], \label{eq:frsi_t0d_gen}\\
    & \qquad i \in \{1,\ldots,n_{\mathrm{0d}}^{\mathrm{b}}\},&&\nonumber \\
    \bs{t}_{0}^{\mathrm{\fF\mhyphen\fSr}} &= -h_0\left(\rho_{0,\fS} \frac{\partial \bs{v}}{\partial t} - \tilde{\nabla}_{0}\cdot\tilde{\bs{P}}(\bs{u}_{\fF}(\bs{v}))\right) \quad &&\text{on} \; \Gm_{0}^{\mathrm{\fF\mhyphen\fSr}} \times [0,T], \label{eq:frsi_tsolid_gen}\\
    \bs{t}_{0}^{N} &= \hat{\bs{t}}_{N}(t,\bs{v}) \quad && \text{on}\; \Gm_{0}^{N} \times [0,T],\label{eq:frsi_nbc_gen}\\
    \bs{v} &= \hat{\bs{v}} \quad && \text{on}\; \Gm_{0}^{D} \times [0,T],\label{eq:frsi_dbc_gen}\\
    \int\limits_{\Gm_{0,j}^{\mathrm{\fF\mhyphen 0d}}} (\bs{v}-\widehat{\bs{w}})\cdot\widehat{J}\widehat{\bs{F}}^{-\mathrm{T}}\bs{n}_{0}\,\mathrm{d}A_0 &= \alpha_j\, q_j(\{\lmz\}_{n_{\mathrm{0d}}^{\mathrm{b}}}) \quad &&\text{in} \; [0,T],\label{eq:frsi_constr_gen}\\
    & \qquad j \in \{1,\ldots,n_{\mathrm{0d}}^{\mathrm{b}}\},&&\nonumber
\end{align}
where $\nabla_{0}$ denotes the gradient operator with respect to reference space ($\nabla_{0}\bs{v}:=\frac{\partial v_{i}}{\partial x_{0,j}}\bs{e}_{i}\otimes\bs{e}_{j}$ or $\nabla_{0}\bs{\sigma}:=\frac{\partial \sigma_{ij}}{\partial x_{0,k}}\bs{e}_{i}\otimes\bs{e}_{j}\otimes\bs{e}_{k}$), and $\left.\frac{\partial (\bullet)}{\partial t}\right|_{\bs{x}_{0}}$ is the time derivative in the ALE frame. Further, $\rho$ and $\rho_{0,\fS}$ are the fluid's and reduced solid's density, respectively, $\bs{n}_{0}$ a unit outward normal vector in the reference frame, and $\bs{t}_{0}^{N}(t,\bs{v})$ a generic, possibly time- or velocity-dependent Neumann or Robin-like traction vector. The fluid is Newtonian (dynamic viscosity $\mu$), with the ALE version of the constitutive equation for the Cauchy stress
\begin{align}
    \bs{\sigma}(\bs{v},p,\bs{d}) = -p \bs{I} + \mu \left(\nabla_0 \bs{v}\,\widehat{\bs{F}}(\bs{d})^{-1} + \widehat{\bs{F}}(\bs{d})^{-\mathrm{T}}(\nabla_0 \bs{v})^{\mathrm{T}}\right). \label{eq:cauchy_ale}
\end{align}
The constraint Eq.~(\ref{eq:frsi_constr_gen}) enforces consistency between the flux over a fluid boundary and a flux $q_{j}$ emerging from the solution of the 0D model problem, possibly dependent on one or more multiplier variables. Parameters $\alpha_j$ are introduced similar to the Eulerian case, cf. Eq.~(\ref{eq:fluid_constr_gen}). Specific forms of possible 0D models are given in Sec.~\ref{app:0d} and will be introduced along with the example-specific problems.\\

The physics-based reduction component of the FrSI method entails a finite strain solid dynamics model living on the 2-dimensional manifold of the fluid-solid interface $\Gm_{0}^{\mathrm{\fF\mhyphen\fSr}}$. Its governing equation is the balance of linear momentum of nonlinear elastodynamics, eventually resulting in the boundary traction given by Eq.~(\ref{eq:frsi_tsolid_gen}). Therein, $h_0$ represents a wall thickness parameter, and $\tilde{\nabla}_{0}$ is the gradient operator on the 2D manifold (with vanishing surface normal components) \cite{colciago2014,hirschvogel2024-frsi}. The first Piola-Kirchhoff stress is mapped from its material counterpart, the second Piola-Kirchhoff stress, as follows:
\begin{align}
    \tilde{\bs{P}} = (\bs{F}_{\fF} - \bs{F}_{\fF}\,\bs{n}_{0}\otimes\bs{n}_{0}) \tilde{\bs{S}}, \label{eq:PK1map}
\end{align}
with the fluid deformation gradient
\begin{align}
    \bs{F}_{\fF} = \bs{I} + \nabla_{0}\bs{u}_{\fF}, \label{eq:Ffluid}
\end{align}
using Eq.~(\ref{eq:ufluid}). Making use of Eq.~(\ref{eq:Ffluid}), eliminating the normal components, and re-defining the out-of-plane stretch on assumptions of incompressibility, the membrane right Cauchy-Green tensor $\tilde{\bs{C}}$ of the 2-dimensional manifold as well as its time derivative can be defined \cite{hirschvogel2024-frsi}. The specific constitutive equation for the stress tensor $\tilde{\bs{S}}$ in Eq.~(\ref{eq:PK1map}) will be introduced along with the examples presented later. Eventually, the projection-based component of FrSI is performed post spatio-temporal discretization and detailed in Sec.~\ref{subsec:frsi}. 

\subsection{Weak Forms and Time-Discrete Representations}\label{app:weakforms_discr}

Here, we state the generic time-discrete weak representations of the strong governing equations for Eulerian and ALE fluid dynamics coupled to 0D models, cf. Sec.~\ref{app:eulerian_strong} and Sec.~\ref{app:frsi_strong}, respectively.
We use an implicit single-step time integration method based on the generalized midpoint scheme, where a quantity at time instance $t^{n+\theta}$ is a linear combination of the value of the current time step at $t^{n+1}$ and the one of the old step at $t^{n}$:
\begin{align}
    (\bullet)^{n+\theta} :=\theta (\bullet)^{n+1} + (1-\theta) (\bullet)^{n}, \label{eq:theta}
\end{align}
with $\theta \in (0, 1]$. For the subsequent finite element approximation, let $\elemset^h(\Om_0)$ denote the set of elements corresponding to the 3D discretization and let
\begin{equation}
    \Om^{h}_0 = \bigcup\limits_{\tau\in\elemset^h(\Om_0)}  \tau
\end{equation}
indicate the meshed representation of the domain $\Om_0$. All discrete function spaces are based on continuous piecewise polynomial spaces in the form of:
\begin{align}
    S^k(\Om^{h}_0) = \left\{ f : \Om^{h}_0 \rightarrow \realset\, |\, f\in C^0({\bar{\Om}^{h}_0}),\, \left. f \right|_{\tau} \in \mathbb{P}^k(\tau)\, \hspace{3mm} \forall\, \tau \in \elemset^{h}(\Om_0)\right\},
\end{align}
representing the general continuous $k$th-order piecewise polynomial spaces defined on $\Om^{h}_0$. We define the velocity, domain displacement, and pressure spaces as follows:
\begin{equation*}
    \vsp^h = \left[ S^1 \left( \Om^{h}_0 \right) \right]^3
    \text{, } 
    \dsp^h = \left[ S^1 \left( \Om^{h}_0 \right) \right]^3,
    \text{ and }
    \psp^h = S^1(\tilde{\Om}^{h}_0).
\end{equation*}
The sub-spaces incorporating the Dirichlet and homogeneous boundary read:
\begin{align}
    \vsp^{D,h}  & = \left\{ \bs{v} \in \vsp^{h}\, |\, \bs{v} = \bs{v}^d \text{ on } \Gm_0^{D,h}  \right\},
    \\
    \vsp^{0,h}  & = \left\{ \bs{v} \in \vsp^{h}\, |\, \bs{v} = \bs{0} \text{ on } \Gm_0^{D,h}  \right\},
    \\
    \dsp^{D,h} & = \left\{ \bs{d} \in \dsp^h\, |\, \bs{d} = \bs{d}^d \text{ on } \Gm_0^{D,h} \right\},
    \\
    \dsp^{0,h} & = \left\{ \bs{d} \in \dsp^h\, |\, \bs{d} = \bs{0} \text{ on } \Gm_0^{D,h} \right\}.
\end{align}

Here, we allow that the domain on which the pressure spaces are defined, $\tilde{\Om}^{h}_0$, may differ from the one on which velocity (and domain displacement) live, $\Om^{h}_0$. This allows a split pressure domain across valve planes (with joint interfaces, using a matching discretization) in order to facilitate modeling of a pressure jump.

\subsubsection{Eulerian Fluid Coupled to 0D Models}\label{app:eulerian_weak}
The general Eulerian fluid problem with $n_{\mathrm{0d}}^{\mathrm{b}}$ coupling constraints to (one or multiple) 0D lumped-parameter network model(s), cf. Sec.~\ref{app:eulerian_strong}, at the current unknown time step indexed by $n+1$, can be stated as follows:\\

Find fluid velocity and pressure variables, $\left(\bs{v}^{n+1},\, p^{n+1}\right)\in \vsp^{D,h} \times \psp^{h}$, as well as multiplier variables $\{\lmz^{n+1}\}_{n_{\mathrm{0d}}^{\mathrm{b}}}\in \realset^{n_{\mathrm{0d}}^{\mathrm{b}}}$ such that conservation of linear momentum,
\begin{equation}
\begin{aligned}
    R_{\delta v}\left(\bs{v}^{n+1},p^{n+1},\{\lmz^{n+1}\}_{n_{\mathrm{0d}}^{\mathrm{b}}}; \delta\bs{v}\right) := \int\limits_{\Om^h} \rho & \left(\frac{\bs{v}^{n+1} - \bs{v}^{n}}{\Delta t} + (\nabla\bs{v}^{n+\theta})\,\bs{v}^{n+\theta}\right) \cdot \delta\bs{v} \,\mathrm{d}V \\
    &+ \int\limits_{\Om^h} \bs{\sigma}\left(\bs{v}^{n+\theta},p^{n+\theta}\right) : \nabla \delta\bs{v} \,\mathrm{d}V \\ 
    &+ \sum\limits_{i=1}^{n_{\mathrm{0d}}^{\mathrm{b}}} \lmz_{i}^{n+\theta} \int\limits_{\Gm_{i}^{\mathrm{\fF\mhyphen 0d},h}} \bs{n}\cdot\delta\bs{v}\,\mathrm{d}A \\
    &+ R_{\delta v}^{N}\left(t^{n+\theta};\delta\bs{v}\right) + R_{\delta v}^{R}\left(\bs{v}^{n+\theta};\delta\bs{v}\right) \\
    &+ S_{\delta v}^{\mathrm{D}}\left(\bs{v}^{n+\theta},p^{n+\theta}; \delta\bs{v}) + S_{\delta v}^{\mathrm{out}}(\bs{v}^{n+\theta}; \delta\bs{v}\right) = 0, \label{eq:fluid_weakform_v}
\end{aligned}
\end{equation}
conservation of mass,
\begin{equation}
\begin{aligned}
    R_{\delta p}(\bs{v}^{n+1},p^{n+1};\delta p) := 
    \int\limits_{\tilde{\Om}^h} (\nabla\cdot\bs{v}^{n+1})\,\delta p\,\mathrm{d}V + S_{\delta p}^{\mathrm{D}}(\bs{v}^{n+1},p^{n+1};\delta p) = 0, \label{eq:fluid_weakform_p}
\end{aligned}
\end{equation}
as well as constraints enforcing consistency between 0D models and the 3D domain,
\begin{equation}
\begin{aligned}
    R_{\lmzt} \left(\bs{v}^{n+1}, \{\lmz^{n+1}\}_{n_{\mathrm{0d}}^{\mathrm{b}}}; \{\lmzt\}_{n_{\mathrm{0d}}^{\mathrm{b}}}\right) := \sum\limits_{i=1}^{n_{\mathrm{0d}}^{\mathrm{b}}} \left(\int\limits_{\Gm_{i}^{\mathrm{\fF\mhyphen 0d},h}} \bs{v}^{n+1}\cdot\bs{n}\,\mathrm{d}A - \alpha_i\,\zd\left(\{\lmz^{n+1}\}_{n_{\mathrm{0d}}^{\mathrm{b}}}\right) \cdot \bsf{e}_{i}\right)\lmzt_{i} = 0,
    \label{eq:fluid_3d0d_coupling_weakform}
\end{aligned}
\end{equation}
hold true, for all fluid velocity and pressure test functions $\left(\delta\bs{v},\, \delta p\right)\in \vsp^{0,h} \times \psp^{h}$ as well as multiplier test functions $\{\lmzt\}_{n_{\mathrm{0d}}^{\mathrm{b}}}\in \realset^{n_{\mathrm{0d}}^{\mathrm{b}}}$.\\

In order to fulfill Eq.~(\ref{eq:fluid_3d0d_coupling_weakform}), 0D variables $\zd \in \realset^{n_{\mathrm{0d}}^{\mathrm{e}}}$, governed by a system of $n_{\mathrm{0d}}^{\mathrm{e}}$ algebraic and first-order differential equations, have to be found such that
\begin{equation}
\begin{aligned}
    &R_{\mathrm{0d}} \left(\zd^{n+1},\{\lmz^{n+1}\}_{n_{\mathrm{0d}}^{\mathrm{b}}}; \zdt\right) := \\
    &\Biggl(\frac{\bsf{g}\left(\bsf{y}^{n+1},\{\lmz^{n+1}\}_{n_{\mathrm{0d}}^{\mathrm{b}}}\right)-\bsf{g}\left(\bsf{y}^{n},\{\lmz^{n}\}_{n_{\mathrm{0d}}^{\mathrm{b}}}\right)}{\Delta t} \; + \\
    &\qquad\theta_{\mathrm{0d}}\,\bsf{f}\left(\bsf{y}^{n+1},\{\lmz^{n+1}\}_{n_{\mathrm{0d}}^{\mathrm{b}}}\right) + (1-\theta_{\mathrm{0d}})\,\bsf{f}\left(\bsf{y}^{n},\{\lmz^{n}\}_{n_{\mathrm{0d}}^{\mathrm{b}}}\right)\Biggr) \cdot \zdt = 0,
    \label{eq:0d_weakform_discr}
\end{aligned}
\end{equation}
for all 0D test functions $\zdt \in \realset^{n_{\mathrm{0d}}^{\mathrm{e}}}$, where $\bsf{g}$ is a linear (``left-hand side'') and $\bsf{f}$ a possibly nonlinear (``right-hand side'') function in the variable vector $\bsf{y}$ and/or multipliers $\{\lmz\}_{n_{\mathrm{0d}}^{\mathrm{b}}}$, with $\theta_{\mathrm{0d}} \in (0, 1]$.\\

The constitutive law for the Cauchy stress tensor is given by Eq.~(\ref{eq:cauchy_eulerian}). Generic Neumann and Robin terms may be introduced in Eq.~(\ref{eq:fluid_weakform_v}), i.e.
\begin{align}
    R_{\delta v}^{N}(t;\delta\bs{v}) = -\int\limits_{\Gm^{N,h}} \bs{t}_{N}(t)\cdot\delta\bs{v}\,\mathrm{d}A \quad \text{and} \quad R_{\delta v}^{R}(\bs{v};\delta\bs{v}) = \int\limits_{\Gm^{R,h}} c\,\bs{v} \cdot\delta\bs{v}\,\mathrm{d}A, \label{eq:fluid_generic_neumann_robin}
\end{align}
where $\bs{t}_{N}$ represents a traction vector and $c$ a viscosity parameter.\\

Here, we use a variant of the G2 stabilization method \cite{johnson1998,hoffman2003,hessenthaler2017} suitable for equal-order $\mathbb{P}_1\mhyphen\mathbb{P}_1$ finite element approximations for the velocity and pressure space. The respective operators which are introduced in Eq.~(\ref{eq:fluid_weakform_v}) and in Eq.~(\ref{eq:fluid_weakform_p}) read
\begin{equation}
\begin{aligned}
    S_{\delta v}^{\mathrm{D}}(\bs{v},p; \delta\bs{v}) &= \int\limits_{\Om^h} d_1 (\nabla\bs{v})\,\bs{v} \cdot \mathrm{sym}(\nabla\delta\bs{v})\,\bs{v}\,\mathrm{d}V + \int\limits_{\Om^h} d_2\,(\nabla\cdot\bs{v}) (\nabla\cdot\delta\bs{v})\,\mathrm{d}V \\ &+ \int\limits_{\Om^h} d_3\,(\nabla p) \cdot \mathrm{sym}(\nabla\delta\bs{v})\,\bs{v}\,\mathrm{d}V, \label{eq:supg}
\end{aligned}
\end{equation}
and
\begin{equation}
\begin{aligned}
    S_{\delta p}^{\mathrm{D}}(\bs{v},p;\delta p) = \frac{1}{\rho}\int\limits_{\tilde{\Om}^h} d_1 (\nabla\bs{v})\,\bs{v} \cdot (\nabla\delta p)\,\mathrm{d}V + \frac{1}{\rho}\int\limits_{\tilde{\Om}^h} d_3\,(\nabla p) \cdot (\nabla\delta p)\,\mathrm{d}V, \label{eq:pspg}
\end{aligned}
\end{equation}
with 
\begin{equation}
\begin{aligned}
    d_1 = \frac{h_{\mathrm{e}}}{v_{\mathrm{max}}}, \quad d_2 = h_{\mathrm{e}} v_{\mathrm{max}}, \quad d_3 = \frac{h_{\mathrm{e}}}{v_{\mathrm{max}}}, \label{eq:stabparams}
\end{aligned}
\end{equation}
where $h_{\mathrm{e}}$ is the cell diameter and $v_{\mathrm{max}}$ a reference velocity. In order to prevent backflow-induced divergence, all Neumann boundaries will be subject to an outflow stabilization \cite{esmailymoghadam2011}:
\begin{equation}
\begin{aligned}
    S_{\delta v}^{\mathrm{out}}(\bs{v};\delta\bs{v}) = -\int\limits_{\Gm^{N,h}}\beta \min(\bs{v}\cdot\bs{n},0)\, \bs{v}\cdot\delta\bs{v}\,\mathrm{d}A\label{eq:sout}.
\end{aligned}
\end{equation}

\subsubsection{ALE Fluid Coupled to 0D and RD Models (Fluid-reduced-Solid Interaction, FrSI)}\label{app:frsi_weak}

A general ALE fluid problem with deformable boundary treated with the FrSI method \cite{hirschvogel2024-frsi} and $n_{\mathrm{0d}}^{\mathrm{b}}$ coupling constraints to (one or multiple) 0D lumped-parameter network model(s), cf. Sec.~\ref{app:frsi_strong}, at the current unknown time step indexed by $n+1$, can be stated as follows:\\

Find fluid velocity and pressure variables, $\left(\bs{v}^{n+1},\, p^{n+1}\right)\in \vsp^{D,h} \times \psp^{h}$, multiplier variables $\{\lmz^{n+1}\}_{n_{\mathrm{0d}}^{\mathrm{b}}}\in \realset^{n_{\mathrm{0d}}^{\mathrm{b}}}$, as well as ALE domain motion variables $\bs{d}^{n+1}\in \dsp^{D,h}$ such that conservation of linear momentum,
\begin{equation}
\begin{aligned}
    &R_{\delta v}\left(\bs{v}^{n+1},p^{n+1},\{\lmz^{n+1}\}_{n_{\mathrm{0d}}^{\mathrm{b}}},\bs{d}^{n+1};\delta\bs{v}\right) := \\
    &\quad\int\limits_{\Om_0^h} \widehat{J}(\bs{d}^{n+\theta}) \rho  \left(\frac{\bs{v}^{n+1}-\bs{v}^{n}}{\Delta t} + \left(\nabla_{0}\bs{v}^{n+\theta}\,\widehat{\bs{F}}(\bs{d}^{n+\theta})^{-1}\right)\,\left(\bs{v}^{n+\theta}-\widehat{\bs{w}}^{n+\theta}\right)\right) \cdot \delta\bs{v} \,\mathrm{d}V_0 \\ &
    + \int\limits_{\Om_0^h} \widehat{J}(\bs{d}^{n+\theta})\,\bs{\sigma}(\bs{v}^{n+\theta},p^{n+\theta},\bs{d}^{n+\theta}) : \nabla_{0} \delta\bs{v}\,\widehat{\bs{F}}(\bs{d}^{n+\theta})^{-1} \,\mathrm{d}V_0 \\
    &+ \sum\limits_{i=1}^{n_{\mathrm{0d}}^{\mathrm{b}}} \lmz_{i}^{n+\theta} \int\limits_{\Gm_{0,i}^{\mathrm{\fF\mhyphen 0d},h}} \widehat{J}(\bs{d}^{n+\theta})\widehat{\bs{F}}(\bs{d}^{n+\theta})^{-\mathrm{T}}\bs{n}_{0}\cdot\delta\bs{v}\,\mathrm{d}A_0 \\
    &+ \int\limits_{\Gm_{0}^{\mathrm{\fF\mhyphen\fSr},h}} h_0 \left(rho_{0,\fS}\,\frac{\bs{v}^{n+1}-\bs{v}^{n}}{\Delta t} \cdot \delta\bs{v} + \tilde{\bs{P}}\left(\uf^{n+\theta},\bs{v}^{n+\theta}\right) : \tilde{\nabla}_{0} \delta\bs{v}\right) \,\mathrm{d}A_0 \\
    &+ R_{\delta v}^{N}(t^{n+\theta};\delta\bs{v}) + R_{\delta v}^{R}(\bs{v}^{n+\theta},\widehat{\bs{w}}^{n+\theta};\delta\bs{v}) \\
    &+ S_{\delta v}^{\mathrm{D}}(\bs{v}^{n+\theta},p^{n+\theta},\bs{d}^{n+\theta};\delta\bs{v}) + S_{\delta v}^{\mathrm{out}}(\bs{v}^{n+\theta},\bs{d}^{n+\theta};\delta\bs{v}) = 0, \label{eq:frsi_weakform_v}
\end{aligned}
\end{equation}
conservation of mass,
\begin{equation}
\begin{aligned}
    &R_{\delta p}(\bs{v}^{n+1},p^{n+1},\bs{d}^{n+1};\delta p) := \\
    &\quad\int\limits_{\tilde{\Om}_0^h} \widehat{J}(\bs{d}^{n+1})\,\nabla_{0}\bs{v}^{n+1} : \widehat{\bs{F}}(\bs{d}^{n+1})^{-\mathrm{T}}\delta p\,\mathrm{d}V_0 + S_{\delta p}^{\mathrm{D}}(\bs{v}^{n+1},p^{n+1},\bs{d}^{n+1};\delta p) = 0,\label{eq:frsi_weakform_p}
\end{aligned}
\end{equation}
constraints enforcing consistency between 0D models and the 3D realm,
\begin{equation}
\begin{aligned}
    &R_{\lmzt} \left(\bs{v}^{n+1}, \{\lmz^{n+1}\}_{n_{\mathrm{0d}}^{\mathrm{b}}}, \bs{d}^{n+1}; \{\lmzt\}_{n_{\mathrm{0d}}^{\mathrm{b}}}\right) := \\
    &\sum\limits_{i=1}^{n_{\mathrm{0d}}^{\mathrm{b}}} \left(\int\limits_{\Gm_{0,i}^{\mathrm{\fF\mhyphen 0d},h}} (\bs{v}^{n+1}-\widehat{\bs{w}}^{n+1})\cdot\widehat{J}(\bs{d}^{n+1})\widehat{\bs{F}}(\bs{d}^{n+1})^{-\mathrm{T}}\bs{n}_{0}\,\mathrm{d}A_0 - \alpha_i\,\zd\left(\{\lmz^{n+1}\}_{n_{\mathrm{0d}}^{\mathrm{b}}}\right) \cdot \bsf{e}_{i}\right)\lmzt_{i} = 0,
    \label{eq:frsi_3d0d_coupling_weakform}
\end{aligned}
\end{equation}
as well as ALE domain motion
\begin{align}
    R_{\delta d}(\bs{d}^{n+1},\bs{v}^{n+1};\delta\bs{d}) &:= \int\limits_{\Om_0^h}\bs{\sigma}_{\mathrm{g}}(\bs{d}^{n+1}) : \nabla_{0}\delta\bs{d}\,\mathrm{d}V_0 = 0,\label{eq:ale_weakform}
\end{align}
hold true, for all fluid velocity and pressure test functions $\left(\delta\bs{v},\, \delta p\right)\in \vsp^{0,h} \times \psp^{h}$, ALE domain motion test functions $\delta\bs{d}\in \dsp^{0,h}$, as well as multiplier test functions $\{\lmzt\}_{n_{\mathrm{0d}}^{\mathrm{b}}}\in \realset^{n_{\mathrm{0d}}^{\mathrm{b}}}$. The ALE problem is further subject to the essential boundary condition Eq.~(\ref{eq:ale_dbc_gen}) at the deformable interface where the reduced solid is defined:
\begin{align}
    \bs{d}^{n+1} = \uf^{n+1}
    \quad \text{on}\; \Gm_{0}^{\mathrm{\fF\mhyphen\fSr},h} \times [0,T].   
\end{align}
The time-discrete representations of the fluid displacement Eq.~(\ref{eq:ufluid}) and domain velocity Eq.~(\ref{eq:defgrad_ale})$_{\text{1}}$ are
\begin{align}
    \uf^{n+1} = \theta \Delta t\,\bs{v}^{n+1} + (1-\theta)\Delta t\,\bs{v}^{n} + \bs{u}_{\mathrm{f}}^{n} \label{eq:ufluid_discr}
\end{align}
and
\begin{align}
    \widehat{\bs{w}}^{n+1} = \frac{\bs{d}^{n+1}-\bs{d}^{n}}{\theta\Delta t} - \frac{1-\theta}{\theta}\widehat{\bs{w}}^{n}, \label{eq:wale_discr}
\end{align}
respectively.\\

The fluid is of Newtonian type, and the constitutive equation for the Cauchy stress is given by Eq.~(\ref{eq:cauchy_ale}).
Generic Neumann and Robin terms may be introduced according to Eq.~(\ref{eq:fluid_generic_neumann_robin}), with the Robin term reading
\begin{align}
    \quad R_{\delta v}^{R}(\bs{v},\widehat{\bs{w}};\delta\bs{v}) = \int\limits_{\Gm_{0}^{R,h}} c\,(\bs{v}-\widehat{\bs{w}}) \cdot\delta\bs{v}\,\mathrm{d}A_0 \label{eq:fluidale_generic_robin}.
\end{align}

The ALE versions of the G2 stabilization operators, cf. their Eulerian counterparts Eq.~(\ref{eq:supg}) and Eq.~(\ref{eq:pspg}), respectively, read
\begin{equation}
\begin{aligned}
    S_{\delta v}^{\mathrm{D}}(\bs{v},p,\bs{d};\delta\bs{v}) &= \int\limits_{\Om_0^h} \widehat{J}(\bs{d})\,d_1 (\nabla_{0}\bs{v}\,\widehat{\bs{F}}(\bs{d})^{-1})\,\bs{v} \cdot \mathrm{sym}(\nabla_{0}\delta\bs{v}\,\widehat{\bs{F}}(\bs{d})^{-1})\,\bs{v}\,\mathrm{d}V_0 \\
    & + \int\limits_{\Om_0^h} \widehat{J}(\bs{d})\,d_2\,(\nabla_{0}\bs{v} : \widehat{\bs{F}}(\bs{d})^{-\mathrm{T}}) (\nabla_{0}\delta\bs{v} : \widehat{\bs{F}}(\bs{d})^{-\mathrm{T}})\,\mathrm{d}V_0\\
    &+ \int\limits_{\Om_0^h} \widehat{J}(\bs{d})\,d_3\,(\widehat{\bs{F}}(\bs{d})^{-\mathrm{T}}\nabla_{0}p) \cdot \mathrm{sym}(\nabla_{0}\delta\bs{v}\,\widehat{\bs{F}}(\bs{d})^{-1})\,\bs{v}\,\mathrm{d}V_0 \label{eq:supg_ale}
\end{aligned}
\end{equation}
and
\begin{equation}
\begin{aligned}
    S_{\delta p}^{\mathrm{D}}(\bs{v},p,\bs{d};\delta p) &= \frac{1}{\rho}\int\limits_{\tilde{\Om}_0^h} \widehat{J}(\bs{d})\,d_1 (\nabla_{0}\bs{v}\,\widehat{\bs{F}}(\bs{d})^{-1})\,\bs{v}\cdot (\widehat{\bs{F}}(\bs{d})^{-\mathrm{T}}\nabla_{0}\delta p)\,\mathrm{d}V_0 \\
    &+ \frac{1}{\rho}\int\limits_{\tilde{\Om}_0^h} \widehat{J}(\bs{d})\,d_3\,(\widehat{\bs{F}}(\bs{d})^{-\mathrm{T}}\nabla_{0}p) \cdot (\widehat{\bs{F}}(\bs{d})^{-\mathrm{T}}\nabla_{0}\delta p)\,\mathrm{d}V_0, \label{eq:pspg_ale}
\end{aligned}
\end{equation}
with the stabilization parameters defined by Eq.~(\ref{eq:stabparams}).\\

Similarly, in order to fulfill Eq.~(\ref{eq:frsi_3d0d_coupling_weakform}), 0D variables $\zd \in \realset^{n_{\mathrm{0d}}^{\mathrm{e}}}$ have to be found such that Eq.~(\ref{eq:0d_weakform_discr}) holds true for all 0D test functions $\zdt \in \realset^{n_{\mathrm{0d}}^{\mathrm{e}}}$.

The ALE counterpart of the outflow stabilization, Eq.~(\ref{eq:sout}), preventing backflow-induced divergence, reads:
\begin{equation}
\begin{aligned}
    S_{\delta v}^{\mathrm{out}}(\bs{v},\bs{d};\delta\bs{v}) = -\int\limits_{\Gm_{0}^{N,h}}\beta \min((\bs{v}-\widehat{\bs{w}})\cdot \widehat{J}(\bs{d})\widehat{\bs{F}}(\bs{d})^{-\mathrm{T}}\bs{n}_{0},0)\, \bs{v}\cdot\delta\bs{v}\,\mathrm{d}A_{0}\label{eq:sout_ale}.
\end{aligned}
\end{equation}

\subsection{Examples: Governing Equations}\label{app:examples}

\subsubsection{Fluid-0D: Blocked Pipe Flow with Bypass}\label{app:examples_fluidpipe}

Here, the governing equations for the blocked pipe flow model, Sec.~\ref{sec:examples_fluidpipe}, Fig.~\ref{fig:models_all}A, are elaborated, which are the incompressible transient Navier-Stokes equations in Eulerian description. Balance of linear momentum and conservation of mass are given by Eq.~(\ref{eq:fluid_mom_gen}) and Eq.~(\ref{eq:fluid_mass_gen}), respectively. The set of multipliers for the coupling to the lumped-parameter model is $\{\lmz\}_{n_{\mathrm{0d}}^{\mathrm{b}}}=\left\{\lmz_{1},\lmz_{2}\right\}$. The boundary tractions from the constraints, no-slip as well as inflow Dirichlet conditions read:
\begin{align}
    \bs{t}_{1}^{\mathrm{\fF\mhyphen 0d}} &= -\lmz_{1} \,\bs{n} \quad &&\text{on} \; \Gm_{\mathrm{out}}^{\mathrm{\fF\mhyphen 0d}} \times [0,T], \\
    \bs{t}_{2}^{\mathrm{\fF\mhyphen 0d}} &= -\lmz_{2} \,\bs{n} \quad &&\text{on} \; \Gm_{\mathrm{in}}^{\mathrm{\fF\mhyphen 0d}} \times [0,T], \\
    \bs{v} &= \bs{0} \quad && \text{on}\; \Gm_{\mathrm{wall}}^{D} \cup \Gm_{\mathrm{v}}^{D} \times [0,T],\\
    \bs{v} &= \bar{v}(t,x,y) \bs{e}_{z} \quad &&  
    \text{on}\; \Gm_{\mathrm{in}}^{D} \times [0,T],
\end{align}
where $\bs{e}_z$ is a Cartesian basis vector. The time- and space-dependent parabolic inflow velocity is given by
\begin{equation}
\begin{aligned}
    \bar{v}(t,x,y) = \frac{1}{2}\hat{v}_{\max}\left(1-\cos\frac{2\pi t}{\hat{T}}\right) \frac{1}{r^4}\left(x^2 - r^2\right)\left(y^2 - r^2\right),
    \label{eq:fluidpipe_dbc_vbar}
\end{aligned}
\end{equation}
plotted in Fig.~\ref{fig:pipefluid_results}A. The constitutive equation for the Cauchy stress is given by Eq.~(\ref{eq:cauchy_eulerian}). The constraints that enforce consistency between the 3D domain and the 0D model here are
\begin{align}
    \int\limits_{\Gm_{\mathrm{out}}^{\mathrm{\fF\mhyphen 0d}}} \bs{v}\cdot\bs{n}\,\mathrm{d}A &= q_{\mathrm{in}}(\{\lmz\}_{n_{\mathrm{0d}}^{\mathrm{b}}}) \quad &&\text{in} \; [0,T],\\
    \int\limits_{\Gm_{\mathrm{in}}^{\mathrm{\fF\mhyphen 0d}}} \bs{v}\cdot\bs{n}\,\mathrm{d}A &= -q_{\mathrm{out}}(\{\lmz\}_{n_{\mathrm{0d}}^{\mathrm{b}}}) \quad &&\text{in} \; [0,T]. \label{eq:fluid0d_qout}
\end{align}
The 0D model equations governing $q_{\mathrm{in}}$ and $q_{\mathrm{out}}$ are detailed in Sec.~\ref{app:crlink}, with $p_{\mathrm{i}} \leftarrow \lmz_1$ and $p_{\mathrm{o}} \leftarrow \lmz_2$. We note the minus sign in Eq.~(\ref{eq:fluid0d_qout}) since an inflow into the fluid domain is enforced. Table \ref{tab:params_pipefluid} summarizes all parameters and their values.

\begin{table}[!htp]
\begin{center}
\caption{Parameters for fluid-0D problem of pipe flow.}\label{tab:params_pipefluid}
\begin{tabular}{rl|c|l}\hline
\multicolumn{4}{l}{Time parameters} \\\hline
$T$ & $[\mathrm{s}]$ & $0.2$ & Simulation time \\
$\Delta t$ & $[\mathrm{s}]$ & $0.002$ & Simulation time step \\
$\theta$ & $[-]$ & $1.0$ & Time-integration parameter, Eq.~(\ref{eq:theta}) \\
$\theta_{\mathrm{0d}}$ & $[-]$ & $1.0$ & Time-integration parameter for 0D model, Eq.~(\ref{eq:0d_weakform_discr}) \\\hline
\multicolumn{4}{l}{Geometry and boundary conditions} \\\hline
$l$ & $[\mathrm{mm}]$ & $100$ & Pipe length \\
$r$ & $[\mathrm{mm}]$ & $15$ & Pipe radius \\
$\hat{v}_{\max}$ & $[\frac{\mathrm{mm}}{\mathrm{s}}]$ & $10^{3}$ & Peak inflow velocity, Eq.~(\ref{eq:fluidpipe_dbc_vbar}) \\
$\hat{T}$ & $[\mathrm{s}]$ & $0.4$ & Period of velocity profile, Eq.~(\ref{eq:fluidpipe_dbc_vbar})  \\\hline
\multicolumn{4}{l}{Constitutive parameters} \\\hline
$\mu$ & $[\mathrm{kPa}\cdot\mathrm{s}]$ & $4\cdot 10^{-6}$ & Fluid viscosity, Eq.~(\ref{eq:cauchy_eulerian}) \\
$\rho$ & $[\frac{\mathrm{kg}}{\mathrm{mm}^{3}}]$ & $1.025\cdot 10^{-6}$ & Fluid density, Eq.~(\ref{eq:fluid_mom_gen}) \\\hline
\multicolumn{4}{l}{Stabilization parameters} \\\hline
$v_{\max}$ & $[\frac{\mathrm{mm}}{\mathrm{s}}]$ & $5\cdot 10^{3}$ & Velocity scale for stabilization, Eq.~(\ref{eq:stabparams}) \\
$\beta$ & $[\frac{\mathrm{kPa} \cdot \mathrm{s}^2}{\mathrm{mm}^3}]$ & $0.205\cdot 10^{-6}$ & Outflow stabilization parameter, Eq.~(\ref{eq:sout}) \\\hline
\multicolumn{4}{l}{0D model parameters} \\\hline
$C_{\mathrm{in}}$ & $[\frac{\mathrm{mm}^3}{\mathrm{kPa}}]$ & $10^{3}$ & Inflow compliance \\
$R_{\mathrm{in}}$ & $[\frac{\mathrm{kPa} \cdot \mathrm{s}}{\mathrm{mm}^3}]$ & $160\cdot 10^{-6}$ & Inflow resistance \\
$C_{\mathrm{out}}$ & $[\frac{\mathrm{mm}^3}{\mathrm{kPa}}]$ & $0.01$ & Outflow compliance \\
$R_{\mathrm{out}}$ & $[\frac{\mathrm{kPa} \cdot \mathrm{s}}{\mathrm{mm}^3}]$ & $10^{-6}$ & Outflow resistance \\
\end{tabular}
\end{center}
\end{table}

\subsubsection{ALE Fluid-RD/0D (FrSI): Deformable Pipe Flow with Resistive BC}\label{app:examples_frsipipe}

The governing equations for the fluid-reduced-solid interaction (FrSI) model of flow through a deformable pipe are given, Sec.~\ref{sec:examples_frsipipe}, Fig.~\ref{fig:models_all}B, which are the transient Navier-Stokes equations in Arbitrary Lagrangian-Eulerian (ALE) description. Balance of linear momentum and conservation of mass are given by Eq.~(\ref{eq:frsi_mom_gen}) and Eq.~(\ref{eq:frsi_mass_gen}), respectively. The domain motion is governed by the PDE Eq.~(\ref{eq:ale_strong_gen}) with DBC Eq.~(\ref{eq:ale_dbc_gen}). The domain is further clamped in axial $z$-direction:
\begin{align}
    \bs{d}\cdot\bs{e}_{z} &= 0
    \quad &&  
    \text{on}\; \Gm_{0}^{\mathrm{f\mhyphen 0d}} \cup \Gm_{0}^{N} \times [0,T].
    \label{eq:ale_dbc_pipe2}
\end{align}
A simple diffusion model is used as constitutive equation:
\begin{align}
    \bs{\sigma}_{\mathrm{g}}(\bs{d}) = \nabla_{0}\bs{d}.
\end{align}

The boundary conditions for this example read:
\begin{align}
    \bs{t}_{0}^{\mathrm{\fF\mhyphen 0d}} &= -\lmz \,\widehat{J}\widehat{\bs{F}}^{-\mathrm{T}}\bs{n}_{0} \quad &&\text{on} \; \Gm_{0}^{\mathrm{\fF\mhyphen 0d}} \times [0,T], \\
    \bs{t}_{0}^{\mathrm{\fF\mhyphen\fSr}} &= -h_0\left(\rho_{0,\fS} \frac{\partial \bs{v}}{\partial t} - \tilde{\nabla}_{0}\cdot\tilde{\bs{P}}(\bs{u}_{\fF}(\bs{v}))\right) \quad &&\text{on} \; \Gm_{0}^{\mathrm{\fF\mhyphen\fSr}} \times [0,T], \label{eq:Pred_pipe}\\
    \bs{t}_{0}^{N} &= -\hat{p}(t)\widehat{J}\widehat{\bs{F}}^{-\mathrm{T}}\bs{n}_{0} \quad && \text{on}\; \Gm_{0}^{N} \times [0,T],\label{eq:frsipipe_nbc_in}
\end{align}
with the prescribed time-dependent inflow pressure
\begin{equation}
\begin{aligned}
    \hat{p}(t) = \frac{1}{2}\hat{p}_{\max}\left(1-\cos\frac{2\pi t}{\hat{T}}\right),
    \label{eq:frsipipe_nbc_phat}
\end{aligned}
\end{equation}
plotted in Fig.~\ref{fig:pipefrsi_results}A. The constitutive equation for the Cauchy stress is given by Eq.~(\ref{eq:cauchy_ale}). The constraint that links the 3D fluid domain to the 0D model is
\begin{align}
    \int\limits_{\Gm_{0}^{\mathrm{\fF\mhyphen 0d}}} (\bs{v}-\widehat{\bs{w}})\cdot\widehat{J}\widehat{\bs{F}}^{-\mathrm{T}}\bs{n}_{0}\,\mathrm{d}A_0 &= q_{\mathrm{in}}(\lmz) \quad &&\text{in} \; [0,T].
\end{align}
The 0D model equations that govern the variable $q_{\mathrm{in}}$ are described in Sec.~\ref{app:resist}, where the 0D model pressure takes the value of the multiplier, $p_{\mathrm{o}} \leftarrow \lmz$.\\

The constitutive equation for the second Piola-Kirchhoff stress of the reduced solid here is
\begin{align}
\tilde{\bs{S}} = 2\frac{\partial\mathit{\Psi}(\tilde{\bs{C}})}{\partial \tilde{\bs{C}}} + 2\frac{\partial\mathit{\Psi}_{\mathrm{v}}(\dot{\tilde{\bs{C}}})}{\partial \dot{\tilde{\bs{C}}}},
\end{align}
with an isotropic-exponential strain energy function \cite{demiray1972}
\begin{align}
    \mathit{\Psi}(\tilde{\bs{C}}) = \frac{a_0}{2b_0}\left(e^{b_0(\mathrm{tr}\tilde{\bs{C}} - 3)} - 1\right)\label{eq:psi_red_pipe}
\end{align}
and a viscous pseudo-potential \cite{chapelle2012} of the form
\begin{equation}
\begin{aligned}
    \mathit{\Psi}_{\mathrm{v}}(\dot{\tilde{\bs{C}}}) = \frac{\eta}{8}\dot{\tilde{\bs{C}}}:\dot{\tilde{\bs{C}}}.\label{eq:psi_visc_red_pipe}
\end{aligned}
\end{equation}

Table \ref{tab:params_pipersi} summarizes all parameters.

\begin{table}[!htp]
\begin{center}
\caption{Parameters for ALE fluid-0D/FrSI problem of pipe flow.}\label{tab:params_pipersi}
\begin{tabular}{rl|c|l}\hline
\multicolumn{4}{l}{Time parameters} \\\hline
$T$ & $[\mathrm{s}]$ & $0.5$ & Simulation time \\
$\Delta t$ & $[\mathrm{s}]$ & $0.005$ & Simulation time step \\
$\theta$ & $[-]$ & $1.0$ & Time-integration parameter, Eq.~(\ref{eq:theta}) \\
$\theta_{\mathrm{0d}}$ & $[-]$ & $1.0$ & Time-integration parameter for 0D model, Eq.~(\ref{eq:0d_weakform_discr}) \\\hline
\multicolumn{4}{l}{Geometry and boundary conditions} \\\hline
$l$ & $[\mathrm{mm}]$ & $100$ & Pipe length (undeformed) \\
$r$ & $[\mathrm{mm}]$ & $15$ & Pipe radius (undeformed) \\
$\hat{p}_{\max}$ & $[\mathrm{kPa}]$ & $0.1$ & Peak inflow pressure, Eq.~(\ref{eq:frsipipe_nbc_phat}) \\
$\hat{T}$ & $[\mathrm{s}]$ & $\frac{2}{3}$ & Period of pressure profile, Eq.~(\ref{eq:frsipipe_nbc_phat})  \\\hline
\multicolumn{4}{l}{Fluid constitutive parameters} \\\hline
$\mu$ & $[\mathrm{kPa}]$ & $4\cdot 10^{-6}$ & Viscosity, Eq.~(\ref{eq:cauchy_ale}) \\
$\rho$ & $[\frac{\mathrm{kg}}{\mathrm{mm}^{3}}]$ & $1.025\cdot 10^{-6}$ & Density, Eq.~(\ref{eq:frsi_mom_gen}) \\\hline
\multicolumn{4}{l}{Reduced solid constitutive parameters} \\\hline
$h_0$ & $[\mathrm{mm}]$ & $1.0$ & Wall thickness, Eq.~(\ref{eq:Pred_pipe}) \\
$a_0$ & $[\mathrm{kPa}]$ & $1.0$ & Stiffness, Eq.~(\ref{eq:psi_red_pipe}) \\
$b_0$ & $[-]$ & $6.0$ & Exponential, Eq.~(\ref{eq:psi_red_pipe}) \\
$\eta$ & $[\mathrm{kPa}\cdot\mathrm{s}]$ & $0.1$ & Viscosity, Eq.~(\ref{eq:psi_visc_red_pipe}) \\
$\rho_{0,\fS}$ & $[\frac{\mathrm{kg}}{\mathrm{mm}^{3}}]$ & $10^{-6}$ & Density, Eq.~(\ref{eq:Pred_pipe}) \\\hline
\multicolumn{4}{l}{Stabilization parameters} \\\hline
$v_{\max}$ & $[\frac{\mathrm{mm}}{\mathrm{s}}]$ & $0.5\cdot 10^{3}$ & Velocity scale for stabilization, Eq.~(\ref{eq:stabparams}) \\
$\beta$ & $[\frac{\mathrm{kPa} \cdot \mathrm{s}^2}{\mathrm{mm}^3}]$ & $0.205\cdot 10^{-6}$ & Outflow stabilization parameter, Eq.~(\ref{eq:sout_ale}) \\\hline
\multicolumn{4}{l}{0D model parameters} \\\hline
$R$ & $[\frac{\mathrm{kPa} \cdot \mathrm{s}}{\mathrm{mm}^3}]$ & $10^{3}$ & Resistance\\
$p_{\mathrm{ref}}$ & $[\mathrm{kPa}]$ & $0$ & Downstream reference pressure 
\end{tabular}
\end{center}
\end{table}

\subsubsection{Fluid-0D: Blood Flow in Patient-Specific Aortic Arch}\label{app:examples_fluidaort}

In this section, the governing equations for blood flow in a rigid patient-specific aortic arch model, Sec.~\ref{sec:examples_fluidaort}, Fig.~\ref{fig:models_all}C, are elaborated. These are the incompressible transient Navier-Stokes equations in Eulerian description. Balance of linear momentum and conservation of mass are given by Eq.~(\ref{eq:fluid_mom_gen}) and Eq.~(\ref{eq:fluid_mass_gen}), respectively. The set of multipliers for the coupling to the lumped-parameter model is $\{\lmz\}_{n_{\mathrm{0d}}^{\mathrm{b}}}=\left\{\lmz_{1},\ldots,\lmz_{4}\right\}$. The boundary tractions from the constraints, the Robin traction at the aortic valve plane, as well as no-slip Dirichlet conditions read:
\begin{align}
    \bs{t}_{1}^{\mathrm{\fF\mhyphen 0d}} &= -\lmz_{1} \,\bs{n} \quad &&\text{on} \; \Gm_{\mathrm{AO}_{\mathrm{in}}}^{\mathrm{\fF\mhyphen 0d}} \times [0,T], \\
    \bs{t}_{i}^{\mathrm{\fF\mhyphen 0d}} &= -\lmz_{i} \,\bs{n} \quad &&\text{on} \; \Gm_{\mathrm{AO}_{\mathrm{out}}^{i-1}}^{\mathrm{\fF\mhyphen 0d}} \times [0,T], \\
    &i \in\{2,\ldots,4\},\nonumber\\
    \bs{t}^{R} &= -c(\Delta\bar{p})\,\bs{v} \quad &&\text{on} \; \Gm_{\mathrm{av}}^{R} \times [0,T], \label{eq:av_aort}\\
    \bs{v} &= \bs{0} \quad && \text{on}\; \Gm_{\mathrm{wall}}^{D} \times [0,T].
\end{align}
The constitutive equation for the Cauchy stress is given by Eq.~(\ref{eq:cauchy_eulerian}). The model for the pressure difference-dependent viscosity coefficient in Eq.~(\ref{eq:av_aort}) is detailed in Sec.~\ref{app:3dvalve}.

The four constraints that enforce consistency between the closed-loop circulation model and the 3D domain are: 
\begin{align}
    \int\limits_{\Gm_{\mathrm{AO}_{\mathrm{in}}}^{\mathrm{\fF\mhyphen 0d}}} \bs{v}\cdot\bs{n}\,\mathrm{d}A &= -q_{\mathrm{v,out}}^{\ell}(\{\lmz\}_{n_{\mathrm{0d}}^{\mathrm{b}}}) \quad &&\text{in} \; [0,T], \label{eq:aort0d_qout}\\
    \int\limits_{\Gm_{\mathrm{AO}_{\mathrm{out}}^{1}}^{\mathrm{\fF\mhyphen 0d}}} \bs{v}\cdot\bs{n}\,\mathrm{d}A &= q_{\mathrm{cor,p,in}}^{\mathrm{sys},\ell}(\{\lmz\}_{n_{\mathrm{0d}}^{\mathrm{b}}}) \quad &&\text{in} \; [0,T],\\
    \int\limits_{\Gm_{\mathrm{AO}_{\mathrm{out}}^{2}}^{\mathrm{\fF\mhyphen 0d}}} \bs{v}\cdot\bs{n}\,\mathrm{d}A &= q_{\mathrm{cor,p,in}}^{\mathrm{sys},r}(\{\lmz\}_{n_{\mathrm{0d}}^{\mathrm{b}}}) \quad &&\text{in} \; [0,T],\\
    \int\limits_{\Gm_{\mathrm{AO}_{\mathrm{out}}^{3}}^{\mathrm{\fF\mhyphen 0d}}} \bs{v}\cdot\bs{n}\,\mathrm{d}A &= q_{\mathrm{ar,p}}^{\mathrm{sys}}(\{\lmz\}_{n_{\mathrm{0d}}^{\mathrm{b}}}) \quad &&\text{in} \; [0,T].
\end{align}

The circulatory system model equations that govern the aortic inflow (left ventricular out-flux $q_{\mathrm{v,out}}^{\ell}$), left and right aortic coronary outflow (coronary circulation in-fluxes $q_{\mathrm{cor,p,in}}^{\mathrm{sys},\ell}$ and $q_{\mathrm{cor,p,in}}^{\mathrm{sys},r}$), as well as aortic main outflow (proximal aortic in-flux $q_{\mathrm{ar,p}}^{\mathrm{sys}}$) are detailed in Sec.~\ref{app:syspul} and Sec.~\ref{app:coronary}. The 0D pressures that take the multiplier values are $p_{\mathrm{v,o}}^{\ell} \leftarrow \lmz_1$, $p_{\mathrm{ar}}^{\mathrm{sys},\ell} \leftarrow \lmz_2$, $p_{\mathrm{ar}}^{\mathrm{sys},r} \leftarrow \lmz_3$, and $p_{\mathrm{ar}}^{\mathrm{sys}} \leftarrow \lmz_4$. We note the minus sign in Eq.~(\ref{eq:aort0d_qout}) since an inflow into the fluid domain is enforced. Table \ref{tab:params_aort} summarizes all parameters.

\begin{table}[!htp]
\begin{center}
\caption{Parameters for fluid-0D problem of aortic arch flow.}\label{tab:params_aort}
\begin{tabular}{rl|c|l}\hline
\multicolumn{4}{l}{Time parameters} \\\hline
$T$ & $[\mathrm{s}]$ & $1.0$ & Simulation time \\
$\Delta t$ & $[\mathrm{s}]$ & $0.002$ & Simulation time step \\
$\theta$ & $[-]$ & $1.0$ & Time-integration parameter, Eq.~(\ref{eq:theta}) \\
$\theta_{\mathrm{0d}}$ & $[-]$ & $1.0$ & Time-integration parameter for 0D model, Eq.~(\ref{eq:0d_weakform_discr}) \\\hline
\multicolumn{4}{l}{Constitutive parameters} \\\hline
$\mu$ & $[\mathrm{kPa}\cdot\mathrm{s}]$ & $4\cdot 10^{-6}$ & Fluid viscosity, Eq.~(\ref{eq:cauchy_eulerian}) \\
$\rho$ & $[\frac{\mathrm{kg}}{\mathrm{mm}^{3}}]$ & $1.025\cdot 10^{-6}$ & Fluid density, Eq.~(\ref{eq:fluid_mom_gen}) \\\hline
\multicolumn{4}{l}{Stabilization parameters} \\\hline
$v_{\max}$ & $[\frac{\mathrm{mm}}{\mathrm{s}}]$ & $4\cdot 10^{3}$ & Velocity scale for stabilization, Eq.~(\ref{eq:stabparams}) \\
$\beta$ & $[\frac{\mathrm{kPa} \cdot \mathrm{s}^2}{\mathrm{mm}^3}]$ & $0.205\cdot 10^{-6}$ & Outflow stabilization parameter, Eq.~(\ref{eq:sout}) \\\hline
\multicolumn{4}{l}{0D model parameters} \\\hline
$R_{\mathrm{ar}}^{\mathrm{sys}}$ & $[\frac{\mathrm{kPa} \cdot \mathrm{s}}{\mathrm{mm}^3}]$ & $90\cdot 10^{-6}$ & Systemic arterial resistance \\
$C_{\mathrm{ar}}^{\mathrm{sys}}$ & $[\frac{\mathrm{mm}^3}{\mathrm{kPa}}]$ & $19^{3}$ & Systemic arterial compliance \\
$Z_{\mathrm{ar}}^{\mathrm{sys}}$ & $[\frac{\mathrm{kPa} \cdot \mathrm{s}}{\mathrm{mm}^3}]$ & $4.5\cdot 10^{-6}$ & Systemic arterial impedance \\
$L_{\mathrm{ar}}^{\mathrm{sys}}$ & $[\frac{\mathrm{kPa} \cdot \mathrm{s}^2}{\mathrm{mm}^3}]$ & $0$ & Systemic arterial inertance \\
$R_{\mathrm{ven}}^{\mathrm{sys}}$ & $[\frac{\mathrm{kPa} \cdot \mathrm{s}}{\mathrm{mm}^3}]$ & $24\cdot 10^{-6}$ & Systemic venous resistance \\
$C_{\mathrm{ven}}^{\mathrm{sys}}$ & $[\frac{\mathrm{mm}^3}{\mathrm{kPa}}]$ & $413105.83$ & Systemic venous compliance \\
$L_{\mathrm{ven}}^{\mathrm{sys}}$ & $[\frac{\mathrm{kPa} \cdot \mathrm{s}^2}{\mathrm{mm}^3}]$ & $0$ & Systemic venous inertance \\
$R_{\mathrm{ar}}^{\mathrm{pul}}$ & $[\frac{\mathrm{kPa} \cdot \mathrm{s}}{\mathrm{mm}^3}]$ & $15\cdot 10^{-6}$ & Pulmonary arterial resistance \\
$C_{\mathrm{ar}}^{\mathrm{pul}}$ & $[\frac{\mathrm{mm}^3}{\mathrm{kPa}}]$ & $20^{3}$ & Pulmonary arterial compliance \\
$L_{\mathrm{ar}}^{\mathrm{pul}}$ & $[\frac{\mathrm{kPa} \cdot \mathrm{s}^2}{\mathrm{mm}^3}]$ & $0$ & Pulmonary arterial inertance \\
$R_{\mathrm{ven}}^{\mathrm{pul}}$ & $[\frac{\mathrm{kPa} \cdot \mathrm{s}}{\mathrm{mm}^3}]$ & $15\cdot 10^{-6}$ & Pulmonary venous resistance \\
$C_{\mathrm{ven}}^{\mathrm{pul}}$ & $[\frac{\mathrm{mm}^3}{\mathrm{kPa}}]$ & $50^{3}$ & Pulmonary venous compliance \\
$L_{\mathrm{ven}}^{\mathrm{pul}}$ & $[\frac{\mathrm{kPa} \cdot \mathrm{s}^2}{\mathrm{mm}^3}]$ & $0$ & Pulmonary venous inertance \\
$E_{\mathrm{at},\max}^{\ell}$, $E_{\mathrm{at},\min}^{\ell}$ & $[\frac{\mathrm{kPa}}{\mathrm{mm}^3}]$ & $29^{-6}$, $9^{-6}$ & Maximum, minimum left atrial elastance\\
$E_{\mathrm{at},\max}^{r}$, $E_{\mathrm{at},\min}^{r}$ & $[\frac{\mathrm{kPa}}{\mathrm{mm}^3}]$ & $18^{-6}$, $8^{-6}$ & Maximum, minimum right atrial elastance\\
$E_{\mathrm{v},\max}^{\ell}$, $E_{\mathrm{v},\min}^{\ell}$ & $[\frac{\mathrm{kPa}}{\mathrm{mm}^3}]$ & $600^{-6}$, $12^{-6}$ & Maximum, minimum left ventricular elastance\\
$E_{\mathrm{v},\max}^{r}$, $E_{\mathrm{v},\min}^{r}$ & $[\frac{\mathrm{kPa}}{\mathrm{mm}^3}]$ & $400^{-6}$, $10^{-6}$ & Maximum, minimum right ventricular elastance\\
$R_{\mathrm{mv},\min}$, $R_{\mathrm{mv},\max}$ & $[\frac{\mathrm{kPa} \cdot \mathrm{s}}{\mathrm{mm}^3}]$ & $10^{-6}$, $10^{1}$ & Minimum, maximum mitral valve resistance \\
$R_{\mathrm{av},\min}$, $R_{\mathrm{av},\max}$ & $[\frac{\mathrm{kPa} \cdot \mathrm{s}}{\mathrm{mm}^3}]$ & $10^{-6}$, $10^{-6}$ & Minimum, maximum aortic valve resistance\footnote{The aortic valve here lies within the 3D fluid domain and is controlled via a pressure jump-dependent Robin condition. The 0D valve model hence is ``disabled'' by setting its maximum resistance to the same as the minimum (``open'') resistance value.} \\
$R_{\mathrm{tv},\min}$, $R_{\mathrm{tv},\max}$ & $[\frac{\mathrm{kPa} \cdot \mathrm{s}}{\mathrm{mm}^3}]$ & $10^{-6}$, $10^{1}$ & Minimum, maximum tricuspid valve resistance \\
$R_{\mathrm{pv},\min}$, $R_{\mathrm{pv},\max}$ & $[\frac{\mathrm{kPa} \cdot \mathrm{s}}{\mathrm{mm}^3}]$ & $10^{-6}$, $10^{1}$ & Minimum, maximum pulmonary valve resistance \\
$R_{\mathrm{cor,p}}^{\mathrm{sys},\ell}$, $R_{\mathrm{cor,p}}^{\mathrm{sys},r}$ & $[\frac{\mathrm{kPa} \cdot \mathrm{s}}{\mathrm{mm}^3}]$ & $6.55\cdot 10^{-3}$ & Left, right coronary proximal resistance \\
$C_{\mathrm{cor,p}}^{\mathrm{sys},\ell}$, $C_{\mathrm{cor,p}}^{\mathrm{sys},r}$ & $[\frac{\mathrm{mm}^3}{\mathrm{kPa}}]$ & $4.5$ & Left, right coronary proximal compliance\\
$Z_{\mathrm{cor,p}}^{\mathrm{sys},\ell}$, $Z_{\mathrm{cor,p}}^{\mathrm{sys},r}$ & $[\frac{\mathrm{kPa} \cdot \mathrm{s}}{\mathrm{mm}^3}]$ & $3.2\cdot 10^{-3}$ & Left, right coronary proximal impedance \\
$R_{\mathrm{cor,d}}^{\mathrm{sys},\ell}$, $R_{\mathrm{cor,d}}^{\mathrm{sys},r}$ & $[\frac{\mathrm{kPa} \cdot \mathrm{s}}{\mathrm{mm}^3}]$ & $1.45\cdot 10^{-1}$ & Left, right coronary distal resistance \\
$C_{\mathrm{cor,d}}^{\mathrm{sys},\ell}$, $C_{\mathrm{cor,d}}^{\mathrm{sys},r}$ & $[\frac{\mathrm{mm}^3}{\mathrm{kPa}}]$ & $27$ & Left, right coronary distal compliance\\
\end{tabular}
\end{center}
\end{table}

\subsubsection{ALE Fluid-RD/0D (FrSI): Hemodynamics in the Left Heart}\label{app:examples_frsiheart}

Fluid-reduced-solid interaction (FrSI) in the left heart is considered, Sec.~\ref{sec:examples_frsiheart}, Fig.~\ref{fig:models_all}D, governed by transient Navier-Stokes equations in Arbitrary Lagrangian-Eulerian (ALE) description. Balance of linear momentum and conservation of mass are given by Eq.~(\ref{eq:frsi_mom_gen}) and Eq.~(\ref{eq:frsi_mass_gen}), respectively. The domain motion is governed by the PDE Eq.~(\ref{eq:ale_strong_gen}) with DBC Eq.~(\ref{eq:ale_dbc_gen}) on $\Gm_{0}^{\mathrm{\fF\mhyphen\fSr}} = \Gm_{0,\mathrm{LA}}^{\mathrm{\fF\mhyphen\fSr}} \cup \Gm_{0,\mathrm{LV}}^{\mathrm{\fF\mhyphen\fSr}} \cup \Gm_{0,\mathrm{AO}}^{\mathrm{\fF\mhyphen\fSr}}$. The constitutive model for the ALE domain motion problem here needs to be chosen such that large deformations do not compromise the quality of the spatial discretization. This may happen in regions of high curvature close to the boundary layers, where mesh resolutions typically are refined and more anisotropic than in regions more distant from the boundary. Here, for the monolithic FrSI approach (solver Ambit), we use a fully nonlinear coupled NeoHookean model \cite{holzapfel2000}, which, on the discrete space, is scaled by the inverse of the reference cell's Jacobian determinant \cite{shamanskiy2021}, effectively allocating stiffness to more anisotropic boundary elements and allowing the more regularly shaped bulk elements to bear most of the deformation. For the partitioned approach (solver $\mathcal{C}\mathrm{Heart}$), we use a linear elastic constitutive model with a nonlinear dependence on the current mesh deformation, similar to approaches in \cite{balmus2020-pufem,spuehler2018}. The set of multipliers for the coupling to the lumped-parameter model is $\{\lmz\}_{n_{\mathrm{0d}}^{\mathrm{b}}}=\left\{\lmz_{1},\ldots,\lmz_{7}\right\}$.\\

The boundary conditions for this example read:
\begin{align}
    \bs{t}_{0,i}^{\mathrm{\fF\mhyphen 0d}} &= -\lmz_{i} \,\widehat{J}\widehat{\bs{F}}^{-\mathrm{T}}\bs{n}_{0} \quad &&\text{on} \; \Gm_{0,\mathrm{LA}_{\mathrm{in}}^{i}}^{\mathrm{\fF\mhyphen 0d}} \times [0,T], \\
    &i \in\{1,\ldots,4\},\nonumber\\
    \bs{t}_{0,j}^{\mathrm{\fF\mhyphen 0d}} &= -\lmz_{j} \,\widehat{J}\widehat{\bs{F}}^{-\mathrm{T}}\bs{n}_{0} \quad &&\text{on} \; \Gm_{0,\mathrm{AO}_{\mathrm{out}}^{j-4}}^{\mathrm{\fF\mhyphen 0d}} \times [0,T], \\
    &j \in\{5,\ldots,7\},\nonumber\\
    \bs{t}_{0,k}^{\mathrm{\fF\mhyphen\fSr}} &= -h_0^{k}\left(\rho_{0,\fS}^{k} \frac{\partial\bs{v}}{\partial t} - \tilde{\nabla}_{0}\cdot\tilde{\bs{P}}(\bs{u}_{\fF}(\bs{v})\!+\!\bs{u}_{\mathrm{pre}}, \bs{v})\right) \quad &&\text{on} \; \Gm_{0,k}^{\mathrm{\fF\mhyphen\fSr}} \times [0,T],\label{eq:Pred_lalvao} \\
    &k \in\{\mathrm{LA},\mathrm{LV},\mathrm{AO}\},\nonumber\\
    \bs{t}_{0,m}^{R} &= -c(\Delta\bar{p})\,(\bs{v}-\widehat{\bs{w}}) \quad &&\text{on} \; \Gm_{0,m}^{R} \times [0,T],\label{eq:mvav_lalvao}\\
    &m \in\{\mathrm{mv},\mathrm{av}\}.\nonumber
\end{align}
The constitutive equation for the Cauchy stress is given by Eq.~(\ref{eq:cauchy_ale}). The model for the pressure difference-dependent viscosity coefficients in Eq.~(\ref{eq:mvav_lalvao}) is detailed in Sec.~\ref{app:3dvalve}.\\

The seven constraints enforcing consistency between the closed-loop circulation model and the 3D domain are:
\begin{align}
    \int\limits_{\Gm_{0,\mathrm{LA}_{\mathrm{in}}^{i}}^{\mathrm{\fF\mhyphen 0d}}} (\bs{v}-\widehat{\bs{w}})\cdot\widehat{J}\widehat{\bs{F}}^{-\mathrm{T}}\bs{n}_{0}\,\mathrm{d}A_0 &= -q_{\mathrm{ven},i}^{\mathrm{pul}}(\{\lmz\}_{n_{\mathrm{0d}}^{\mathrm{b}}}) \quad &&\text{in} \; [0,T],\label{eq:qvenpul}\\
    &i \in\{1,\ldots,4\},\nonumber\\
    \int\limits_{\Gm_{0,\mathrm{AO}_{\mathrm{out}}^{1}}^{\mathrm{\fF\mhyphen 0d}}} (\bs{v}-\widehat{\bs{w}})\cdot\widehat{J}\widehat{\bs{F}}^{-\mathrm{T}}\bs{n}_{0}\,\mathrm{d}A_0 &= q_{\mathrm{cor,p,in}}^{\mathrm{sys},\ell}(\{\lmz\}_{n_{\mathrm{0d}}^{\mathrm{b}}}) \quad &&\text{in} \; [0,T],\\
    \int\limits_{\Gm_{0,\mathrm{AO}_{\mathrm{out}}^{2}}^{\mathrm{\fF\mhyphen 0d}}} (\bs{v}-\widehat{\bs{w}})\cdot\widehat{J}\widehat{\bs{F}}^{-\mathrm{T}}\bs{n}_{0}\,\mathrm{d}A_0 &= q_{\mathrm{cor,p,in}}^{\mathrm{sys},r}(\{\lmz\}_{n_{\mathrm{0d}}^{\mathrm{b}}}) \quad &&\text{in} \; [0,T],\\
    \int\limits_{\Gm_{0,\mathrm{AO}_{\mathrm{out}}^{3}}^{\mathrm{\fF\mhyphen 0d}}} (\bs{v}-\widehat{\bs{w}})\cdot\widehat{J}\widehat{\bs{F}}^{-\mathrm{T}}\bs{n}_{0}\,\mathrm{d}A_0 &= q_{\mathrm{ar,p}}^{\mathrm{sys}}(\{\lmz\}_{n_{\mathrm{0d}}^{\mathrm{b}}}) \quad &&\text{in} \; [0,T].
\end{align}

The circulatory system model equations that govern the four left atrial inflows (pulmonary venous out-fluxes $q_{\mathrm{ven},i}^{\mathrm{pul}}$), left and right aortic coronary outflow (coronary circulation in-fluxes $q_{\mathrm{cor,p,in}}^{\mathrm{sys},\ell}$ and $q_{\mathrm{cor,p,in}}^{\mathrm{sys},r}$), as well as aortic main outflow (proximal aortic in-flux $q_{\mathrm{ar,p}}^{\mathrm{sys}}$) are detailed in Sec.~\ref{app:syspul} and Sec.~\ref{app:coronary}. The 0D pressures that take the multiplier values are the four left atrial inlet pressures, $p_{\mathrm{at,i},j}^{\ell} \leftarrow \lmz_j$ ($j=1,\hdots,4$), as well as the outlet pressures at the aortic arch, $p_{\mathrm{ar}}^{\mathrm{sys},\ell} \leftarrow \lmz_5$, $p_{\mathrm{ar}}^{\mathrm{sys},r} \leftarrow \lmz_6$, and $p_{\mathrm{ar}}^{\mathrm{sys}} \leftarrow \lmz_7$. We note the minus signs in Eq.~(\ref{eq:qvenpul}) since inflows into the fluid domain are enforced.\\

The constitutive equations for the second Piola-Kirchhoff stress are, for the left atrium,
\begin{align}
\tilde{\bs{S}}^{(\mathrm{LA})} = 2\frac{\partial\mathit{\Psi}(\tilde{\bs{C}})}{\partial \tilde{\bs{C}}} + 2\frac{\partial\mathit{\Psi}_{\mathrm{v}}(\dot{\tilde{\bs{C}}})}{\partial \dot{\tilde{\bs{C}}}} + \tau_{\mathrm{a}}^{\mathrm{LA}}(t) \bs{I}, \label{eq:S_la}
\end{align}
the left ventricle,
\begin{align}
\tilde{\bs{S}}^{(\mathrm{LV})} = 2\frac{\partial\mathit{\Psi}(\tilde{\bs{C}})}{\partial \tilde{\bs{C}}} + 2\frac{\partial\mathit{\Psi}_{\mathrm{v}}(\dot{\tilde{\bs{C}}})}{\partial \dot{\tilde{\bs{C}}}} + \tau_{\mathrm{a}}^{\mathrm{LV}}(t) \tilde{\bs{M}}_{0}, \label{eq:S_lv}
\end{align}
and the aortic arch:
\begin{align}
\tilde{\bs{S}}^{(\mathrm{AO})} = 2\frac{\partial\mathit{\Psi}(\tilde{\bs{C}})}{\partial \tilde{\bs{C}}} + 2\frac{\partial\mathit{\Psi}_{\mathrm{v}}(\dot{\tilde{\bs{C}}})}{\partial \dot{\tilde{\bs{C}}}}, \label{eq:S_ao}
\end{align}
with the previously defined isotropic-exponential strain energy, Eq.~(\ref{eq:psi_red_pipe}), and viscous pseudo-potential, Eq.~(\ref{eq:psi_visc_red_pipe}). Parameters differ for each of the three regions, cf. Tab.~\ref{tab:params_lalvao} for their values. Further, $\tau_{\mathrm{a}}^{\mathrm{LA}}(t)$ and $\tau_{\mathrm{a}}^{\mathrm{LV}}(t)$ are time-dependent active stresses stemming from the solution of the evolution equation
\begin{equation}
    \dot{\tau}_{\mathrm{a}}=-|u|\tau_{\mathrm{a}} + \sigma_{0} |u|_{+},\label{eq:tau}
\end{equation}
with $|u|$ for the absolute value of $u$ and $|u|_{+} := \max(0,u)$ \cite{bestel2001}. The function $u$ is a scaling of the user-defined activation function $\hat{f}(t) \in [0,1]$:
\begin{equation}
    u=\hat{f}(t)\cdot \alpha_{\max} + (1-\hat{f}(t))\cdot \alpha_{\min}.\label{eq:u}
\end{equation}
The contractility of the muscle is denoted by $\sigma_{0}$, and the upstroke and relaxation rates are $\alpha_{\max}$ and $\alpha_{\min}$, respectively.
$\tilde{\bs{M}}_{0}$ is a reduced structural tensor, cf. \cite{hirschvogel2024-frsi}. Prior to any transient simulation, the reduced solid model is \textit{prestressed} \cite{schein2021,gee2011} to the initial pressure values, cf. the methods described in \cite{hirschvogel2024-frsi}, where $\bs{u}_{\mathrm{pre}}$ is the prestress displacement. The projection-based component of FrSI is performed post spatio-temporal discretization and detailed in Sec.~\ref{subsec:frsi}. The construction of the POD modes of the wall along with region-wise decomposition and boundary conditions in reduced space is detailed in Sec.~\ref{app:frsi_heart_pod}. The wall thickness is further modified to translate smoothly around the valve rims, Eq.~(\ref{eq:h0la})---(\ref{eq:h0ao}). Table \ref{tab:params_lalvao} summarizes all parameters.

\begin{table}[!htp]
\begin{center}
\caption{Parameters for ALE fluid-0D problem of left heart hemodynamics.}\label{tab:params_lalvao}
\resizebox{1.0\columnwidth}{!}{%
\begin{tabular}{rl|c|l}\hline
\multicolumn{4}{l}{Time parameters} \\\hline
$T$ & $[\mathrm{s}]$ & $1.0$ & Simulation time \\
$\Delta t$ & $[\mathrm{s}]$ & $0.00125$ & Simulation time step \\
$\theta$ & $[-]$ & $1.0$ & Time-integration parameter, Eq.~(\ref{eq:theta}) \\
$\theta_{\mathrm{0d}}$ & $[-]$ & $1.0$ & Time-integration parameter for 0D model, Eq.~(\ref{eq:0d_weakform_discr}) \\\hline
\multicolumn{4}{l}{Fluid constitutive parameters} \\\hline
$\mu$ & $[\mathrm{kPa}\cdot\mathrm{s}]$ & $4\cdot 10^{-6}$ & Fluid viscosity, Eq.~(\ref{eq:cauchy_ale}) \\
$\rho$ & $[\frac{\mathrm{kg}}{\mathrm{mm}^{3}}]$ & $1.025\cdot 10^{-6}$ & Fluid density, Eq.~(\ref{eq:frsi_mom_gen}) \\\hline
\multicolumn{4}{l}{Reduced solid constitutive parameters} \\\hline
 &  & LA \; LV \;  AO &  \\\hline
$h_{0,\mathrm{b}}^{i}$ & $[\mathrm{mm}]$ & $5.0$ \; $10.0$ \; $1.0$ & Base wall thickness for atrium, ventricle, aorta, Eq.~(\ref{eq:Pred_lalvao}) \\
$a_0^{i}$ & $[\mathrm{kPa}]$ & $1.7$ \; $1.7$ \; $44.2$ & Stiffness for atrium, ventricle \cite{hirschvogel2024-frsi}, aorta \cite{delfino1997,holzapfel2000-artery} \\
$b_0^{i}$ & $[-]$ & $3.9$ \; $3.9$ \; $8.35$ & Exponential for atrium, ventricle \cite{hirschvogel2024-frsi}, aorta \cite{delfino1997,holzapfel2000-artery}, Eq.~(\ref{eq:psi_red_pipe}) \\
$\eta^{i}$ & $[\mathrm{kPa}\cdot\mathrm{s}]$ & $0.1$ \; $0.1$ \; $0.1$ & Viscosity for atrium, ventricle, aorta, Eq.~(\ref{eq:psi_visc_red_pipe})\\
$\rho_{0,\fS}^{i}$ & $[\frac{\mathrm{kg}}{\mathrm{mm}^{3}}]$ & $10^{-6}$ \; $10^{-6}$ \; $10^{-6}$ & Density for atrium, ventricle, aorta, Eq.~(\ref{eq:Pred_lalvao}) \\
$\sigma_0^{i}$ & $[\mathrm{kPa}]$ & $3$ \;\;\; $50$ \;\;\; $-$ & Contractility for atrium, ventricle, Eq.~(\ref{eq:tau}) \\
$\alpha_{\max}^{i}$ & $[\frac{1}{\mathrm{s}}]$ & $10$ \;\; $10$ \;\; $-$ & Upstroke rate for atrium, ventricle, Eq.~(\ref{eq:tau}) \\
$\alpha_{\min}^{i}$ & $[\frac{1}{\mathrm{s}}]$ & \!\!\!\!$-30$ \; \!\!$-30$ \; $-$ & Relaxation rate for atrium, ventricle, Eq.~(\ref{eq:tau}) \\\hline
\multicolumn{4}{l}{Stabilization parameters} \\\hline
$v_{\max}$ & $[\frac{\mathrm{mm}}{\mathrm{s}}]$ & $10^{3}$ & Velocity scale for stabilization, Eq.~(\ref{eq:stabparams}) \\
$\beta$ & $[\frac{\mathrm{kPa} \cdot \mathrm{s}^2}{\mathrm{mm}^3}]$ & $0.205\cdot 10^{-6}$ & Outflow stabilization parameter, Eq.~(\ref{eq:sout_ale}) \\\hline
\multicolumn{4}{l}{0D model parameters} \\\hline
$R_{\mathrm{ar}}^{\mathrm{sys}}$ & $[\frac{\mathrm{kPa} \cdot \mathrm{s}}{\mathrm{mm}^3}]$ & $120\cdot 10^{-6}$ & Systemic arterial resistance \\
$C_{\mathrm{ar}}^{\mathrm{sys}}$ & $[\frac{\mathrm{mm}^3}{\mathrm{kPa}}]$ & $13770.19$ & Systemic arterial compliance \\
$Z_{\mathrm{ar}}^{\mathrm{sys}}$ & $[\frac{\mathrm{kPa} \cdot \mathrm{s}}{\mathrm{mm}^3}]$ & $6\cdot 10^{-6}$ & Systemic arterial impedance \\
$L_{\mathrm{ar}}^{\mathrm{sys}}$ & $[\frac{\mathrm{kPa} \cdot \mathrm{s}^2}{\mathrm{mm}^3}]$ & $0$ & Systemic arterial inertance \\
$R_{\mathrm{ven}}^{\mathrm{sys}}$ & $[\frac{\mathrm{kPa} \cdot \mathrm{s}}{\mathrm{mm}^3}]$ & $24\cdot 10^{-6}$ & Systemic venous resistance \\
$C_{\mathrm{ven}}^{\mathrm{sys}}$ & $[\frac{\mathrm{mm}^3}{\mathrm{kPa}}]$ & $413105.83$ & Systemic venous compliance \\
$L_{\mathrm{ven}}^{\mathrm{sys}}$ & $[\frac{\mathrm{kPa} \cdot \mathrm{s}^2}{\mathrm{mm}^3}]$ & $0$ & Systemic venous inertance \\
$R_{\mathrm{ar}}^{\mathrm{pul}}$ & $[\frac{\mathrm{kPa} \cdot \mathrm{s}}{\mathrm{mm}^3}]$ & $15\cdot 10^{-6}$ & Pulmonary arterial resistance \\
$C_{\mathrm{ar}}^{\mathrm{pul}}$ & $[\frac{\mathrm{mm}^3}{\mathrm{kPa}}]$ & $20^{3}$ & Pulmonary arterial compliance \\
$L_{\mathrm{ar}}^{\mathrm{pul}}$ & $[\frac{\mathrm{kPa} \cdot \mathrm{s}^2}{\mathrm{mm}^3}]$ & $0$ & Pulmonary arterial inertance \\
$R_{\mathrm{ven}}^{\mathrm{pul}}$ & $[\frac{\mathrm{kPa} \cdot \mathrm{s}}{\mathrm{mm}^3}]$ & $15\cdot 10^{-6}$ & Pulmonary venous resistance \\
$C_{\mathrm{ven}}^{\mathrm{pul}}$ & $[\frac{\mathrm{mm}^3}{\mathrm{kPa}}]$ & $50^{3}$ & Pulmonary venous compliance \\
$L_{\mathrm{ven}}^{\mathrm{pul}}$ & $[\frac{\mathrm{kPa} \cdot \mathrm{s}^2}{\mathrm{mm}^3}]$ & $0$ & Pulmonary venous inertance \\
$E_{\mathrm{at},\max}^{r}$, $E_{\mathrm{at},\min}^{r}$ & $[\frac{\mathrm{kPa}}{\mathrm{mm}^3}]$ & $18^{-6}$, $8^{-6}$ & Maximum, minimum right atrial elastance\\
$E_{\mathrm{v},\max}^{r}$, $E_{\mathrm{v},\min}^{r}$ & $[\frac{\mathrm{kPa}}{\mathrm{mm}^3}]$ & $400^{-6}$, $10^{-6}$ & Maximum, minimum right ventricular elastance\\
$R_{\mathrm{tv},\min}$, $R_{\mathrm{tv},\max}$ & $[\frac{\mathrm{kPa} \cdot \mathrm{s}}{\mathrm{mm}^3}]$ & $10^{-6}$, $10^{1}$ & Minimum, maximum tricuspid valve resistance \\
$R_{\mathrm{pv},\min}$, $R_{\mathrm{pv},\max}$ & $[\frac{\mathrm{kPa} \cdot \mathrm{s}}{\mathrm{mm}^3}]$ & $10^{-6}$, $10^{1}$ & Minimum, maximum pulmonary valve resistance \\
$R_{\mathrm{cor,p}}^{\mathrm{sys},\ell}$, $R_{\mathrm{cor,p}}^{\mathrm{sys},r}$ & $[\frac{\mathrm{kPa} \cdot \mathrm{s}}{\mathrm{mm}^3}]$ & $6.55\cdot 10^{-3}$ & Left, right coronary proximal resistance \\
$C_{\mathrm{cor,p}}^{\mathrm{sys},\ell}$, $C_{\mathrm{cor,p}}^{\mathrm{sys},r}$ & $[\frac{\mathrm{mm}^3}{\mathrm{kPa}}]$ & $4.5$ & Left, right coronary proximal compliance\\
$Z_{\mathrm{cor,p}}^{\mathrm{sys},\ell}$, $Z_{\mathrm{cor,p}}^{\mathrm{sys},r}$ & $[\frac{\mathrm{kPa} \cdot \mathrm{s}}{\mathrm{mm}^3}]$ & $3.2\cdot 10^{-3}$ & Left, right coronary proximal impedance \\
$R_{\mathrm{cor,d}}^{\mathrm{sys},\ell}$, $R_{\mathrm{cor,d}}^{\mathrm{sys},r}$ & $[\frac{\mathrm{kPa} \cdot \mathrm{s}}{\mathrm{mm}^3}]$ & $1.45\cdot 10^{-1}$ & Left, right coronary distal resistance \\
$C_{\mathrm{cor,d}}^{\mathrm{sys},\ell}$, $C_{\mathrm{cor,d}}^{\mathrm{sys},r}$ & $[\frac{\mathrm{mm}^3}{\mathrm{kPa}}]$ & $27$ & Left, right coronary distal compliance\\
\end{tabular}
}%
\end{center}
\end{table}

\subsection{0D Model Equations}\label{app:0d}

Here, we present the time-continuous ODEs of the 0D models that are used in our examples. The simple resistive model is described in Sec.~\ref{app:resist}, the 2-element Windkessel models in series that link an out- to an inflow in Sec.~\ref{app:crlink}, and the circulatory system as well as coronary circulation model in Sec.~\ref{app:syspul} and Sec.~\ref{app:coronary}, respectively. The set of algebraic and first-order ordinary differential equations is discretized with an implicit one-step $\theta$-scheme, cf. Eq.~(\ref{eq:0d_weakform_discr}).

\subsubsection{Resistive Model}\label{app:resist}
The resistive 0D model, cf. example in Sec.~\ref{sec:examples_frsipipe}, represents the simplest form of a flux-dependent condition and mimics the steady Poiseuille's law \cite{sutera1993} of flow through a rigid vessel:
\begin{align}
0 = q_{\mathrm{in}} - q_{\mathrm{out}}, \label{eq:0d_res_flowb} \\
R\,q_{\mathrm{out}} = p_{\mathrm{o}} - p_{\mathrm{ref}},
\end{align}
with the resistance $R$, the downstream reference pressure $p_{\mathrm{ref}}$, fluxes $q$, and the pressure $p_{\mathrm{o}}$. Due to the absence of compliance, Eq.~(\ref{eq:0d_res_flowb}) may well be omitted.

\subsubsection{Resistive-Compliant Link}\label{app:crlink}
For the example in Sec.~\ref{sec:examples_fluidpipe}, a 0D model consisting of two 2-element Windkessel models in series, connecting two regions of the 3D domain, is used. Its governing equations read:
\begin{align}
C_{\mathrm{in}}\frac{\mathrm{d}p_{\mathrm{i}}}{\mathrm{d}t} = q_{\mathrm{in}} - q_{\mathrm{d}}, \\
R_{\mathrm{in}}\,q_{\mathrm{d}} = p_{\mathrm{i}} - p_{\mathrm{d}}, \\
C_{\mathrm{out}}\frac{\mathrm{d}p_{\mathrm{d}}}{\mathrm{d}t} = q_{\mathrm{d}} - q_{\mathrm{out}}, \\
R_{\mathrm{out}}\,q_{\mathrm{out}} = p_{\mathrm{d}} - p_{\mathrm{o}},
\end{align}
where subscript $\mathrm{d}$ stands for ``distal'', and $\mathrm{in}$ and $\mathrm{out}$ refer to the in- and outflow with respect to the 0D model, receiving an out-flux of the 3D domain or linking to the inflow of the 3D domain, respectively.

\subsubsection{Closed-Loop Circulation Model}\label{app:syspul}

The closed-loop circulation model used to simulate aortic arch blood flow as well as left heart hemodynamics, Sec.~\ref{sec:examples_fluidaort} and Sec.~\ref{sec:examples_frsiheart}, respectively, is an extension of the model presented in \cite{hirschvogel2017}. We present the model in its general form for $n_{\mathrm{ven}}^{\mathrm{sys}}$ systemic and $n_{\mathrm{ven}}^{\mathrm{pul}}$ pulmonary veins as well as two coronary arteries. Its state variables are compartment fluxes $Q$, in-/out-fluxes $q$, and pressures $p$. It is parameterized by elastances $E$, compliances $C$, resistances $R$, impedances $Z$ (proximal resistances), and proximal (distal) inertances $I$ ($L$). Superscripts $\ell$, $r$, $\mathrm{sys}$, and $\mathrm{pul}$ refer to ``left'', ``right'', ``systemic'', and ``pulmonary'', respectively. Subscripts $\mathrm{v}$, $\mathrm{at}$, $\mathrm{ar}$, $\mathrm{ven}$, and $\mathrm{cor}$ refer to ``ventricular'', ``atrial'', ``arterial'', ``venous'', and ``coronary'', respectively. Secondary subscripts $\mathrm{in}$ ($\mathrm{i}$), $\mathrm{out}$ ($\mathrm{o}$), $\mathrm{d}$, and $\mathrm{p}$ refer to ``in-flux'' (``inlet''), ``out-flux'' (``outlet''), ``distal'', and ``proximal'', respectively. The governing equations read:

\begin{align}
&-Q_{\mathrm{at}}^{\ell} = \sum\limits_{j=1}^{n_{\mathrm{ven}}^{\mathrm{pul}}}q_{\mathrm{ven},j}^{\mathrm{pul}} - q_{\mathrm{v,in}}^{\ell} && \text{Left atrium flow balance}\\
&q_{\mathrm{v,in}}^{\ell} = q_{\mathrm{mv}}(p_{\mathrm{at,o}}^{\ell}-p_{\mathrm{v,i}}^{\ell}) && \text{Mitral valve momentum}\label{eq:0d_mv}\\
&-Q_{\mathrm{v}}^{\ell} = q_{\mathrm{v,in}}^{\ell} - q_{\mathrm{v,out}}^{\ell} && \text{Left ventricle flow balance}\\
&q_{\mathrm{v,out}}^{\ell} = q_{\mathrm{av}}(p_{\mathrm{v,o}}^{\ell}-p_{\mathrm{ar}}^{\mathrm{sys}}) && \text{Aortic valve momentum}\label{eq:0d_av}\\
&-Q_{\mathrm{aort}}^{\mathrm{sys}} = q_{\mathrm{v,out}}^{\ell} - q_{\mathrm{ar,p}}^{\mathrm{sys}} - \sum\limits_{j\in\{\ell,r\}} q_{\mathrm{cor,p,in}}^{\mathrm{sys},j} && \text{Aortic root flow balance}\\
&I_{\mathrm{ar}}^{\mathrm{sys}} \frac{\mathrm{d}q_{\mathrm{ar,p}}^{\mathrm{sys}}}{\mathrm{d}t} + Z_{\mathrm{ar}}^{\mathrm{sys}}\,q_{\mathrm{ar,p}}^{\mathrm{sys}}=p_{\mathrm{ar}}^{\mathrm{sys}}-p_{\mathrm{ar,d}}^{\mathrm{sys}} && \text{Aortic root inertia}\\
&C_{\mathrm{ar}}^{\mathrm{sys}} \frac{\mathrm{d}p_{\mathrm{ar,d}}^{\mathrm{sys}}}{\mathrm{d}t} = q_{\mathrm{ar,p}}^{\mathrm{sys}} - q_{\mathrm{ar}}^{\mathrm{sys}} && \text{Systemic arterial flow balance}\\
&L_{\mathrm{ar}}^{\mathrm{sys}} \frac{\mathrm{d}q_{\mathrm{ar}}^{\mathrm{sys}}}{\mathrm{d}t} + R_{\mathrm{ar}}^{\mathrm{sys}}\,q_{\mathrm{ar}}^{\mathrm{sys}}=p_{\mathrm{ar,d}}^{\mathrm{sys}}-p_{\mathrm{ven}}^{\mathrm{sys}} && \text{Systemic arterial momentum}\\
&C_{\mathrm{ven}}^{\mathrm{sys}} \frac{\mathrm{d}p_{\mathrm{ven}}^{\mathrm{sys}}}{\mathrm{d}t} = q_{\mathrm{ar}}^{\mathrm{sys}}-\sum\limits_{j=1}^{n_{\mathrm{ven}}^{\mathrm{sys}}}q_{\mathrm{ven},j}^{\mathrm{sys}}\ && \text{Systemic venous flow balance}\\
&L_{\mathrm{ven},j}^{\mathrm{sys}} \frac{\mathrm{d}q_{\mathrm{ven},j}^{\mathrm{sys}}}{\mathrm{d}t} + R_{\mathrm{ven},j}^{\mathrm{sys}}\, q_{\mathrm{ven},j}^{\mathrm{sys}} = p_{\mathrm{ven}}^{\mathrm{sys}} - p_{\mathrm{at,i},j}^{r} && \text{Systemic venous momentum}\\
&\qquad\qquad j \in \{1,...,n_{\mathrm{ven}}^{\mathrm{sys}}\}\nonumber && 
\end{align}

\begin{align}
&-Q_{\mathrm{at}}^{r} = \sum\limits_{j=1}^{n_{\mathrm{ven}}^{\mathrm{sys}}}q_{\mathrm{ven},j}^{\mathrm{sys}} - q_{\mathrm{v,in}}^{r} + q_{\mathrm{cor,d,out}}^{\mathrm{sys}} && \text{Right atrium flow balance}\\
&q_{\mathrm{v,in}}^{r} = q_{\mathrm{tv}}(p_{\mathrm{at,o}}^{r}-p_{\mathrm{v,i}}^{r}) && \text{Tricuspid valve momentum}\label{eq:0d_tv}\\
&-Q_{\mathrm{v}}^{r} = q_{\mathrm{v,in}}^{r} - q_{\mathrm{v,out}}^{r} && \text{Right ventricle flow balance}\\
&q_{\mathrm{v,out}}^{r} = q_{\mathrm{pv}}(p_{\mathrm{v,o}}^{r}-p_{\mathrm{ar}}^{\mathrm{pul}}) && \text{Pulmonary valve momentum}\label{eq:0d_pv}\\
&C_{\mathrm{ar}}^{\mathrm{pul}} \frac{\mathrm{d}p_{\mathrm{ar}}^{\mathrm{pul}}}{\mathrm{d}t} = q_{\mathrm{v,out}}^{r} - q_{\mathrm{ar}}^{\mathrm{pul}} && \text{Pulmonary arterial flow balance}\\
&L_{\mathrm{ar}}^{\mathrm{pul}} \frac{\mathrm{d}q_{\mathrm{ar}}^{\mathrm{pul}}}{\mathrm{d}t} + R_{\mathrm{ar}}^{\mathrm{pul}}\,q_{\mathrm{ar}}^{\mathrm{pul}}=p_{\mathrm{ar}}^{\mathrm{pul}} -p_{\mathrm{ven}}^{\mathrm{pul}} && \text{Pulmonary arterial momentum}\\
&C_{\mathrm{ven}}^{\mathrm{pul}} \frac{\mathrm{d}p_{\mathrm{ven}}^{\mathrm{pul}}}{\mathrm{d}t} = q_{\mathrm{ar}}^{\mathrm{pul}} - \sum\limits_{j=1}^{n_{\mathrm{ven}}^{\mathrm{pul}}}q_{\mathrm{ven},j}^{\mathrm{pul}} && \text{Pulmonary venous flow balance}\\
&L_{\mathrm{ven},j}^{\mathrm{pul}} \frac{\mathrm{d}q_{\mathrm{ven},j}^{\mathrm{pul}}}{\mathrm{d}t} + R_{\mathrm{ven},j}^{\mathrm{pul}}\, q_{\mathrm{ven},j}^{\mathrm{pul}}=p_{\mathrm{ven}}^{\mathrm{pul}}-p_{\mathrm{at,i},j}^{\ell} && \text{Pulmonary venous momentum}\\
&\qquad\qquad j \in \{1,...,n_{\mathrm{ven}}^{\mathrm{pul}}\}\nonumber && 
\end{align}

For heart chambers modeled as 0D time-varying elastance models \cite{sagawa1988}, the flux is expressed as
\begin{equation}
\begin{aligned}
Q_{c}^{k} = -\frac{\mathrm{d}V_{c}^{k}}{\mathrm{d}t}, \quad c \in\{\mathrm{at},\mathrm{v}\}, \; k \in\{\ell,r\},
\end{aligned}
\end{equation}
where the chamber volume is governed by
\begin{equation}
\begin{aligned}
V_{c}^{k}(t) = \frac{p_{c}^{k}}{E_{c}^{k}(t)} + V_{c,\mathrm{u}}^{k},
\end{aligned}
\end{equation}
with the unstressed volume $V_{\mathrm{u}}$ and the time-varying elastance
\begin{equation}
\label{equation-at-elast}
\begin{aligned}
E_{c}^{k}(t)=\left(E_{c,\mathrm{max}}^{k}-E_{c,\mathrm{min}}^{k}\right)\cdot \hat{y}_{c}^{k}(t)+E_{c,\mathrm{min}}^{k},
\end{aligned}
\end{equation}
where $E_{c,\mathrm{max}}^{k}$ and $E_{c,\mathrm{min}}^{k}$ denote the maximum and minimum elastance, respectively. The normalized activation function $\hat{y}_{c}^{k}(t)$ is input by the user. Further, chamber pressures at the inlet (``$\mathrm{i}$'') coincide with the ones at the outlet (``$\mathrm{o}$''), hence $p_{c,\mathrm{i}}^{k} \equiv p_{c,\mathrm{o}}^{k} =: p_{c}^{k}$.\\

Valve laws for the mitral, aortic, tricuspid, and pulmonary valve, Eq.~(\ref{eq:0d_mv}), Eq.~(\ref{eq:0d_av}), Eq.~(\ref{eq:0d_tv}), and Eq.~(\ref{eq:0d_pv}), respectively, here are governed by the following piecewise-linear pressure-flow relationship:
\begin{equation}
\begin{aligned}
q_{i}(p_{i}-p_{\mathrm{open},i}) := \frac{p_{i}-p_{\mathrm{open},i}}{\tilde{R}_{i}}, \quad \text{with}\; \tilde{R}_{i} = \begin{cases} R_{i,\max}, & p_{i} < p_{\mathrm{open},i} \\
R_{i,\min}, & p_{i} \geq p_{\mathrm{open},i} \end{cases}, \qquad i\in\{\mathrm{mv},\mathrm{av},\mathrm{tv},\mathrm{pv}\}.
\end{aligned}
\end{equation}

In case of heart chambers that are isolated from any 3D-0D coupling boundary, i.e. the left ventricle domain from the example in Sec.~\ref{sec:examples_frsiheart}, mitral and aortic valve laws Eq.~(\ref{eq:0d_mv}) and Eq.~(\ref{eq:0d_av}) become obsolete and fluxes $q_{\mathrm{v,in}}^{\ell}$ and $q_{\mathrm{v,out}}^{\ell}$ are instead computed from the valve planes in the 3D domain. An explicit update of these fluxes (hence from the previous converged time step) is favored here for the sake of algorithmic convenience.

\subsubsection{Coronary Circulation}\label{app:coronary}

The coronary circulation consists of both a left and a right coronary branch, each comprised of a (proximal) 3-element Windkessel model \cite{westerhof2009} in series with a (distant) 2-element Windkessel. The distant model, thereby, receives the difference in distal and left ventricular pressure, cf. \cite{arthurs2016}, in order to reproduce the typical heart phase-dependent compliance of the coronary vessels. These exhibit the majority of perfusion during ventricular diastole but are compressed in ventricular systole \cite{barral2011}, hence become less compliant. The coronary circulation governing equations read:
\begin{align}
&C_{\mathrm{cor,p}}^{\mathrm{sys},\ell} \left(\frac{\mathrm{d}p_{\mathrm{ar}}^{\mathrm{sys},\ell}}{\mathrm{d}t}-Z_{\mathrm{cor,p}}^{\mathrm{sys},\ell}\frac{\mathrm{d}q_{\mathrm{cor,p,in}}^{\mathrm{sys},\ell}}{\mathrm{d}t}\right) = q_{\mathrm{cor,p,in}}^{\mathrm{sys},\ell} - q_{\mathrm{cor,p}}^{\mathrm{sys},\ell} && \text{Left coronary proximal flow balance}\\
&R_{\mathrm{cor,p}}^{\mathrm{sys},\ell}\,q_{\mathrm{cor,p}}^{\mathrm{sys},\ell}=p_{\mathrm{ar}}^{\mathrm{sys},\ell}-p_{\mathrm{cor,d}}^{\mathrm{sys},\ell} - Z_{\mathrm{cor,p}}^{\mathrm{sys},\ell}\,q_{\mathrm{cor,p,in}}^{\mathrm{sys},\ell} && \text{Left coronary proximal momentum}\\
&C_{\mathrm{cor,d}}^{\mathrm{sys},\ell} \frac{\mathrm{d}(p_{\mathrm{cor,d}}^{\mathrm{sys},\ell}-p_{\mathrm{v,o}}^{\ell})}{\mathrm{d}t} = q_{\mathrm{cor,p}}^{\mathrm{sys},\ell} - q_{\mathrm{cor,d}}^{\mathrm{sys},\ell} && \text{Left coronary distal flow balance}\label{eq:0d_lcd_flow}\\
&R_{\mathrm{cor,d}}^{\mathrm{sys},\ell}\,q_{\mathrm{cor,d}}^{\mathrm{sys},\ell}=p_{\mathrm{cor,d}}^{\mathrm{sys},\ell}-p_{\mathrm{at,i}}^{r} && \text{Left coronary distal momentum}\\
&C_{\mathrm{cor,p}}^{\mathrm{sys},r} \left(\frac{\mathrm{d}p_{\mathrm{ar}}^{\mathrm{sys},r}}{\mathrm{d}t}-Z_{\mathrm{cor,p}}^{\mathrm{sys},r}\frac{\mathrm{d}q_{\mathrm{cor,p,in}}^{\mathrm{sys},r}}{\mathrm{d}t}\right) = q_{\mathrm{cor,p,in}}^{\mathrm{sys},r} - q_{\mathrm{cor,p}}^{\mathrm{sys},r} && \text{Right coronary proximal flow balance}\\
&R_{\mathrm{cor,p}}^{\mathrm{sys},r}\,q_{\mathrm{cor,p}}^{\mathrm{sys},r}=p_{\mathrm{ar}}^{\mathrm{sys},r}-p_{\mathrm{cor,d}}^{\mathrm{sys},r} - Z_{\mathrm{cor,p}}^{\mathrm{sys},r}\,q_{\mathrm{cor,p,in}}^{\mathrm{sys},r} && \text{Right coronary proximal momentum}\\
&C_{\mathrm{cor,d}}^{\mathrm{sys},r} \frac{\mathrm{d}(p_{\mathrm{cor,d}}^{\mathrm{sys},r}-p_{\mathrm{v,o}}^{\ell})}{\mathrm{d}t} = q_{\mathrm{cor,p}}^{\mathrm{sys},r} - q_{\mathrm{cor,d}}^{\mathrm{sys},r} && \text{Right coronary distal flow balance}\label{eq:0d_rcd_flow}\\
&R_{\mathrm{cor,d}}^{\mathrm{sys},r}\,q_{\mathrm{cor,d}}^{\mathrm{sys},r}=p_{\mathrm{cor,d}}^{\mathrm{sys},r}-p_{\mathrm{at,i}}^{r} && \text{Right coronary distal momentum}\\
&0=q_{\mathrm{cor,d}}^{\mathrm{sys},\ell}+q_{\mathrm{cor,d}}^{\mathrm{sys},r}-q_{\mathrm{cor,d,out}}^{\mathrm{sys}} && \text{Distal coronary junction flow balance}
\end{align}

In case the left ventricle is isolated from any 3D-0D coupling boundary, cf. the example in Sec.~\ref{sec:examples_frsiheart}, the 0D model does not provide any governing equation for its pressure $p_{\mathrm{v,o}}^{\ell}$, since Eq.~(\ref{eq:0d_mv}) and Eq.~(\ref{eq:0d_av}) are replaced by prescribed flows from the 3D domain, cf. Sec.~\ref{app:syspul}. However, since distal coronary flow balances, Eq.~(\ref{eq:0d_lcd_flow}) and Eq.~(\ref{eq:0d_rcd_flow}), need to be informed of left ventricular pressure, we provide the fluid pressure averaged over the ventricular aortic outflow plane $\Gm_{i}(\tilde{\Om}_{u})$ to the 0D system, cf. Eq.~(\ref{eq:pbar_valve})$_{\text{1}}$. For the sake of algorithmic convenience, an explicit update of $p_{\mathrm{v,o}}^{\ell}$ (hence from the previous converged time step) from the 3D world is performed.

\subsection{Valves Immersed in 3D Fluid Space}\label{app:3dvalve}
Immersed resistive valve models for computational fluid dynamics of the heart are numerous \cite{zingaro2023,this2020,tagliabue2017,takizawa2014,griffith2012}, where most approaches have in common that natural or essential boundary conditions are imposed at an interface between the left ventricle and left atrium (mitral valve) or left ventricle and aortic outflow tract (aortic valve). There, a discontinuous representation of the fluid pressure field has to be accounted for, which either can be inherent to the numerical approximation method (discontinuous Galerkin methods \cite{cockburn2017}) or has to be considered by means of split domains across these interfaces (if continuous Lagrange finite element approximations are used). These boundary conditions may be time-controlled or driven by a pressure jump across that interface.\\

We denote the pressure jump across the mitral (``$\mathrm{mv}$'') or aortic valve (``$\mathrm{av}$'') plane as follows:
\begin{align}
\Delta\bar{p}_{i} = \bar{p}_{i} - \bar{p}_{i,\mathrm{open}}, \quad i \in \{\mathrm{mv},\mathrm{av}\}, \label{eq:deltap_valve}
\end{align}
with the surface-averaged pressures
\begin{align}
\qquad \bar{p}_{i} = \frac{1}{\int_{\Gm_{i}(\tilde{\Om}_{u})} \mathrm{d}A}\;\int\limits_{\Gm_{i}(\tilde{\Om}_{u})} p\, \mathrm{d}A, \qquad \bar{p}_{i,\mathrm{open}} = \frac{1}{\int_{\Gm_{i}(\tilde{\Om}_{d})} \mathrm{d}A}\;\int\limits_{\Gm_{i}(\tilde{\Om}_{d})} p\, \mathrm{d}A,\label{eq:pbar_valve}
\end{align}
where $u$ and $d$ refer to the up- and downstream compartments (mitral valve: $u = \mathrm{LA}$, $d = \mathrm{LV}$, aortic valve: $u = \mathrm{LV}$, $d = \mathrm{AO}$), respectively. Here, the downstream pressure is denoted by subscript ``$\mathrm{open}$'', indicating the pressure threshold to be surpassed in order to open the valve.\\

The valves then are controlled by a Robin condition, Eq.~(\ref{eq:fluid_generic_neumann_robin})$_{\text{2}}$ for Eulerian or Eq.~(\ref{eq:fluidale_generic_robin}) for ALE fluid dynamics, where the resistive coefficient depends on the pressure difference across the valve, Eq.~(\ref{eq:deltap_valve}), as follows:
\begin{align}
c_{i}(\Delta\bar{p}_{i}) := \begin{cases} c_{i,\max}, & \mathrm{if}\;\Delta\bar{p}_{i} > 0, \\ c_{i,\min}, & \mathrm{if}\;\Delta\bar{p}_{i} \leq 0. \end{cases} \label{eq:c_valve}
\end{align}
For all valves and models reported, we use $c_{i,\max}=10^3\;\frac{\mathrm{kPa}\cdot\mathrm{s}^2}{\mathrm{mm}}$ and $c_{i,\min}=10^{-3}\;\frac{\mathrm{kPa}\cdot\mathrm{s}^2}{\mathrm{mm}}$. In this study, for the sake of algorithmic efficiency, we discretize the immersed valve law in an explicit manner. Hence, the value of the previous, converged time step of Eq.~(\ref{eq:deltap_valve}) is used to evaluate Eq.~(\ref{eq:c_valve}).\\

\subsection{Construction of Wall Motion Space for Left Heart FrSI}\label{app:frsi_heart_pod}

\subsubsection{Imaging Snapshots and POD}\label{app:snap_pod}

\begin{figure}[!htp]
\centering
\includegraphics[width=1.0\textwidth]{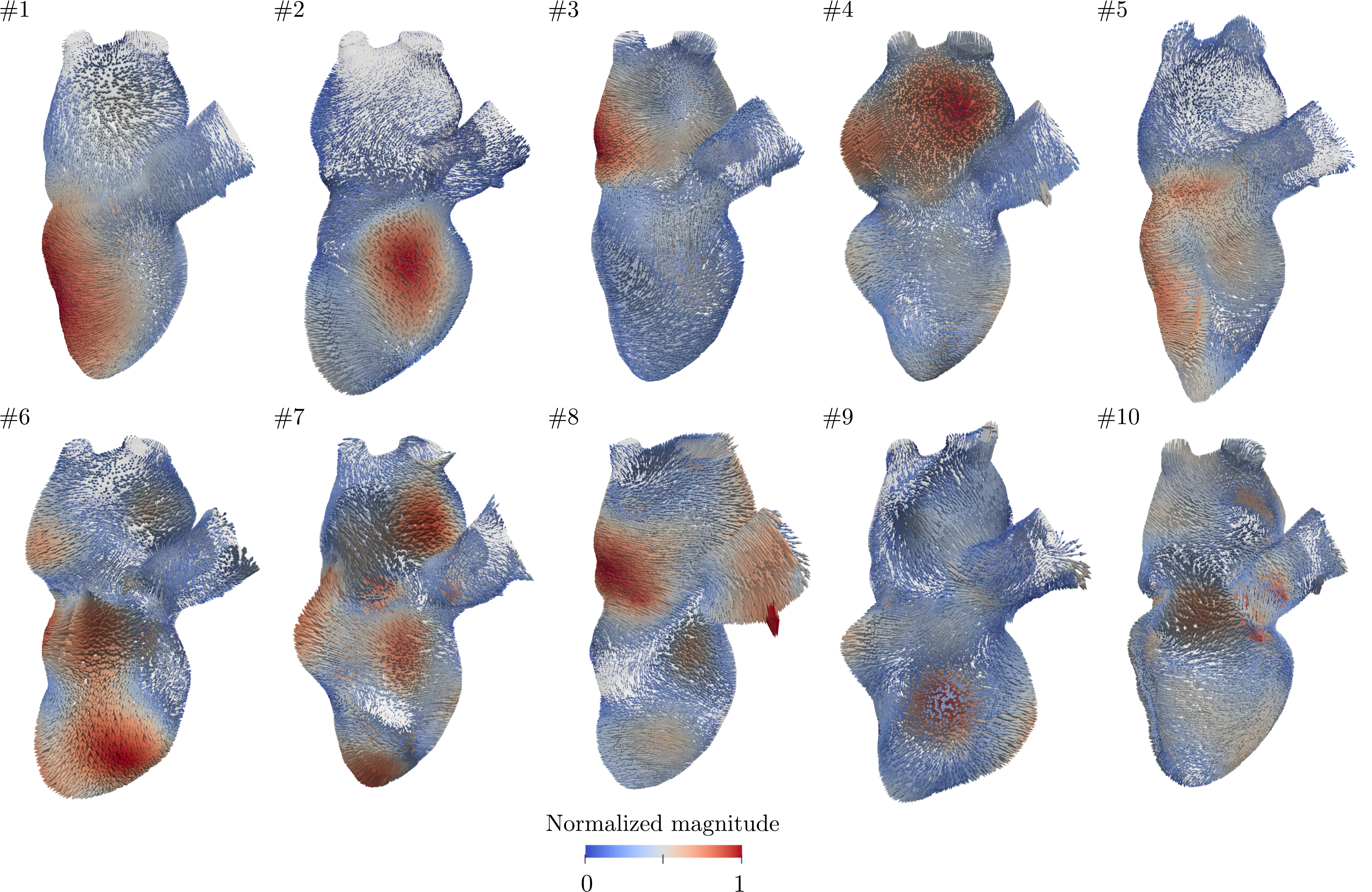}
\caption{First 10 POD modes of deformation states retrieved from a time series of imaging data.}\label{fig:lalvao_modes}
\end{figure}

Modeling blood flow in the left heart using the FrSI method \cite{hirschvogel2024-frsi} described in Sec.~\ref{subsec:frsi} requires information about the motion of the endocardial heart wall in order to construct a suitable space for projection. Here, we use time-resolved data retrieved from CT imaging that provides a time series of 19 wall motion states over one cardiac cycle. These states subsequently are compressed and decomposed by means of Proper Orthogonal Decomposition (POD) \cite{rathinam2003}, yielding a set of $r_{v}$ POD modes that are used for projecting the high-dimensional boundary model to a lower-dimensional space. Here, we consider the first $r_{v}=10$ POD modes, shown in Fig.~\ref{fig:lalvao_modes} sorted in descending order with respect to their relative energy content.

\subsubsection{Region-wise Locality of POD Subspace}\label{app:reg_local_pod}

\begin{figure}[!htp]
\centering
\includegraphics[width=0.8\textwidth]{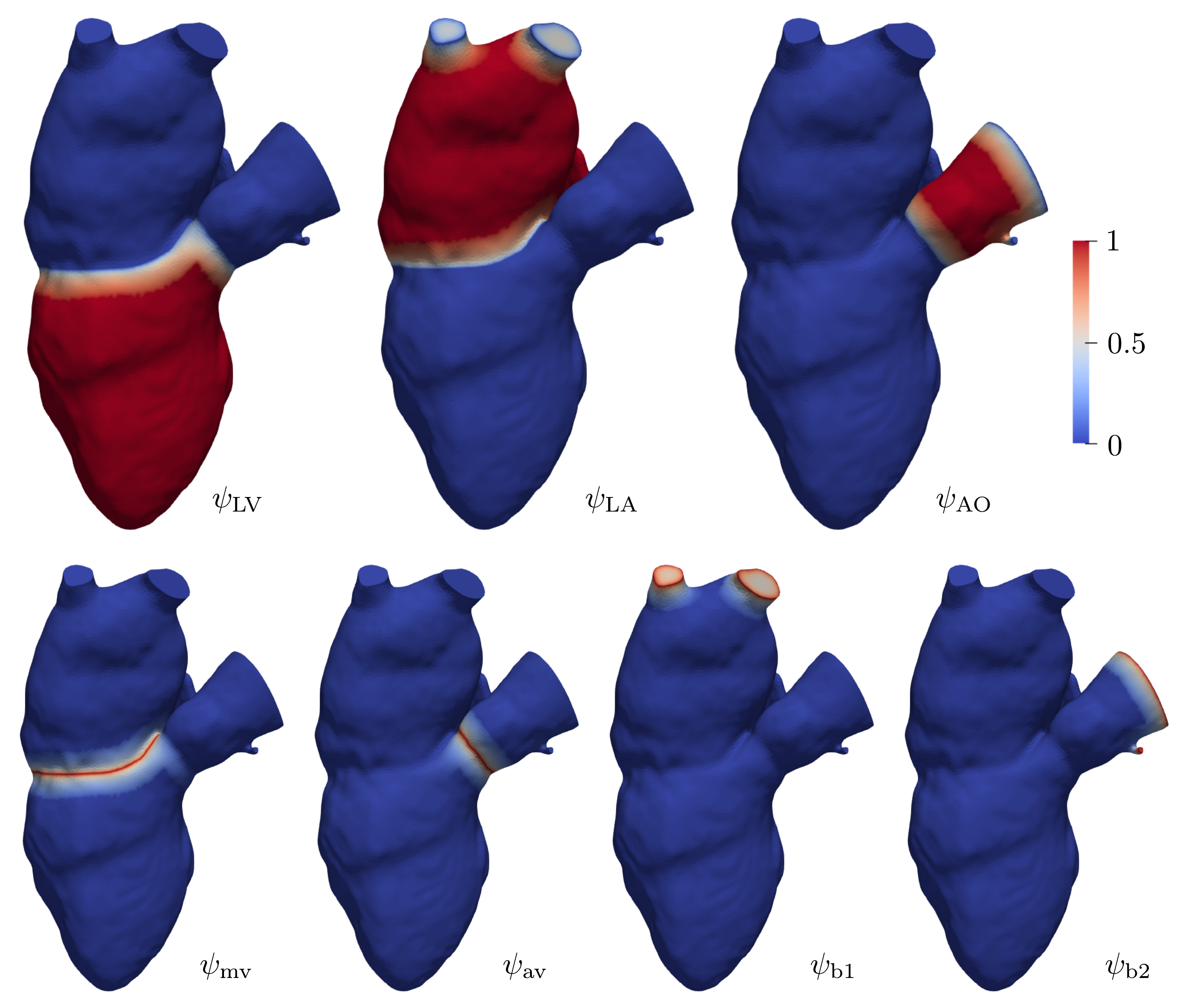}
\caption{Partition of unity decomposition of POD subspace for LA-LV-AO model.}\label{fig:lalvao_partitions}
\end{figure}

The global nature of the modes constructed as described in Sec.~\ref{app:snap_pod} may impose additional constraints on the model's kinematics and may limit adaptive wall motion in response to a change in hemodynamic loads. Therefore, we propose a method to decompose the POD subspace such that the left atrium, the left ventricle, and the aortic outflow tract can experience independent dynamics. By means of a partition of unity approach, we provide regionally local POD spaces for each of the three wall regions, and additionally at their junctions (mitral and aortic valve rims) as well as at the rims of pulmonary veins and aortic outflows. These partitions are shown in Fig.~\ref{fig:lalvao_partitions}. Hence, here, we have $n_{\mathrm{loc}}=7$ local POD subspaces, yielding a total of $n_{\mathrm{loc}} \cdot r_{v} = 70$ degrees of freedom for the reduced wall model.

Furthermore, we use the partition of unity fields to smoothly translate between the different wall thicknesses of atrium, ventricle, and aortic arch, assuming that the wall is thicker around the rims of mitral and aortic valve:
\begin{align}
    & h_0^{\mathrm{LA}} = h_{0,\mathrm{b}}^{\mathrm{LA}} (1 - \psi_{\mathrm{mv}}) + h_{0,\mathrm{v}} \psi_{\mathrm{mv}}, \label{eq:h0la}\\
    & h_0^{\mathrm{LV}} = h_{0,\mathrm{b}}^{\mathrm{LV}} (1 - \psi_{\mathrm{mv}}) + h_{0,\mathrm{v}} \psi_{\mathrm{mv}} + h_{0,\mathrm{v}} \psi_{\mathrm{av}} - h_{0,\mathrm{b}}^{\mathrm{LV}} \psi_{\mathrm{av}}, \quad\text{and} \label{eq:h0lv}\\
    & h_0^{\mathrm{AO}} = h_{0,\mathrm{b}}^{\mathrm{AO}} (1 - \psi_{\mathrm{av}}) + h_{0,\mathrm{v}} \psi_{\mathrm{av}},\label{eq:h0ao}
\end{align}
with the wall thickness at the valve rims $h_{0,\mathrm{v}} = 20 \,\mathrm{mm}$ and the base thicknesses $h_{0,\mathrm{b}}^{i}$.

\subsubsection{Boundary Conditions in Reduced Space}\label{app:bc_pod}
Suitable boundary conditions that mimic the mechanical tissue properties of adjacent portions of the vascular beds truncated from the model region of interest have to be accounted for. Here, these would be at the upstream locations of the pulmonary veins or at locations downstream, i.e. the aortic outflow tract or the coronary arteries. In order to localize the effect of the adjacent tissue, Robin conditions are imposed in the POD space to the modes associated to the pulmonary veins and aortic root rims. Hence, the following term is added to the first row of block vector Eq.~(\ref{eq:res_nonlin_frsi}):
\begin{align}
    \hdots + \bsf{C}_{\mathrm{R}} \tilde{\bsf{v}} + \bsf{K}_{\mathrm{R}} (\tilde{\bsf{u}}_{\fF} + \tilde{\bsf{u}}_{\mathrm{pre}}),
\end{align}
where $\bsf{C}_{\mathrm{R}}, \bsf{K}_{\mathrm{R}} \in \mathbb{R}^{(r_v + n_{v}^{\Om})\times(r_v + n_{v}^{\Om})}$ are diagonal damping and stiffness matrices, with entries $C_{\mathrm{R},ii}, K_{\mathrm{R},ii} \neq 0$ at locations that link to the subset of modes belonging to the pulmonary vein and aortic outflow boundary partitions defined by $\psi_{\mathrm{b1}}$ and $\psi_{\mathrm{b2}}$, respectively, cf. Fig. \ref{fig:lalvao_partitions}. Vectors $\tilde{\bsf{u}}_{\fF}$ and $\tilde{\bsf{u}}_{\mathrm{pre}}$ represent the projected discrete fluid and prestress displacements, respectively, cf. the space-continuous time-discrete counterpart Eq.~(\ref{eq:ufluid_discr}). Pulmonary vein and aortic boundaries are damped by $0.05\;\frac{\mathrm{mN}}{\mathrm{mm}}$ and $0.5\;\frac{\mathrm{mN}}{\mathrm{mm}}$, and the stiffnesses are $1\;\frac{\mathrm{mN}}{\mathrm{mm}\cdot\mathrm{s}}$ and $10\;\frac{\mathrm{mN}}{\mathrm{mm}\cdot\mathrm{s}}$, respectively.

\subsection{Meshes}\label{app:meshes}
For the sake of reproducibility, we share all meshes (of pipe, aorta, and heart---each in rf0, rf1, and rf2 discretization, cf. Fig.~\ref{fig:meshes_dofs_all}) that are used throughout this paper. We use the \verb|.xdmf/.h5| file format and separately store domain and boundary meshes (\verb|*_domain.xdmf/.h5|, \verb|*_boundary.xdmf/.h5|) for easier visualization. The elements of all relevant subdomains and boundaries are labeled with respective attributes such that they can be viewed and used in a straightforward manner.

\clearpage
\bibliography{ref}

\end{document}